\documentclass[11pt,a4paper]{article}
\usepackage{jheppub}
\usepackage{color}
\usepackage{amssymb,amsmath,bm}
\usepackage{epsf}
\usepackage{epsfig}
\usepackage{afterpage}
\usepackage{longtable}
\usepackage{graphics}
\usepackage{url}
\usepackage{paralist}
\usepackage{subcaption}
\usepackage{float}
\usepackage{booktabs}
\usepackage[table]{xcolor}

\usepackage{adjustbox}
\newcommand{\opL}[3][{}]{{\mathcal{O}_{\! \underset{#3}{ #2}}^{{\tiny #1}}} }
\newcommand{\wcL}[3][{}]{{C_{\! \underset{#3}{ #2}}^{{\tiny #1}}} }

\usepackage{fancyhdr}
\advance \headheight by 3.0truept       % for 12pt mandatory...
\definecolor{red}{cmyk}{0,1,1,0.4}

\newcommand{\beq}{\begin{equation}}
\newcommand{\eeq}{\end{equation}}
\newcommand{\be}{\begin{equation}}
\newcommand{\ee}{\end{equation}}
\newcommand{\bi}{\begin{itemize}}
\newcommand{\ei}{\end{itemize}}
\newcommand{\ba}{\begin{array}}
\newcommand{\ea}{\end{array}}
\newcommand{\beqa}{\begin{eqnarray}}
\newcommand{\eeqa}{\end{eqnarray}}
\newcommand{\bea}{\begin{eqnarray}}
\newcommand{\eea}{\end{eqnarray}}
\newcommand{\beqn}{\begin{eqnarray}}
\newcommand{\eeqn}{\end{eqnarray}}

\newcounter{TODO}

\newcommand{\wc}[3][{}]{\big[{C}_{#2}^{#1}\big]_{#3}}

\newcommand{\op}[3][{}]{[{ O}_{#2}^{#1}]_{#3}}

{\boldmath
\title{Electric Dipole Moments in 5+3 Flavor Weak Effective Theory}
}
\author{Jacky Kumar,}
\author{Emanuele Mereghetti}

\affiliation{
Theoretical Division, Los Alamos National Laboratory, Los Alamos, NM 87545, USA}

\emailAdd{jacky.kumar@lanl.gov}
\emailAdd{emereghetti@lanl.gov}

\begin{document}

\begin{flushleft}
{\em Version of \today}
\end{flushleft}

\vspace{-14mm}
\begin{flushright}
  {LA-UR-24-20122}
\end{flushright}

\abstract{
A fully generic treatment of electric dipole moments (EDMs) is presented in  
the CP-violating and flavor-conserving weak effective field theory (WET) with five flavors
of quarks and three flavors of leptons.
We systematically analyze leading contributions to EDMs originating 
from QCD and QED renormalization group running between 
the electroweak scale and low energy scales of about 2 GeV.
We include the full one-loop anomalous dimension and a subset of two-loop corrections,
as well as threshold corrections at the bottom, charm and $\tau$ masses. 
This allows us to derive master formulae in the space of generic WET for the neutron and proton EDMs,
for EDMs of diamagnetic atoms, and for the precession frequencies constrained in molecular EDM experiments, from which 
bounds on the electron EDM are extracted. 
In particular, our master formulae capture the contributions of WET CP-violating operators with heavy quark and lepton flavors.
As an application, we study EDM constraints on the Yukawa couplings of the 
Higgs boson, in both the linear and non-linear realizations of electroweak 
symmetry breaking. 
}

\maketitle

%\tableofcontents

%\newpage
\section{Introduction}
\label{sec:1}
The origin of the baryon asymmetry in the Universe (BAU) is one of the most pressing open problems in particle physics.
While the Standard Model (SM) satisfies the three Sakharov conditions \cite{Sakharov:1967dj} for the dynamical generation of an asymmetry, for a Higgs mass of $m_h= 125$ GeV the electroweak (EW) phase transition does not provide sufficient deviation from thermal equilibrium to explain the observed BAU.  
Even if the SM were able to induce a first-order phase transition, 
the violation of the symmetry under a charge conjugation and parity (CP) transformation induced by the phase of the Cabibbo-Kobayashi-Maskawa (CKM) matrix would be far too weak, underpredicting the BAU by several orders of magnitude \cite{Gavela:1994ds,Gavela:1993ts,Gavela:1994dt,Huet:1994jb}.
The generation of the BAU thus quite generally requires physics beyond the Standard Model (BSM) to have new sources of CP violation. 
Permanent electric dipole moments (EDMs) of leptons, nucleons, atoms, and molecules are extremely sensitive probes 
of flavor-diagonal CP-violation, as they can be constrained very accurately and receive negligible contributions from the phase of the CKM matrix. For example, the current bound on the neutron EDM is
$|d_n| < 1.8 \cdot 10^{-26}$ $e$ cm, naively probing BSM physics at scales far above 10 TeV. On the other hand,  this bound 
is five to six orders of magnitude away from the contribution of the SM \cite{Khriplovich:1985jr,Czarnecki:1997bu,Mannel:2012qk,Seng:2014lea,Pospelov:2005pr}. 
An observation of a neutron EDM in the next generation of experiments would then be a clear indication of CP violation beyond the CKM paradigm, and observations in multiple systems will be able to rule out a QCD $\bar\theta$ term and demonstrate the existence of BSM CP violation.
For this reason, a rich experimental program is in place, with ongoing and planned searches in different systems, from the muon and tau leptons to the neutron, 
to atomic and molecular systems, aiming at improving current bounds by at least one or two orders of magnitude \cite{Chupp:2017rkp,Alarcon:2022ero}. 

Because of the strength of existing and upcoming constraints, it is important to address 
the question of what existing EDM experiments teach us about CP violation in generic BSM scenarios.
In addition, assuming multiple observations of non-zero EDMs in the next generation of experiments, 
we will be faced with the ``inverse problem'' of identifying the fundamental mechanism
of CP violation from a set of low-energy measurements. 
Effective Field Theories (EFT) provide the theoretical tools to answer these questions with minimal reliance on specific models of BSM physics \cite{Buras:2020xsm}.
First of all, EFTs such as  the Weak  EFT (WET) \cite{Jenkins:2017jig, Jenkins:2017dyc}
allow one to organize all possible flavor-diagonal ($\Delta F=0$) CP-violating operators of a given dimension. 
Secondly, in WET it is possible to systematically consider loop effects, such as 
renormalization group (RG) evolution and matching corrections induced by heavy quark thresholds,
and to connect low-scale observables with physics at the electroweak scale.
Here one can match WET onto other EFTs,  such as the Standard Model EFT (SMEFT) \cite{Buchmuller:1985jz,Grzadkowski:2010es} or the Higgs EFT (HEFT) \cite{Buchalla:2013rka,Buchalla:2017jlu}, and study collider phenomenology.  This tower of EFTs provides a general and model-independent link between EDMs, other precision flavor observables, and collider experiments, which will be necessary to solve the aforementioned inverse problem.

The advantages of parameterizing EDMs in terms of a minimal set of low-energy operators have long been recognized 
\cite{Pospelov:2005pr,Khriplovich:1997ga}, and there is a vast literature on EDM calculations in specific BSM models 
(see \cite{Pospelov:2005pr,Engel:2013lsa,Chupp:2017rkp} and references therein).
In EFTs,  EDMs in the SMEFT at tree level and the matching onto hadronic EFTs were considered in Ref. 
\cite{deVries:2012ab}. Beyond the tree level, the contributions to EDMs induced  by classes of SMEFT operators, including in the 
 Yukawa \cite{Brod:2013cka,Chien:2015xha, Buras:2020xsm,Brod:2022bww,Bahl:2022yrs,Fuchs:2020uoc,Brod:2023wsh}, top  \cite{Cirigliano:2016nyn,Fuyuto:2017xup}, 
 heavy flavor \cite{Ema:2022pmo,Haisch:2021hcg},
Higgs-gauge  \cite{Dekens:2013zca,Cirigliano:2019vfc,Haisch:2019xyi}, right-handed charged current \cite{Alioli:2017ces}
and flavor \cite{Endo:2019mxw,Fajfer:2023gie} sectors, have been studied in a mostly piecemeal approach. 
A more systematic analysis of the 1-loop contributions to EDMs in SMEFT was carried out in Ref. \cite{Kley:2021yhn},
but mostly with a focus on the electron and nucleon EDMs. This paper is a first step towards a complete study of $\Delta F=0$ CP violation in EFT extensions of the SM,
and its correlations with other sectors of the theory (such as the Higgs and flavor sectors).
We focus here on CP-violating (CPV) $\Delta F=0$ WET operators, with $n_f=5$ quark flavors and $n_\ell = 3$ lepton flavors.
The main new results of this work are:
\begin{itemize}
\item We derive master formulae for the neutron, proton, and diamagnetic EDMs, and for the precession frequencies in molecular EDM experiments, in terms of the full
set of $\Delta F=0$ WET operators at the electroweak scale. 
The master formulae for $\omega_{\rm HfF}$, and  $\omega_{\rm ThO}$,  $\omega_{\rm YbF}$
are given in Tables \ref{tab:masterHfF} and \ref{tab:masterThOYbF}, respectively.
Contributions to $d_{\rm Hg}$ are listed in Tables \ref{tab:masterdHg1} and \ref{tab:masterdHg2}, 
while the neutron and proton EDMs, $d_n$ and $d_p$,  are given in Tables \ref{tab:masterdn} and \ref{tab:masterdp}, respectively.
\item The expressions use up-to-date hadronic and nuclear matrix elements, which are however still affected by large and often uncontrolled theoretical uncertainties. 
For hadronic matrix elements, including the neutron and proton EDMs and CPV pion-nucleon couplings, we rely as much as possible on information from Lattice QCD,
using either direct EDM calculations or indirect calculations (such as Lattice QCD calculations of mesonic matrix elements which contribute to pion-nucleon couplings).
The implementation of the master formulae in the package \texttt{flavio} \cite{Straub:2018kue} allows for a quick implementation of new matrix elements, as soon as they become available.
\item For  WET operators involving only the $u$, $d$ and $s$ quark flavors ($n_f=3$) and the electron ($n_\ell =1$), our formulae have leading logarithmic accuracy.
We present master formulae in the Jenkins, Manohar, and Stoffer (JMS) basis for WET \cite{Jenkins:2017jig, Jenkins:2017dyc},
which can be easily connected to the SMEFT.
\item We consider the contributions of operators with heavy quark ($c$ and $b$)
and lepton ($\mu$ and $\tau$) flavors, where renormalization
group mixing and matching at the heavy fermion thresholds are crucial.
In the case of  scalar four-quark operators, all operators  with two heavy and two light quarks 
contribute to EDMs at leading logarithmic accuracy. 
Scalar operators with four heavy quarks do not contribute at the leading log.  
In this case, we identify next-to-leading logarithmic contributions, arising from 1-loop running and 1-loop matching at the heavy flavor threshold. Since the renormalization group mixing is mostly driven by QCD,  
$d_n$ and $d_{\rm Hg}$ provide strong constraints even on scalar operators with only heavy quarks.
\item 
Four-quark vector operators with two heavy and two light quarks
also contribute at the next-to-leading log. In this case, the dominant contributions arise from 2-loop running and
1-loop matching. 
Four-quark vector operators with two $b$ and two $c$ quarks contribute at the next-to-next-to-leading log.
EDM contributions from scalar and vector operators with the same flavors are comparable.
\item Leptonic operators with two heavy leptons and two electrons 
contribute to the electron EDM at the leading log and next-to-leading log for scalar and vector operators, respectively.
We also find that all semileptonic operators with two muons or $\tau$ leptons contribute to EDMs of diamagnetic atoms
at the order we are working. Since the mixing is driven by QED and proportional to the lepton masses, however, some of the constraints are weak,
especially for muonic operators.
\end{itemize}
As we illustrate in Section \ref{sec:Higgs},  the WET master formulae can be easily adapted to EFTs at the electroweak scale, by calculating the matching coefficients in SMEFT, HEFT, or in any given BSM model.

The paper is organized as follows. In Section \ref{sec:wet} we introduce the WET basis  of $\Delta F=0$ 
CPV operators. In Section \ref{sec:edms} we provide expressions for the neutron, atomic, and molecular EDMs in terms of WET operators at low energy. 
In Section \ref{sec:runmatch} we discuss the renormalization group evolution and threshold effects and in Section \ref{sec:master} we provide master formulae for EDMs in terms of WET coefficients at the electroweak scale. In Section \ref{sec:Higgs} we discuss one example of application of the master formulae, EDM constraints on non-standard Higgs couplings. We conclude in Section \ref{sec:concl}.

\section{Flavor-conserving CP-violation in WET}
\label{sec:wet}
There are numerous sources of CP-violation in WET at the dim-4, 5, and 6  
levels. Since our main goal is to study EDMs, we will restrict ourselves to flavor-conserving WET operators.
We can identify two kinds of contributions. Operators constructed out of light fermion flavors, like $u, d$, or $s$ quarks or electrons,
generate EDMs at the tree level. As we will see, they will receive the strongest constraints.
Operators that contain heavier quark and lepton generations 
can also contribute via loop effects originating 
either from the renormalization group running or from threshold effects. 
Below we consider the full list of WET operators with five quark and three lepton flavors that can contribute 
to EDMs.
Very schematically, WET Wilson coefficients at low energy are obtained by solving renormalization group equations (RGEs), which give rise to expressions of the form
\begin{equation}\label{eq:RGEsolution}
C(\mu) =\left[ \sum_{n=0} \left(\frac{g^2}{(4\pi)^2}\right)^n \log^n \frac{\mu}{\mu_{ew}}    + \sum_{n=1} \left(\frac{g^2}{(4\pi)^2}\right)^n \log^{n-1} \frac{\mu}{\mu_{ew}} + \ldots \right] \times   C(\mu_{ew}),  
\end{equation}
where $g$ here stands for any coupling in WET, and $\mu_{ew}$ is a scale close to the electroweak.
The first and second terms in the bracket on the r.h.s of Eq. \eqref{eq:RGEsolution} denote the leading logarithmic (LL) and next-to-leading logarithmic (NLL) series. 
As we will discuss, the great majority of $\Delta F=0$ CPV operators
feeds into light flavor operators at LL.
These contributions are captured by the solution of the WET 1-loop 
RGEs, which are known \cite{Buchalla:1995vs, Aebischer:2015fzz, Jenkins:2017dyc}. 
Certain classes of operators, including, for example, vector left-right operators, do not contribute at LL. In these cases, we identify NLL corrections, arising from 2-loop running and 1-loop matching at the heavy quark threshold.
{
A summary of the order in which each WET operator contributes is given in Table \ref{tab:wet4},
where operators contributing at LL, NLL, and next-to-next-to-leading log (NNLL) are indicated by no asterisk, or single and double asterisks, respectively.
We stress that, while the logarithmic series expansion in Eq. \eqref{eq:RGEsolution} provides useful guidance 
to organize WET contributions to EDMs, and we will therefore use it throughout the paper, the RGEs involve small couplings 
(e.g. the QED coupling), for which the suppression from additional loops is not fully offset by the large logarithms. In these cases, 
it is possible that higher loop corrections to matching and running will be more important than those included here. 
These can only be captured by pushing the calculations of WET anomalous dimensions and matching corrections to higher order.}
For the sake of generality, we also include the operators that cannot be generated 
in SMEFT at leading order, i.e. the ones that violate the hypercharge quantum number.

In the JMS basis  complete 1-loop 
anomalous dimension matrices (ADMs) are known \cite{Jenkins:2017jig, Jenkins:2017dyc}.
Moreover,  the complete 1-loop matching conditions of WET onto 
SMEFT has been recently worked out  \cite{Dekens:2019ept} in the JMS basis. 
Given these developments,  it is useful to express the EDMs in terms of the 
 Wilson coefficients (WCs) of the JMS basis so that their connection to UV physics can be easily established.
The translation of the JMS WCs into several other popular WET bases or vice versa can be numerically 
carried out using the {\tt wilson} \cite{ Aebischer:2018bkb} program, which we also use to perform the LL component of the RG running.
Moreover, throughout this paper, we will follow the WCxf conventions for WET and SMEFT \cite{Aebischer:2017ugx}.
{\boldmath
\subsection{Operator Basis for 3+1 Flavors} 
}
At dim-$4$ in WET, CP can be violated by a complex quark mass term and by 
the  QCD $\bar\theta$ term.
We choose to work in a basis in which the quark mass matrices are diagonal and real, so that, at dim-4, the only CP-odd interactions is
\begin{equation}
\mathcal L_\theta = \bar \theta\, \frac{g_s^2}{32\pi^2} \widetilde G^A_{\mu \nu} G^{A\, \mu \nu}, 
\end{equation}
where $g_s$ is the strong coupling constant,
$G^A_{\mu \nu}$ is the gluon field strength
and $\widetilde{G}^{A}_{\mu\nu} = \frac{1}{2}\varepsilon_{\mu \nu \alpha \beta} G^{A\,\alpha \beta}$.
Throughout this paper, we assume that the contribution of the $\bar\theta$ term to EDMs is canceled by the Peccei-Quinn (PQ) mechanism 
\cite{Peccei:1977ur,Peccei:1977hh}.

In a $n_f=3+1$ flavor theory, the 
dipole operators can contribute to EDMs at the dim-5 level.
At dim-6, there are contributions from the Weinberg three-gluon operator and various four-fermion operators. 
%%%%%%%%%%%%%
\subsubsection{Dipole Operators}
The flavor-conserving QCD and QED dipole operators are given by  
\be
\begin{aligned}
\op[]{u\gamma}{ii} &=  (\bar u_i \sigma^{\mu \nu}P_R u_i) F_{\mu\nu}\,,  \,\ \,\ \,\ \,\ \,\
\op[]{d\gamma}{ii} = (\bar d_i \sigma^{\mu \nu}P_R d_i) F_{\mu\nu}\,,  \\
\op[]{uG}{ii} &=  (\bar u_i \sigma^{\mu \nu} T^A P_R u_i) G_{\mu\nu}^A\,,  \,\ \,\
\op[]{dG}{ii} =  (\bar d_i \sigma^{\mu \nu} T^A P_R d_i) G_{\mu\nu}^A\,, \\
\op[]{ e \gamma}{ii} &=  (\bar e_i \sigma^{\mu \nu}P_R e_i) F_{\mu\nu} .
\end{aligned}
\ee
Here $ii$ indicates the flavor index and $T^A$ are the $SU(3)_c$ generators. Down-type type operators having $ii=11,22$, 
and 
up-type quarks with $ii=11$ directly contribute to the neutron and atomic EDMs. 
Dipole operators with $b$ and $c$ quarks contribute via loops.
In the leptonic sector, $ii=11$, $22$, and $33$ contribute directly to 
electron, muon, and $\tau$ EDMs, respectively. 
Since the limits on $d_\mu$ and $d_\tau$ are much weaker than on the electron EDM, it is important also to consider the
matching and running of heavier leptons into leptonic and semileptonic operators that contribute to molecular EDMs.
\subsubsection{Weinberg Operator}
The CPV Weinberg operator has the form 
\be
\op[]{\widetilde G}{} =  f^{ABC} \widetilde G_{\mu}^{A\nu} G_{\nu}^{B \rho } G_{\rho}^{C\mu}.
\ee
Here  $f^{ABC}$ are the structure constants of the $SU(3)_c$ 
group. Note that the WC $\wc[]{\widetilde G}{}$ is real by definition. 
{\boldmath
\subsubsection{\boldmath Four-fermion operators with 3+1 Flavors}\label{sub:fourquark}
}
In this section, we give a complete list of four-fermion WET operators with $n_f = 3+1$ quark and lepton flavors.
All of these operators enter directly into the EDM expressions, hence these are the leading operators. 
A four-fermion basis in terms of scalar-pseudoscalar and tensor operators was presented in Ref. \cite{Buhler:2023gsg}, where the 1-loop matching to the gradient flow scheme used in Lattice QCD calculations is also provided.
Here, to facilitate the renormalization group evolution and the matching to EFTs at the electroweak scale, we adopt the JMS basis.
In the first category, we consider vector operators having $(LL)(RR)$ chiralities of the fermion currents. These are given by
\be
\begin{aligned} 
\, [O_{uddu}^{V1,LR}]_{ijkl} &= (\bar u_i \gamma^\mu P_L d_j)(\bar d_k \gamma_\mu P_R u_l)\,, \\
 [O_{uddu}^{V8,LR}]_{ijkl}   &= (\bar u_i \gamma^\mu  P_L T^A d_j)(\bar d_k \gamma_\mu P_R T^A u_l),\\
 \,   [O_{dd}^{V1,LR}]_{ijkl} & = (\bar d_i \gamma^\mu P_L d_j)(\bar d_k \gamma_\mu P_R d_l)\,, \\
    [O_{dd}^{V8,LR}]_{ijkl} &= (\bar d_i \gamma^\mu  P_L T^A d_j)(\bar d_k \gamma_\mu P_R T^A d_l).
 \end{aligned}
\ee
These lead to six flavor structures that  are CP-violating and directly contribute to EDMs: 
\begin{equation}
[O_{uddu}^{V1,LR}]_{1111} \,, ~ [O_{uddu}^{V1,LR}]_{1221}\,,~  [O_{uddu}^{V8,LR}]_{1111} \,, ~[O_{uddu}^{V8,LR}]_{1221}, \, 
[O_{dd}^{V1,LR}]_{1221} \,, ~ [O_{dd}^{V8,LR}]_{1221}. 
\end{equation}
In the next category, there are various scalar operators. First we have $(LR)(LR)$ up-type 
operators
\be
\begin{aligned}
   \, [O_{uu}^{S1,RR}]_{ijkl} &= (\bar u_i  P_R u_j)(\bar u_k  P_R u_l)\,, \\
     [O_{uu}^{S8,RR}]_{ijkl} &= (\bar u_i P_R T^A u_j)(\bar u_k P_R T^A u_l).
\end{aligned}
\ee
The following two operators are CP-violating and enter at the tree-level
\begin{equation}
[O_{uu}^{S1,RR}]_{1111} \,, ~ [O_{uu}^{S8,RR}]_{1111}. 
\end{equation}
The down-type operators are given by
\be
\begin{aligned}
  \,  [O_{dd}^{S1,RR}]_{ijkl} &= (\bar d_i  P_R d_j)(\bar d_k  P_R d_l)\,, \\
    [O_{dd}^{S8,RR}]_{ijkl} &= (\bar d_i P_R T^A d_j)(\bar d_k P_R T^A d_l).
\end{aligned}
\ee
Here the following eight operators directly contribute 
\be
\begin{aligned}
& [O_{dd}^{S1,RR}]_{1111}  \,, ~ [O_{dd}^{S8,RR}]_{1111}\,, ~
[O_{dd}^{S1,RR}]_{2222} \,, ~ [O_{dd}^{S8,RR}]_{2222}\,, \\
& [O_{dd}^{S1,RR}]_{1122}  \,, ~ [O_{dd}^{S8,RR}]_{1122}\,, ~
[O_{dd}^{S1,RR}]_{1221} \,, ~ [O_{dd}^{S8,RR}]_{1221}.
\end{aligned}
\ee
Then the operators containing both up and down-quark come in two sub-categories. The first one is
\be
\begin{aligned}
  \,  [O_{ud}^{S1,RR}]_{ijkl} &= (\bar u_i  P_R u_j)(\bar d_k  P_R d_l)\,, \\
     [O_{ud}^{S8,RR}]_{ijkl} &= (\bar u_i P_R T^A u_j)(\bar d_k P_R T^A d_l).
\end{aligned}
\ee
The following four flavor structures can contribute 
\be
[O_{ud}^{S1,RR}]_{1111} \,, ~ [O_{ud}^{S8,RR}]_{1111}\,, ~
[O_{ud}^{S1,RR}]_{1122} \,, ~ [O_{ud}^{S8,RR}]_{1122}. 
\ee
In the second sub-category, we have the operators
\be
\begin{aligned}
  \,  [O_{uddu}^{S1,RR}]_{ijkl} &= (\bar u_i  P_R d_j)(\bar d_k  P_R u_l)\,, \\
     [O_{uddu}^{S8,RR}]_{ijkl} &= (\bar u_i P_R T^A d_j)(\bar d_k P_R T^A u_l).
\end{aligned}
\ee
The following four operators contribute 
\begin{equation}
[O_{uddu}^{S1,RR}]_{1111} \,, ~ [O_{uddu}^{S8,RR}]_{1111}\,, ~
[O_{uddu}^{S1,RR}]_{1221} \,, ~ [O_{uddu}^{S8,RR}]_{1221}.
\end{equation}
In total, there are 24 four-quark operators with $n_f=3$, in agreement with Ref. \cite{Buhler:2023gsg}.
%%%%%%
Semileptonic operators contribute to molecular and atomic EDMs at the tree level
\begin{eqnarray}
\op[S,RL]{eu}{ijkl} &= &  (\bar e_i P_R e_j)  (\bar u_k P_L u_l)\,, \,\ \,\ \,\ \,\
\op[S,RL]{ed}{ijkl} =   (\bar e_i  P_R e_j)  (\bar d_k P_L d_l) \,, \\
%%%%%%%%
\op[S,RR]{eu}{ijkl} &= &  (\bar e_i P_R e_j)  (\bar u_k P_R u_l)\,, \,\ \,\ \,\ \,\
\op[S,RR]{ed}{ijkl} =   (\bar e_i  P_R e_j)  (\bar d_k P_R d_l)\,, \notag \\
%%%%%%%%%%
\op[T,RR]{eu}{ijkl} &= &  (\bar e_i \sigma^{\mu \nu} P_R e_j)  (\bar u_k \sigma_{\mu \nu} P_R u_l)\,,  \,\ \,\
\op[T,RR]{ed}{ijkl} =   (\bar e_i  \sigma^{\mu \nu} P_R e_j)  (\bar d_k \sigma_{\mu \nu} P_R d_l).\notag
\end{eqnarray}

\begin{table}[H] 
\begin{center}
 \renewcommand*{\arraystretch}{1.3}
\begin{tabular}{ |cccccc| } 
\hline 
\multicolumn{6}{|c|}{Four-fermion Operators for EDMs with $n_q+n_\ell$ flavors} \\
\hline
\hline
\multicolumn{6}{|c|}{ $u, d, s, e$ operators in 3+1 flavor WET} \\
\hline \hline

 $ [O_{uddu}^{V1,LR}]_{1111}$  & $ [O_{uddu}^{V8,LR}]_{1111}$   & $[O_{uddu}^{V1,LR}]_{1221}$ &  $[O_{uddu}^{V8,LR}]_{1221}$ &
 $[O_{dd}^{V1,LR}]_{1221}$  & $[O_{dd}^{V8,LR}]_{1221}$  \\ 
  $[O_{uu}^{S1,RR}]_{1111}$  & $[O_{uu}^{S8,RR}]_{1111}$ &  $[O_{dd}^{S1,RR}]_{1111}$  & $[O_{dd}^{S8,RR}]_{1111}$   & $[O_{dd}^{S1,RR}]_{2222}$ &  $[O_{dd}^{S8,RR}]_{2222}$ \\  
 $[O_{dd}^{S1,RR}]_{1122}$  & $[O_{dd}^{S8,RR}]_{1122}$  & $[O_{dd}^{S1,RR}]_{1221}$  & $[O_{dd}^{S8,RR}]_{1221}$ & 
 $[O_{ud}^{S1,RR}]_{1111}$  & $[O_{ud}^{S8,RR}]_{1111}$  \\ 
  $[O_{ud}^{S1,RR}]_{1122}$  & $[O_{ud}^{S8,RR}]_{1122}$ &  $[O_{uddu}^{S1,RR}]_{1111}$  & $[O_{uddu}^{S8,RR}]_{1111}$  & $[O_{uddu}^{S1,RR}]_{1221}$  & $[O_{uddu}^{S8,RR}]_{1221}$\\  \hline 
  %%%%%%
   $[O_{eu}^{S,RL}]_{1111}$  & $[O_{eu}^{S,RR}]_{1111}$ & 
$[O_{ed}^{S,RL}]_{1111}$ & $[O_{ed}^{S,RR}]_{1111}$ & 
 $[O_{ed}^{S,RL}]_{1122}$ &  $[O_{ed}^{S,RR}]_{1122} $\\
 $[O_{eu}^{T,RR}]_{1111}$  & $[O_{ed}^{T,RR}]_{1111}$ 
& $[O_{ed}^{T,RR}]_{1122}$  &    $[O_{ee}^{S,RR}]_{1111}^*$     && \\  \hline
  %%%%%%%
  \hline
\multicolumn{6}{|c|}{$c$ operators $\subset$ 4+1 flavor WET  } \\
\hline \hline
 $[O_{uu}^{S1,RR}]_{2222}^*$  & $[O_{uu}^{S8,RR}]_{2222}^*$ &  $[O_{uu}^{S1,RR}]_{1122}$  & $[O_{uu}^{S8,RR}]_{1122}$ &
$[O_{ud}^{S1,RR}]_{2211}$  & $[O_{ud}^{S8,RR}]_{2211}$  \\
 $[O_{ud}^{S1,RR}]_{2222}$  & $[O_{ud}^{S8,RR}]_{2222}$ & $[O_{uddu}^{S1,RR}]_{2112}$  & $[O_{uddu}^{S8,RR}]_{2112}$  & $[O_{uddu}^{S1,RR}]_{2222}$  & $[O_{uddu}^{S8,RR}]_{2222}$\\ 
$[O_{uu}^{S1,RR}]_{1221}$  & $[O_{uu}^{S8,RR}]_{1221}$ &  
$[O_{uddu}^{V1,LR}]_{2112}^*$  & $[O_{uddu}^{V8,LR}]_{2112}^*$   & $[O_{uddu}^{V1,LR}]_{2222}^*$ &  $[O_{uddu}^{V8,LR}]_{2222}^*$ \\  
 $[O_{uu}^{V1,LR}]_{1221}^*$  & $[O_{uu}^{V8,LR}]_{1221}^*$  &  &    & & \\ 
\hline
 $[O_{eu}^{S,RR}]_{1122}$  &   $[O_{eu}^{T,RR}]_{1122}$ & &    & & \\ 
 \hline \hline
\multicolumn{6}{|c|}{ $b, \mu, \tau$ operators $\subset$ 5+3 flavor WET } \\
\hline \hline
%%%%
$[O_{dd}^{S1,RR}]_{3333}^*$  & $[O_{dd}^{S8,RR}]_{3333}^*$   & $[O_{dd}^{S1,RR}]_{1133}$  & $[O_{dd}^{S8,RR}]_{1133}$ & 
 $[O_{dd}^{S1,RR}]_{2233}$  & $[O_{dd}^{S8,RR}]_{2233}$  \\ 
  $[O_{dd}^{S1,RR}]_{1331}$  & $[O_{dd}^{S8,RR}]_{1331}$ & 
 $[O_{dd}^{S1,RR}]_{2332}$  & $[O_{dd}^{S8,RR}]_{2332}$ & $[O_{ud}^{S1,RR}]_{1133}$  & $[O_{ud}^{S8,RR}]_{1133}$ \\
 $[O_{uddu}^{S1,RR}]_{1331}$  & $[O_{uddu}^{S8,RR}]_{1331}$  &  $[O_{ud}^{S1,RR}]_{2233}^*$  & $[O_{ud}^{S8,RR}]_{2233}^*$  & 
 $[O_{uddu}^{S1,RR}]_{2332}^*$  & $[O_{uddu}^{S8,RR}]_{2332}^*$  \\
 %%%%%%%%%%%%%%
$[O_{uddu}^{V1,LR}]_{1331}^*$  & $[O_{uddu}^{V8,LR}]_{1331}^*$ &  
$[O_{dd}^{V1,LR}]_{1331}^*$  & $[O_{dd}^{V8,LR}]_{1331}^*$ &   $[O_{dd}^{V1,LR}]_{2332}^*$  & $[O_{dd}^{V8,LR}]_{2332}^*$  \\
$[O_{uddu}^{V1,LR}]_{2332}^{**}$  & $[O_{uddu}^{V8,LR}]_{2332}^{**}$   &   &   &   &      \\
\hline
%%%%%%%%%%%%
$[O^{T,RR}_{ed}]_{1133}$  & $[O^{T,RR}_{ed}]_{2211}$   & $[O_{ed}^{T,RR}]_{2222}$  & $[O_{ed}^{T,RR}]_{2233}^*$   &$[O_{ed}^{T,RR}]_{3311}$    &  $[O_{ed}^{T,RR}]_{3322}$    \\
 $[O^{T,RR}_{ed}]_{3333}^*$  & $[O^{T,RR}_{eu}]_{2211}$   & $[O_{eu}^{T,RR}]_{2222}^*$    &$[O_{eu}^{T,RR}]_{3311}$    &  $[O_{eu}^{T,RR}]_{3322}^*$   & \\
%%%%%%%

$[O^{S,RR}_{ed}]_{1133}$  & $[O^{S,RR}_{ed}]_{2211}$   & $[O_{ed}^{S,RR}]_{2222}$  & $[O_{ed}^{S,RR}]_{2233}^*$   &$[O_{ed}^{S,RR}]_{3311}$    &  $[O_{ed}^{S,RR}]_{3322}$    \\
 $[O^{S,RR}_{ed}]_{3333}^*$  & $[O^{S,RR}_{eu}]_{2211}$   & $[O_{eu}^{S,RR}]_{2222}^*$    &$[O_{eu}^{S,RR}]_{3311}$    &  $[O_{eu}^{S,RR}]_{3322}^*$   & \\

 $[O_{ee}^{V,LR}]_{1221}^*$  & $[O_{ee}^{V,LR}]_{1331}^*$   &$[O_{ee}^{S,RR}]_{1221}$    &   $[O_{ee}^{S,RR}]_{1331}$  & $[O_{ee}^{S,RR}]_{1122}$  & $[O_{ee}^{S,RR}]_{1133}$   \\ \hline
\end{tabular}
\caption{The complete list of four-fermion flavor conserving CP-violating operators relevant for EDMs in $n_q+ n_\ell$ =5+3 Flavor WET below the EW scale.
Here the LL, NLL, and NNLL operators are marked by no asterisk, single asterisk, and double asterisk signs, respectively. }
\label{tab:wet4}
\end{center}
\end{table}
The relevant flavor structures for semi-leptonic operators are
\be
\begin{aligned}
& [O_{eu}^{S,RL}]_{1111}  \,, ~ [O_{eu}^{S,RR}]_{1111}\,, ~
[O_{ed}^{S,RL}]_{1111} \,, ~ [O_{ed}^{S,RR}]_{1111}\,, \\
& [O_{ed}^{S,RL}]_{1122} \,, ~ [O_{ed}^{S,RR}]_{1122}\,, 
 [O_{eu}^{T,RR}]_{1111}  \,, ~ [O_{ed}^{T,RR}]_{1111}\,,  \\
& [O_{ed}^{T,RR}]_{1122} .
\end{aligned}
\ee
Finally, we note that there are no leptonic operators that can contribute at the tree-level to any of the EDMs.
As we will see, the purely leptonic operator can contribute at one loop. In particular, the scalar operator 
\begin{equation}
\op[S,RR]{ee}{ijkl} =   (\bar e_i  P_R e_j)  (\bar e_k  P_R e_l)
\end{equation}
can mix onto the electron EDM.
In the $3+1$ theory, the relevant  flavor structure is 
\begin{equation}
[O_{ee}^{S,RR}]_{1111}.
\end{equation}

\subsection{Four-fermion Operators with 5+3 Flavors}

In this section, we list the leading heavy flavor (5+3) four-fermion operators that 
contribute up to the 2-loop level via matching and running effects.   The only new Lorentz structure 
appearing due to loop effects is the pure leptonic operator
\be
[O_{ee}^{V,LR}]_{ijkl}  = (\bar e_i \gamma^\mu P_L e_j)(\bar e_k \gamma_\mu P_R e_l)\,.
\ee
This operator can break CP if $i \neq j$, so that the CP-odd coefficients are 
\begin{equation}
[O_{ee}^{V,LR}]_{1221} \,, ~ [O_{33}^{V,LR}]_{1331}. 
\end{equation}
These two operators contribute to the electron EDM at the 2-loop QED level. Otherwise, only new   flavor structures
are added to the EDM basis for the 5+3 theory.
The complete list of relevant WET operators containing 3+1, 4+1, and 5+3 quark and lepton flavors ($n_q+ n_\ell $) is given in Table~\ref{tab:wet4}.
In this table, the operators are categorized based on their contributions to the EDMs at the  LL, NLL,  and NNLL order 
in RG improved perturbation theory. The order is indicated by the asterisk signs.

\section{Electric dipole moments}\label{sec:edms}
Electric dipole moments are an extremely sensitive probe of $\Delta F =0$ CP-violation at low energy. 
The strongest constraints currently come from measurements in  ThO \cite{ACME:2018yjb}, HfF 
\cite{Roussy:2022cmp}, and YbF \cite{Kara:2012ay},
which can be interpreted as bounds on the electron EDM,
and from bounds on the neutron EDM \cite{Abel:2020pzs} and on the EDM of $^{199}$Hg \cite{Griffith:2009zz}.
Experiments looking for the EDM of $^{129}$Xe \cite{Sachdeva:2019rkt}
and $^{225}$Ra \cite{Parker:2015yka} are at the moment weaker, but they provide constraints on independent combinations of hadronic parameter and Wilson coefficients \cite{Chupp:2014gka}.  A summary of current EDM bounds is shown in Table \ref{tab:EDMexps}.
In the future, the limits on the electron EDM are expected to improve by at least a factor of ten,
while new molecular EDM experiments sensitive to nuclear Schiff moments could lead to a large improvement in the constraints on hadronic CPV interactions    \cite{Arrowsmith-Kron:2023hcr}. Several experimental collaborations, including n2EDM  at the Paul Scherrer Institute \cite{n2EDM:2021yah}, PanEDM \cite{Wurm:2019yfj}
at the Institute Laue Langevin,  TUCAN EDM at Triumf \cite{Martin:2020lbx}, and  LANL EDM at Los Alamos National Laboratory \cite{Ito:2017ywc},
aim at improving the bound on the neutron EDM by one order of magnitude. Finally, EDMs of light ions such as the proton, deuteron, or $^3$He could be measured in  storage ring experiments \cite{Alexander:2022rmq}. To estimate future EDM sensitivities, we will use the projection on the second row of Table \ref{tab:EDMexps}.
In the following subsections, we will express the EDMs of charged leptons, the neutron and proton EDMs, atomic EDMs, and the CPV precession frequencies measured in molecular experiments in terms of WET operators at low energy, while in Section \ref{sec:runmatch} we will show how these operators are generated from WET operators at the EW scale, via running and matching effects. 
\begin{table}[t]
\begin{center}\small
 \resizebox{1.0\textwidth}{!}{
\begin{tabular}{||c||ccc|ccc||}
\hline
&$d_e$ & $d_n$& $d_{\rm Hg}$ &$d_{p}$  & $d_{\rm Xe}$ & $d_{\rm Ra}$\\
\hline
\rule{0pt}{3ex}
current limit &$4.0 \cdot 10^{-30} $ &$ 1.8 \cdot 10^{-26}$  & $6.2 \cdot 10^{-30}$  & x  & $1.4 \cdot 10^{-27}$  , & $4.2\cdot 10^{-22}$  \\
expected limit &$4.0 \cdot 10^{-29}  $  &$ {1.0 \cdot 10^{-28}} $ &$6.2 \cdot 10^{-30}$&$ 1.0 \cdot 10^{-29}$  & $1.4 \cdot 10^{-28}$ & $1.0 \cdot 10^{-27}$\\
\hline \hline
                        & $\omega_{\rm ThO}$   & $\omega_{\rm HfF}$ & $\omega_{\rm YbF}$ && &\\
                        \hline
 current limit  &  1.3 mrad/s  & $0.17$ mrad/s & 23.5 mrad/s  &&&\\
 \hline \hline
\end{tabular}}
\end{center}
\caption{Current limits on the electron \cite{Roussy:2022cmp}, neutron \cite{Abel:2020pzs}, and mercury \cite{Griffith:2009zz,Graner:2016ses} EDMs in units of $e$ cm ($90\%$ confidence level).
We also show an indication of their prospective limits  \cite{Kumar:2013qya,Chupp:2014gka,Alarcon:2022ero} as well as those of the proton, xenon \cite{Sachdeva:2019rkt}, and radium \cite{Parker:2015yka} EDMs, which could provide interesting constraints in the future. 
The electron EDM bound is extracted from CPV precession frequencies in molecular EDM experiments and thus depends on the assumptions that no other CPV operator contributes. In our analysis, we will use directly the constraints on the precession frequencies in HfF \cite{Roussy:2022cmp}, 
ThO \cite{ACME:2018yjb}, and YbF \cite{Kara:2012ay},
which are shown in the bottom part of the table.  In the future, the precession frequencies are expected to improve by one order.}
\label{tab:EDMexps}  
\end{table}

\subsection{Integrating out heavy flavors}

In this paper, we integrate out heavy quarks and leptons at, or close to, their mass threshold.
In particular, we integrate out the $b$ quark at $\mu_b = m^{\overline{\rm MS}}_b = 4.18$ GeV,
and the charm quark at $\mu_c = 2$ GeV.
We thus do not consider nonperturbative matrix elements of operators with $b$ and $c$ quarks between nucleons or nucleons and pions,
but take their effects into account in perturbation theory, via renormalization group mixing or matching at the threshold.
For the $b$ quark, with $\alpha_s(m^{\overline{\rm MS}}_b)$ well in the perturbative regime, this approach is fully justified.
Since the $c$ quark lies at the boundary between the perturbative and nonperturbative regimes, in this case one could pursue the alternative approach of not integrating out the $c$, and evaluating charm matrix elements in Lattice QCD. 
While most Lattice QCD collaborations use actions with dynamical charm quarks, calculations of nucleon charm matrix elements are very preliminary   
in the case of quark bilinears \cite{Alexandrou:2019brg}, and non-existing for more complicated operators, such as the charm quark chromoelectric dipole moment or four-fermion operators. One would thus have to rely on models, which come with uncontrolled uncertainties.
As Lattice QCD calculations improve, it will become interesting to compare the perturbative and nonperturbative treatment of heavy quarks.

\subsection{Leptonic EDMs}
For the electron, muon, and $\tau$ EDMs, there is a direct connection between the WET dipole operators $C_{e\gamma}$ and the electric dipole moment.
In addition, the semileptonic tensor operators $\left[C^{T, RR}_{eq}\right]$ with light quark flavors induce charged lepton dipole operator non-perturbatively  \cite{Dekens:2018pbu,Aebischer:2021uvt}, with a contribution proportional to the correlator of the vector and tensor currents $\Pi_{VT}({\bf q})$   \cite{Cirigliano:2021img}
\begin{align}\label{eq:eEDM}
& d_\ell = -  2  \textrm{Im} \left[C_{e\gamma} - 2  i \Pi_{VT}(0) \left( q_u \left[C^{T, RR}_{eu} \right]_{1111}
+ q_d \left[C^{T, RR}_{ed} \right]_{1111} + q_s \left[C^{T, RR}_{ed} \right]_{1122}
\right) \right]_{\ell \ell} \nonumber \\ &  =  - 1.3 \cdot 10^{-16}   \left[C_{e\gamma}
- 2.5 \cdot 10^{-5} \, {\rm{TeV}} \left(2  \left[C^{T, RR}_{eu} \right]_{1111}
-   \left[C^{T, RR}_{ed} \right]_{1111} -  \left[C^{T, RR}_{ed} \right]_{1122}
\right) \right]_{\ell \ell}   \, {\rm{TeV}}\, e\, {\rm cm}, 
\end{align}
with $q_u = 2/3$ and $q_d = q_s = -1/3$. In Eq. \eqref{eq:eEDM}, and in the rest of this section, the WET couplings are understood to be evaluated at the scale of $\mu = 2$ GeV.
Here we use a resonance model for $\Pi_{VT}$  and write  \cite{Knecht:2001xc,Mateu:2007tr,Cata:2008zc,Cirigliano:2021img}
\begin{equation}
i \Pi_{VT}(0) = \frac{B_0 F_\pi^2}{M_\rho^2},
\end{equation}
with the quark condensate $B_0 = 2.7$ GeV at the renormalization scale $\mu = 2$ GeV, $F_\pi= 92.2$ MeV and $M_\rho = 0.770$ GeV. 
The above estimate is based on large $N_C$ considerations, and on the truncation of the resonance spectrum to the first state, and it is therefore affected by  
large theoretical uncertainties, which were estimated to be at least 50\% \cite{Knecht:2001xc,Mateu:2007tr,Cata:2008zc,Cirigliano:2021img}.

The limit on the electron EDM, as extracted from 
 HfF and ThO measurements  \cite{ACME:2018yjb,Roussy:2022cmp}, is shown in Table \ref{tab:EDMexps}. 
Direct limits exist on the muon EDM, from $g-2$ storage ring experiments \cite{Muong-2:2008ebm},
 and on the EDM of the tau lepton, from the measurement of spin-momentum correlations in $\tau$ decays at $e^+ e^-$ machines  \cite{Belle:2021ybo}.
 The current limits are \cite{Workman:2022ynf}
\begin{equation}
|d_{\mu} | < 1.8 \cdot 10^{-19} e\, {\rm cm}, \qquad     -0.185 \cdot 10^{-16} e\, {\rm cm} < d_\tau < 0.061  \cdot 10^{-16} e\, {\rm cm}.
\end{equation}
These are much weaker than the electron EDM bound so running effects of $\mu$ and $\tau$ flavor operators onto electron operators can be important.

\subsection{ThO, YbF, and HfF Frequencies}
The best limit on the electron EDM currently comes from experiments with polar molecules.
In this case, the observables are the parity ($P$) and time-reversal ($T$) odd frequency shifts $\omega$, which 
receive contributions from the electron EDM and semileptonic scalar operators.
The frequency shifts are expressed in terms of an isospin invariant and an isospin-breaking 
electron-nucleon coupling, captured by the Lagrangian
\begin{equation}\label{eq:scalar}
\mathcal L = - \frac{G_F}{\sqrt{2}} \bar e  i \gamma_5 e\, \bar N \left( C_S^{(0)} + \tau_3 C^{(1)}_{S} \right) N,
\end{equation}
with $N = (p, n)$ the nucleon doublet and $\tau_3$ a Pauli matrix in isospin space.
The nucleon-level scalar couplings can be expressed in terms of WET four-fermion operators as
\begin{align}
C_S^{(0)}  = &  - \frac{v^2}{2} \left( \textrm{Im}[C^{\rm SRL}_{eu}]_{1111} + \textrm{Im}[C^{\rm SRR}_{eu}]_{1111} + \textrm{Im}[C^{\rm SRL}_{ed}]_{1111} + \textrm{Im}[C^{\rm SRR}_{ed}]_{1111} \right) \frac{\sigma_{\pi N}}{ \bar m } \nonumber  \\
 & - v^2  \left( \textrm{Im}[C^{\rm SRL}_{ed}]_{1122} + \textrm{Im}[C^{\rm SRR}_{ed}]_{1122} \right) \frac{\sigma_s}{m_s}  \\
C_S^{(1)}  = &  - \frac{v^2}{2} \left( \textrm{Im}[C^{\rm SRL}_{eu}]_{1111} + \textrm{Im}[C^{\rm SRR}_{eu}]_{1111} - \textrm{Im}[C^{\rm SRL}_{ed}]_{1111} - \textrm{Im}[C^{\rm SRR}_{ed}]_{1111} \right) g^{u-d}_S,
\end{align}
where $2 \bar m = m_u + m_d$, and $v=246$ GeV.
For the nucleon $\sigma_{\pi N}$ term, we use the extraction from $\pi$-nucleon scattering data \cite{Hoferichter:2015hva}
\begin{equation}
\sigma_{\pi N} = 59.0 \pm 3.5 \, {\rm MeV}.
\end{equation}
For the strange $\sigma$ term and the isovector scalar charge, we use the FLAG lattice averages  \cite{FlavourLatticeAveragingGroupFLAG:2021npn}
\begin{equation}
\sigma_s = 52.9  \pm 7.0 \, {\rm MeV}, \qquad g_S^{u-d}  = 1.02 \pm 0.10,
\end{equation}
where the scalar charge is evaluated at the renormalization scale $\mu = 2$ GeV.
With these definitions, we can express the frequency measured in the HfF, ThO, and YbF experiments as   \cite{Degenkolb:2024eve,Chupp:2017rkp,Dekens:2018bci}
\begin{align}
\omega_{\rm HfF} &= \left[ \left(34.9 \pm 1.4\right)  \left(\frac{d_e}{10^{-27} e\, {\rm cm}}\right)  + (32^{+1}_{-2}) \left(\frac{C_S}{10^{-7}}\right) \right]   \left({\rm mrad/s}\right) \\
\omega_{\rm ThO} &= -\left[ \left(121^{+5}_{-39}\right)  \left(\frac{d_e}{10^{-27} e\, {\rm cm}}\right)  + (182^{+42}_{-27}) \left(\frac{C_S}{10^{-7}}\right) \right]   \left({\rm mrad/s}\right) \\
\omega_{\rm YbF} &= -\left[ \left(19.6 \pm 1.5\right)  \left(\frac{d_e}{10^{-27} e\, {\rm cm}}\right)  + (17.6 \pm 2.0) \left(\frac{C_S}{10^{-7}}\right) \right]   \left({\rm mrad/s}\right)
\end{align}
where 
\begin{equation}
C_S = C_S^{(0)} + \frac{Z-N}{Z+N} C_S^{(1)}.
\end{equation}
Here $(Z, A)= (72,106), (90, 142), (70,103)$ for HfF, ThO, and YbF, respectively. 

\subsection{Nucleon EDMs}
The interpretation of neutron and proton EDMs is complicated by the need to match quark-level operators onto hadronic couplings.
This matching requires nonperturbative techniques and it has been carried out with different levels of sophistication for different WET operators.
In the case of the quark dipole operators $[C_{u\gamma}]_{11}$ and $[C_{d\gamma}]_{11, 22}$, the hadronic input is captured by the quark tensor charges,  
which have been computed on the lattice \cite{Gupta:2018lvp,Alexandrou:2019brg,FlavourLatticeAveragingGroupFLAG:2021npn}.
\begin{align}
d_n &= -1.3 \cdot 10^{-16}\left(  g_T^d \textrm{Im} \left[C_{u \gamma} \right]_{11}  + g_T^u \textrm{Im} \left[C_{d \gamma} \right]_{11} + g_T^s \textrm{Im}\left[C_{d \gamma} \right]_{22}  \right)  \, {\rm{TeV}}\, e\, {\rm cm},\\
d_p &= -1.3 \cdot 10^{-16} \left(  g_T^u \textrm{Im} \left[C_{u \gamma} \right]_{11}  + g_T^d \textrm{Im} \left[C_{d \gamma} \right]_{11} + g_T^s \textrm{Im} \left[C_{d \gamma} \right]_{22} \right)
 \, {\rm{TeV}}\, e\, {\rm cm},
\end{align}
with \cite{FlavourLatticeAveragingGroupFLAG:2021npn} 
\begin{equation}
g_T^{u} = 0.784 \pm 0.030, \quad g_T^{d} = -0.204 \pm 0.015, \quad g_T^s = -0.0027 \pm 0.0016
\end{equation}
at the renormalization scale $\mu = 2$ GeV. 
The $u$ and $d$ tensor charges have better than 10\% uncertainties, while the $s$ tensor charge is harder to compute and zero within two sigma. 
The $c$ tensor charge has been considered in Ref. \cite{Alexandrou:2019brg}, but it is even more uncertain,
\begin{equation}
g^c_T =  - \left(10 \pm 19\right) \cdot 10^{-5}.
\end{equation}
For this reason, will not consider directly charm matrix elements, but integrate out the charm quark.

Lattice QCD calculations of the nucleon EDM induced by the quark chromoelectric dipole moment (qCEDM) operators $\left[C_{u G}\right]_{11}$, $\left[C_{d G}\right]_{11, 22}$ are at the moment only preliminary \cite{Bhattacharya:2023qwf,Abramczyk:2017oxr,Kim:2021qae}.
Currently, the best estimates come from QCD sum rules \cite{Pospelov:2000bw,Pospelov:2005pr,Lebedev:2004va,Fuyuto:2012yf}. In Refs. 
\cite{Pospelov:2000bw,Pospelov:2005pr,Lebedev:2004va}, the $u$ and $d$ qCEDM contribution to the neutron and proton EDMs was estimated to be
\begin{align}\label{eq:qcedm1}
d_n &=  -   \frac{4.3 \cdot 10^{-17}}{g_s}\times  (1 \pm 0.5) \left( \textrm{Im}\left[C_{d G} \right]_{11}  + 0.5 \textrm{Im}\left[C_{u G}\right] _{11}  \right)  \, {\rm{TeV}}\, e\, {\rm cm}, \\
d_p &=  +  \frac{5.2 \cdot 10^{-17}}{g_s} \times  (1 \pm 0.5)   \left(0.5   \textrm{Im}\left[C_{d G} \right]_{11}  +  \textrm{Im}\left[C_{u G}\right] _{11}  \right)  \, {\rm{TeV}}\, e\, {\rm cm},
\end{align}
with roughly 50\% uncertainty.
Here $g_s$ is the strong coupling constant.  We will use the estimates in Eq. \eqref{eq:qcedm1} for our numerics.
 Ref. \cite{Fuyuto:2012yf} finds a larger value and a substantial contribution from the strange qCEDM 
\begin{align}\label{eq:qcedm2}
d_n &=  -  \frac{ 9.8 \cdot 10^{-17} }{g_s} \left( \textrm{Im}\left[C_{d G} \right]_{11}  + 1.2 \textrm{Im}\left[C_{u G}\right] _{11}  + 0.2 \textrm{Im}\left[C_{d G}\right] _{22}  \right)  \, {\rm{TeV}}\, e\, {\rm cm}. 
\end{align}
Estimates based on the dominance of long-distance contributions mediated by pion-nucleon couplings end up in a similar range \cite{Seng:2014pba,deVries:2016jox},
though, of course, they are affected by uncontrolled systematic errors.
The above matrix elements are given in the assumption that a PQ mechanism cancels the $\bar\theta$ term,
including the shift in the $\bar\theta$ term induced by the chromoelectric operators \cite{Pospelov:2005pr}.
The scale dependence of $g_s$ and of the WET coefficients should in principle be compensated by the scale dependence of the hadronic matrix elements. In the sum rule calculations, such scale dependence is not explicitly tracked, and thus, at the moment, the choice of $\mu$ introduces some arbitrariness. 
Since the conversion between Lattice QCD schemes and $\overline{\rm MS}$ is known  \cite{Bhattacharya:2015rsa,Mereghetti:2021nkt}, this issue will be overcome once Lattice QCD calculations will become available.
The contribution of the Weinberg three-gluon operator  $C_{\tilde G}$ has been estimated in Refs.  \cite{Demir:2002gg,Haisch:2019bml}.
The two evaluations are in good agreement, and here we use the results of Ref.  \cite{Haisch:2019bml}
\begin{align}\label{eq:gcedm1}
d_n &= 1.5 \cdot 10^{-21} (1 \pm 0.5) \left[C_{\tilde G}\right] \, {\rm{TeV}}^2\, e\, {\rm cm}, \\
d_p &= -2.1 \cdot 10^{-21} (1 \pm 0.5)  \left[C_{\tilde G}\right] \, {\rm{TeV}}^2\, e\, {\rm cm}.
\end{align}
Also in this case, Lattice QCD calculations are preliminary \cite{Bhattacharya:2022whc}, even though 
progress has been achieved on the matching between Lattice QCD and $\overline{\rm MS}$ schemes \cite{Cirigliano:2020msr,Crosas:2023anw}.
The estimates in Eq. \eqref{eq:qcedm1}, \eqref{eq:qcedm2} and \eqref{eq:gcedm1} clearly show that hadronic matrix elements are affected by large theoretical uncertainties, which need to be considered in any realistic analysis of EDM bounds.

The situation for four-fermion operators is even less well-developed. 
The matching between gradient flow and $\overline{\rm MS}$ renormalization at one loop has been derived in Ref. \cite{Buhler:2023gsg}.
Robust calculations of the nucleon EDM currently do not exist. 
Here we will use the expression derived at next-to-next-to-leading order (N$^2$LO) in two-flavor chiral perturbation theory \cite{Seng:2014pba}
\begin{align}
    d_n &= -  \frac{e g_A}{8 \pi^2 F_\pi^2} \left[\left( \bar g_0 - \frac{\bar{g}_2}{3} \right) \left( \ln \frac{m_\pi^2}{m_N^2} - \frac{\pi}{2} \frac{m_\pi}{m_N} \right) + \frac{\bar g_1}{4} (\kappa_1 - \kappa_0) \frac{m_\pi^2}{m_N^2} \ln \frac{m_\pi^2}{m_N^2}  \right] + e \bar d_n \label{eq:nEDMchi} \\
    d_p &=   \frac{e g_A}{8 \pi^2 F_\pi^2} \left[\left( \bar g_0 - \frac{\bar{g}_2}{3} \right) \left( \ln \frac{m_\pi^2}{m_N^2} - 2\pi \frac{m_\pi}{m_N} \right) - \frac{\bar g_1}{4}
    \left( \frac{2\pi m_\pi}{m_N}
 + \left( \frac{5}{2} +  \kappa_1 + \kappa_0 \right) \frac{m_\pi^2}{m_N^2} \ln \frac{m_\pi^2}{m_N^2} \right)  \right] \nonumber \\ & + e \bar d_p.  \label{eq:pEDMchi}
\end{align}
with $g_A = 1.27$. $\kappa_{1,0}$ are the nucleon isovector and isoscalar anomalous magnetic moments, $\kappa_1 = 3.7$, $\kappa_0 = -0.12$.
$\bar g_0$, $\bar g_1$, $\bar d_n$, and $\bar d_p$ are $PT$-odd low-energy couplings in the nucleon EFT, which implicitly depend on WET Wilson coefficients.
In the following, we will provide a recipe to estimate $\bar g_{0,1}$ in the case of four-fermion operators.

\subsubsection{Pion nucleon couplings}

$P$ and $T$-odd pion-nucleon ($\pi N$) couplings play an important role for both nucleon and nuclear EDMs. 
In the case of the nucleon EDM, $\pi N$ couplings induce logarithmically-enhanced contributions  \cite{Crewther:1979pi},
as can be seen in Eqs. \eqref{eq:nEDMchi} and \eqref{eq:pEDMchi}.
Since they are accompanied by unknown low-energy constants, $\pi N$ couplings cannot completely determine the leading contributions to the nucleon EDM,
and thus expressions as Eq. \eqref{eq:nEDMchi} will be eventually superseded by Lattice QCD calculations.
Knowing the strength of $\bar g_{0,1}$, however, can provide guidance for the chiral and momentum extrapolation \cite{Dragos:2019oxn} and for the
removal of excited state contamination \cite{Bhattacharya:2021lol} in Lattice QCD calculations.
Depending on the chiral properties of the WET operators,
$\pi N$ couplings furthermore provide one of the leading contributions to the $PT$-odd potential \cite{deVries:2012ab,Maekawa:2011vs,deVries:2020iea},
which then feeds into the calculation of Schiff moments and thus atomic EDMs \cite{Haxton:1983dq,Engel:1999np}.

We focus here on the most important $\pi N$ couplings and refer for a more detailed discussion to Ref. \cite{deVries:2020iea}.
We define the $PT$-odd $\pi N$ Lagrangian as
\begin{equation}\label{eq:piNLag}
    \mathcal L_{\pi N} = \bar N \left[ \frac{\bar g_0}{F_\pi}  \boldsymbol{\tau}  \cdot \boldsymbol{\pi}  + \frac{\bar g_1}{F_\pi} \pi_0 + \frac{\bar g_2}{F_\pi} \left( \pi_0 \tau_3 - \frac{1}{3}  \boldsymbol{\tau}  \cdot \boldsymbol{\pi} \right) \right] N,
\end{equation}
where $\boldsymbol{\tau}$ are Pauli matrices in isospin space.
The couplings in Eq. \eqref{eq:piNLag} break chiral symmetry and are thus mostly important for chiral-symmetry-breaking WET operators. 
These usually come in chiral multiplets, with the CP-even components inducing corrections to the baryon spectrum 
while the CP-odd components contribute to $\pi N$ couplings.
At leading order in chiral perturbation theory, chiral symmetry then provides a relation between 
corrections to the baryon spectrum and $\pi N$ couplings \cite{deVries:2012ab}, which survives higher order corrections \cite{deVries:2015una,deVries:2016jox}.
It can be shown that $\bar g_{2}$ only arises at higher order in isospin breaking \cite{deVries:2012ab}. 
The couplings $\bar g_0$ and $\bar g_1$ receive a ``direct'' contribution, proportional to the shift in the baryon masses induced by dim-6 operators, and ``indirect'' contributions, arising from ``vacuum alignment'' \cite{Baluni:1978rf}, i.e. from imposing the condition that $\pi_0$ and $\eta$ mesons 
do not disappear into the vacuum in the presence of $PT$-odd interactions. We will thus write
\begin{eqnarray}
    \bar g_i = \left. \bar g_i \right|_{\rm dir} + \left. \bar g_i \right|_{\rm ind}.  
\end{eqnarray}
The indirect contributions depend on meson matrix elements, which are more controlled.

For the qCEDM, the $\pi N$ couplings were extracted using QCD sum rules \cite{Pospelov:2001ys}, 
\begin{align}
\frac{\bar g_0}{F_\pi} &= - 1.97 \cdot 10^{-3 }\frac{ \left(1 \pm 2\right)}{g_s}  \left(\textrm{Im}\left[C_{u G} \right]_{11} + \textrm{Im}\left[C_{d G} \right]_{11} \right)  \, {\rm{TeV}}, \label{eq:g0sum}\\
\frac{\bar g_1}{F_\pi} &= - 1.97 \cdot 10^{-3 }\frac{ \left(4^{+8}_{-2}\right)}{g_s}  \left(\textrm{Im}\left[C_{u G} \right]_{11} - \textrm{Im}\left[C_{d G} \right]_{11} \right)  \, {\rm{TeV}}. 
\label{eq:g1sum}
\end{align}
Again, we are neglected contributions proportional to $\bar\theta_{\rm ind}$, which are canceled if a PQ mechanism is active.
In chiral perturbation theory, the indirect part of the coupling is given by \cite{deVries:2016jox}
\begin{align}
\left. \frac{\bar g_0}{F_\pi}\right|_{\rm   ind} & =   \frac{r g^{u-d}_S}{F_\pi g_s}   \left(\textrm{Im}\left[C_{u G} \right]_{11} + \textrm{Im}\left[C_{d G} \right]_{11} \right) = 
- \frac{4.4 \cdot 10^{-3}}{g_s}  \left(\textrm{Im}\left[C_{u G} \right]_{11} + \textrm{Im}\left[C_{d G} \right]_{11} \right), \label{eq:g0ind} \\
\left. \frac{\bar g_1}{F_\pi}\right|_{\rm   ind}  &=   \frac{r \sigma_{\pi N}}{ \bar m F_\pi g_s}   \left(\textrm{Im}\left[C_{u G} \right]_{11} - \textrm{Im}\left[C_{d G} \right]_{11} \right)
= - \frac{7.4 \cdot 10^{-2}}{g_s} \left(\textrm{Im}\left[C_{u G} \right]_{11} - \textrm{Im}\left[C_{d G} \right]_{11} \right), \label{eq:g1ind}
\end{align}
where $r$ is the ratio of chromomagnetic and $\bar q q$ condensates, for which we used \cite{Pospelov:2001ys} 
\begin{equation}
r = -\frac{1}{2} \frac{\langle 0 | \bar q g_s \sigma G q | 0 \rangle}{\langle 0 | \bar q  q | 0 \rangle}  = -\left(0.4 \pm 0.1 \right) {\rm GeV}^2.
\end{equation}
The ratio between $\bar g_1/\bar g_0$ is determined by the ratio of isoscalar and isovector charges, which, for the value of the sigma term we adopt, is about $17$. 
The assumption of the dominance of the indirect piece leads to larger values compared to the QCD sum rules calculations,
especially in the case of $\bar g_1$.  For the qCEDM operators we will just use the sum rules results, while for the four-fermion operators, where only the indirect pieces can be estimated, we will capture these effects by assigning a large uncertainty.  
Plugging Eqs. \eqref{eq:g0sum} and \eqref{eq:g1sum} or Eqs. \eqref{eq:g0ind} and \eqref{eq:g1ind} in the chiral perturbation theory expression for the nucleon EDM we recover estimates that agree in magnitude with Eqs. \eqref{eq:qcedm1}--\eqref{eq:qcedm2}.

We discussed in some detail various estimates for the qCEDM operators to justify our choices for four-fermion operators, for which complete evaluations of the couplings, either on the Lattice or from QCD sum rules are not available. 
The structure of the couplings is very similar to qCEDM, with a direct piece that could be extracted from dim-6 corrections to the baryon spectrum 
and an indirect piece determined by the ratio of vacuum matrix elements \cite{Cirigliano:2016yhc,Dekens:2022gha}. 
The operators constructed in  Section \ref{sub:fourquark}
belong to the ${\mathbf{8} \times \mathbf{8}}$, 
$\mathbf{6} \times \mathbf{\bar{6}}$ and $\mathbf{3} \times \mathbf{\bar{3}}$ representations of the $SU(3)$ flavor group \cite{Dekens:2022gha}.
The indirect components of the $\pi N$ couplings are given by
\begin{align}
    \left. \bar g_0 \right|_{\rm ind} &= - \frac{F_\pi^2 \bar m g_S^{u-d}}{m_\pi^2}  \left[
     \boldsymbol{\mathcal A}_{\mathbf{8} \times \mathbf{8}}\,  C^{(1,0)}_{\mathbf{8} \times \mathbf{8}} 
     + \boldsymbol{\mathcal{\bar A}}_{\mathbf{8} \times \mathbf{8}} \, C^{(8,0)}_{\mathbf{8} \times \mathbf{8}}     + \boldsymbol{\mathcal A}_{\mathbf{6} \times \mathbf{6}} C^{(1,0)}_{\mathbf{6} \times \mathbf{6}}
    + \boldsymbol{\mathcal{\bar A}}_{\mathbf{6} \times \mathbf{6}} C^{(8,0)}_{\mathbf{6} \times \mathbf{6}} \nonumber \right. \\
& + \left.  
         \boldsymbol{\mathcal A}_{\mathbf{3} \times \mathbf{3}} C^{(1,0)}_{\mathbf{3} \times \mathbf{3}} 
+       \boldsymbol{\mathcal {\bar A}}_{\mathbf{3} \times \mathbf{3}} C^{(8,0)}_{\mathbf{3} \times \mathbf{3}}
\right], \\
    \left. \bar g_1 \right|_{\rm ind} &= - \frac{F_\pi^2 \sigma_{\pi N}}{m_\pi^2}  \left[
     \boldsymbol{\mathcal A}_{\mathbf{8} \times \mathbf{8}} C^{(1,1)}_{\mathbf{8} \times \mathbf{8}}  
     + \boldsymbol{\mathcal{\bar A}}_{\mathbf{8} \times \mathbf{8}} C^{(8,1)}_{\mathbf{8} \times \mathbf{8}}   + \boldsymbol{\mathcal A}_{\mathbf{6} \times \mathbf{6}}  C^{(1,1)}_{\mathbf{6} \times \mathbf{6}}
    + \boldsymbol{\mathcal{\bar A}}_{\mathbf{6} \times \mathbf{6}}   C^{(8,1)}_{\mathbf{6} \times \mathbf{6}} \nonumber \right. \\
& + \left.  
         \boldsymbol{\mathcal A}_{\mathbf{3} \times \mathbf{3}} C^{(1,1)}_{\mathbf{3} \times \mathbf{3}} +
         \boldsymbol{\mathcal {\bar A}}_{\mathbf{3} \times \mathbf{3}} C^{(8,1)}_{\mathbf{3} \times \mathbf{3}}     
    \right].
\end{align}
The operators in the $LL\, RR$ class belong to the ${\mathbf{8} \times \mathbf{8}}$ representation, and we have
\begin{align}
C^{(1,0)}_{\mathbf{8} \times \mathbf{8}} &=   \frac{1}{2} \left( {\rm Im} \left[C^{V1,\, LR}_{uddu} \right]_{1221} + 
     {\rm Im} \left[C^{V1,\, LR}_{dd} \right]_{1221} \right)~, \\
 C^{(8,0)}_{\mathbf{8} \times \mathbf{8}} &= { \frac{1}{2}} \left( {\rm Im} \left[C^{V8,\, LR}_{uddu} \right]_{1221} + 
     {\rm Im} \left[C^{V8,\, LR}_{dd} \right]_{1221} \right)~, \\
 C^{(1,1)}_{\mathbf{8} \times \mathbf{8}} &=    {\rm Im} \left[C^{V1,\, LR}_{uddu} \right]_{1111}
     + \frac{1}{2}
     {\rm Im} \left[C^{V1,\, LR}_{uddu} \right]_{1221} - \frac{1}{2} 
     {\rm Im} \left[C^{V1,\, LR}_{dd} \right]_{1221}~, \\
  C^{(8,1)}_{\mathbf{8} \times \mathbf{8}} &=    {\rm Im} \left[C^{V8,\, LR}_{uddu} \right]_{1111}
     + \frac{1}{2}
     {\rm Im} \left[C^{V8,\, LR}_{uddu} \right]_{1221} - \frac{1}{2} 
     {\rm Im} \left[C^{V8,\, LR}_{dd} \right]_{1221}. 
\end{align}
The first superscript indicates whether the operators are color singlet or octet, while the second superscript whether they are isospin 0 or isospin 1.
The $LR\, LR$ operators, on the other hand, belong to the 
$\mathbf{6} \times \mathbf{\bar{6}}$ and $\mathbf{3} \times \mathbf{\bar{3}}$ representations.
The $\mathbf{6} \times \mathbf{\bar{6}}$ coefficients are given by
\begin{align} \label{eq:C1066}
 C^{(1,0)}_{\mathbf{6} \times \mathbf{6}}  &= 
     \textrm{Im} \left[C^{S1,\, RR}_{uu} \right]_{1111} + \textrm{Im} \left[C^{S1,\, RR}_{dd} \right]_{1111} 
     + {\frac{1}{2}} \left( \textrm{Im} \left[C^{S1,\, RR}_{uddu} \right]_{1111}+  \textrm{Im} \left[C^{S1,\, RR}_{ud}  \right]_{1111} \right) \nonumber  \\ 
       &+ { \frac{1}{4}} \left(\textrm{Im} \left[C^{S1,\, RR}_{uddu} \right]_{1221} +  \textrm{Im} \left[C^{S1,\, RR}_{ud} \right]_{1122}
     +  \textrm{Im} \left[C^{S1,\, RR}_{dd} \right]_{1221}
    +  \textrm{Im} \left[C^{S1,\, RR}_{dd} \right]_{1122}\right),  \\
 C^{(1,1)}_{\mathbf{6} \times \mathbf{6}} &=
     \textrm{Im} \left[C^{S1,\, RR}_{uu} \right]_{1111} {- } \textrm{Im} \left[C^{S1,\, RR}_{dd} \right]_{1111} 
      \nonumber  \\ 
      & + { \frac{1}{4}} \left( \textrm{Im} \left[C^{S1,\, RR}_{uddu} \right]_{1221} +  \textrm{Im} \left[C^{S1,\, RR}_{ud} \right]_{1122}
     -  \textrm{Im} \left[C^{S1,\, RR}_{dd} \right]_{1221}
    -  \textrm{Im} \left[C^{S1,\, RR}_{dd} \right]_{1122}  \right),
    \label{eq:C1166}
\end{align}
with the same expressions for the color octet operators, with $S1 \rightarrow S8$.
For the $\mathbf{3} \times \mathbf{\bar{3}}$ coefficients we have
\begin{align} \label{eq:C1033}
C^{(1,0)}_{\mathbf{3} \times \mathbf{3}}  &=  { \frac{1}{2} }  \left( - \textrm{Im} \left[C^{S1,\, RR}_{uddu} \right]_{1111}+  \textrm{Im} \left[C^{S1,\, RR}_{ud}  \right]_{1111} \right) \nonumber  \\ 
     &  + { \frac{1}{4} } \left( -\textrm{Im} \left[C^{S1,\, RR}_{uddu} \right]_{1221} +  \textrm{Im} \left[C^{S1,\, RR}_{ud} \right]_{1122}
     -  \textrm{Im} \left[C^{S1,\, RR}_{dd} \right]_{1221}
    +  \textrm{Im} \left[C^{S1,\, RR}_{dd} \right]_{1122}  \right) \\
    C^{(1,1)}_{\mathbf{3} \times \mathbf{3}}  &= { \frac{1}{4} }\left(
    - \textrm{Im} \left[C^{S1,\, RR}_{uddu} \right]_{1221} +  \textrm{Im} \left[C^{S1,\, RR}_{ud} \right]_{1122}
     + \textrm{Im} \left[C^{S1,\, RR}_{dd} \right]_{1221}
    -  \textrm{Im} \left[C^{S1,\, RR}_{dd} \right]_{1122} \right),
    \label{eq:C1133}
\end{align}
and again the color octet coefficients are obtained by $S1 \rightarrow S8$.

The ${\mathbf{8} \times \mathbf{8}}$ and  
$\mathbf{6} \times \mathbf{\bar{6}}$ are related by $SU(3)$ chiral symmetry to 
operators that contribute to $K$-$\bar{K}$ oscillations,  $K \rightarrow \pi \pi$ and 
to short-distance contributions to neutrinoless double beta decay \cite{Cirigliano:2017ymo}.
Here we determine the LECs from calculations of neutrinoless double beta decay matrix elements \cite{Nicholson:2018mwc} and write
\begin{align}
    \boldsymbol{\mathcal A}_{\mathbf{8} \times \mathbf{8}} = - g_4^{\pi\pi}, \qquad 
    \boldsymbol{\mathcal{\bar A}}_{\mathbf{8} \times \mathbf{8}} = - \left[\frac{1}{2} g_5^{\pi\pi} - \frac{1}{2 N_c} g_4^{\pi\pi} \right] \\
    \boldsymbol{\mathcal A}_{\mathbf{6} \times \mathbf{6}} = - g_2^{\pi\pi}, \qquad 
    \boldsymbol{\mathcal{\bar A}}_{\mathbf{6} \times \mathbf{6}} = - \left[\frac{1}{2} g_3^{\pi\pi} - \frac{1}{2 N_c} g_2^{\pi\pi} \right],
\end{align}
with 
\begin{align}
    g_2^{\pi\pi} &= 2.0(2) \; {\rm GeV}^2, & 
g_3^{\pi\pi} &= -0.62(6) \; {\rm GeV}^2, \\
    g_4^{\pi\pi} &= -1.9(2) \; {\rm GeV}^2, &
g_5^{\pi\pi} &= -8.0(6) \; {\rm GeV}^2, 
\end{align}
in the $\overline{\rm MS}$ scheme, at $\mu = 2$ GeV.
The LECs for $\boldsymbol{\mathcal A}_{3\times 3}$ and 
$\boldsymbol{\mathcal{\bar A}}_{3\times 3}$
are at the moment unknown. 
We will set them to zero but stress that this is an uncontrolled assumption.
\begin{equation}
\boldsymbol{\mathcal A}_{3\times 3} = 0 \;  {\rm GeV}^2 \qquad 
\boldsymbol{{\mathcal{\bar A}}}_{3\times 3} = 0 \;  {\rm GeV}^2 
\end{equation}

%%%%%%%%%%%%%%
\subsection{Diamagnetic Atomic EDMs}
We finally consider EDMs of diamagnetic atoms. 
The strongest bound is currently on the EDM of $^{199}$Hg \cite{Graner:2016ses,Griffith:2009zz}, followed by 
$^{129}$Xe \cite{Sachdeva:2019rkt}.  We will also consider $^{225}$Ra,
which has enhanced sensitivity to the nuclear Schiff moment \cite{Engel:1999np,Spevak:1996tu} and could provide important constraints in the future \cite{Alarcon:2022ero}.
We can write the EDM of diamagnetic atoms as a term induced by the nuclear Schiff moment, plus contributions from semileptonic operators
\begin{equation}
d_{^A X} = d_{\rm Schiff} +  \sum_{N = n, p}  \left( \alpha^{N}_S C^{(N)}_S + \alpha^{N}_P C^{(N)}_P + \alpha^N_T C^{(N)}_T \right).
\end{equation}
The Schiff moment component is sensitive to the neutron and proton EDM and to $PT$-odd nucleon-nucleon interactions.
These have been classified in phenomenological one-meson-exchange models \cite{Barton:1961eg,Towner:1994qe,Engel:1999np},
and in chiral EFT and pionless EFT \cite{Maekawa:2011vs,deVries:2012ab,Song:2012yh,deVries:2020iea,Yang:2020ges}.
For chiral-symmetry breaking operators, such as the qCEDM and the four-quark operators, $\bar g_{0,1}$ give the leading contribution to the 
$PT$-odd nucleon-nucleon potential \cite{Maekawa:2011vs,deVries:2012ab}, even though short-range operators in the $^3P_0$-$^1S_0$ channel are required for renormalization
\cite{deVries:2020loy}. For chiral-invariant operators, such as the Weinberg three-gluon operator, $\pi N$ couplings and 
short-range isospin-invariant  $NN$ couplings contribute at the same order \cite{Maekawa:2011vs,deVries:2012ab}. Since little can be said about the size of the $NN$ CP-odd short-range interactions induced by WET operators, we focus here on the contributions to the Schiff moment from $\bar g_{0,1}$. With progress from Lattice QCD and nuclear theory, the Schiff moment expressions could in the future be extended to include additional operators.

We write 
\begin{equation}
    d_{ \rm Schiff} = A_{\rm Schiff} \left( \alpha_n d_n + \alpha_p d_p + a_0 \frac{\bar g_0}{F_\pi}  + a_1 \frac{\bar g_1}{F_\pi} + a_2 \frac{\bar g_2}{F_\pi}   \right). 
\end{equation}
The prefactor $A_{\rm Schiff}$ accounts for the Schiff screening, while the coefficients  $\alpha_{n,p}$ and $a_{0,1,2}$ 
contain the information on the nuclear structure of the systems of interest. We report them in Table \ref{Tab:Schiff}. 
We use the results of Ref. \cite{Hubert:2022pnl} for the screening factors of Hg and Xe, and Ref. \cite{Prasannaa:2020cjx} for $^{225}$Ra.
In the case of $^{199}$Hg,  $\alpha_{n,p}$ are known with good accuracy \cite{Dmitriev:2003sc}, while in the case of Xe we use the calculation of $\alpha_n$ from  
\cite{Yanase:2022atk}.
The contributions of the CP-odd $\pi N$ couplings are affected by substantial uncertainties. Here we take the ranges suggested in Refs. \cite{Engel:2013lsa,Chupp:2017rkp}, which considers results from various nuclear structure calculations. 

\begin{table}[t]
\begin{center}
\renewcommand*{\arraystretch}{1.5}
 \begin{adjustbox}{width=0.9\textwidth}
\begin{tabular}{||c||c|c|c|c|c|c||}
\hline  &  $A_{\rm Schiff}$ & $\alpha_n$ & $\alpha_p$ & $a_0$ ($e$ fm) & $a_1$ ($e$ fm) & $a_2$ ($e$ fm)\\ \hline & & & & & &  \\
   $^{199}$Hg  &  $-( 2.40 \pm 0.24) \cdot 10^{-4}$ & $1.9 \pm 0.1$ & $0.20 \pm 0.06$ & $0.13^{+0.5}_{-0.07}$ & $0.25^{+0.89}_{-0.63}$ & $0.09^{+0.17}_{-0.04}$ \\
    & & & & & & \\
  $^{129}$Xe   & $-(0.364 \pm 0.025) \cdot 10^{-4}$  & $-(0.29 \pm 0.10)$ & -- & $0.10^{+0.53}_{-0.037}$ & $0.076^{+0.55}_{-0.038}$ &   \\
  & & & & & & \\
  $^{225}$Ra   & $\left(6.3 \pm 0.5\right) \cdot 10^{-4}$ & -- & -- & $2.5 \pm 7.5$ & $-65\pm40$ & $14\pm 6.5$ \\
  \hline \hline
\end{tabular}
\end{adjustbox}
\caption{Nuclear and atomic theory input for the EDMs of diamagnetic atoms. }\label{Tab:Schiff}
\end{center}
\end{table}

The contributions from semileptonic operators are usually expressed in terms of nucleon-level operators. In addition to the scalar operators in Eq. \eqref{eq:scalar},
one can write pseudoscalar and tensor terms. In a relativistic notation
\begin{equation}
\mathcal L = - \frac{G_F}{\sqrt{2}} \left( \bar e e  \bar N \left(C^{(0)}_P + \tau_3 C^{(1)}_P \right) i \gamma_5 N  
- \frac{1}{2} \varepsilon_{\mu \nu \alpha \beta}\, \bar e \sigma^{\mu \nu}  e\, \bar N \sigma^{\alpha \beta} \left(C_T^{(0)} + C_T^{(1)}\tau_3 \right) N 
\right),
\end{equation}
where the isoscalar/isovector couplings are related to proton/neutron couplings by
\begin{equation}
C^{(p)}_X = C^{(0)}_X + C^{(1)}_X, \qquad C^{(n)}_X =  C^{(0)}_X - C^{(1)}_X.
\end{equation}
As can be more clearly seen in a non-relativistic notation \cite{Dekens:2018bci}, the pseudoscalar operators vanish at zero momentum 
and are thus proportional to powers of $Q/\Lambda_{\rm had}$, where 
$\Lambda_{\rm had}$ is an hadronic scale. Here we only retain the isovector component, which is dominated by the pion pole contribution 
and thus relatively large, with $\Lambda_{\rm had} \sim m_u + m_d$.
\begin{equation}
C^{(1)}_P =  - \frac{v^2}{2} \left( \textrm{Im}[C^{\rm SRR}_{eu}]_{1111} - \textrm{Im}[C^{\rm SRL}_{eu}]_{1111} - \textrm{Im}[C^{\rm SRR}_{ed}]_{1111} + \textrm{Im}[C^{\rm SRL}_{ed}]_{1111} \right) \frac{g_A m_N}{\bar m}.
\end{equation}
The isoscalar component is suppressed by the $\eta$ mass, rather than the pion mass, and thus we neglect it. 
Its contribution was considered, for example, in Ref. \cite{Degenkolb:2024eve}.
In terms of WET operators, the tensor coefficients are
\begin{align}
C_T^{(0)}  &=  -\frac{v^2}{2} \left( \frac{g_T^u + g_T^d}{2}  \left(  \textrm{Im}[C^{\rm TRR}_{eu}]_{1111}  + \textrm{Im}[ C^{\rm TRR}_{ed}]_{1111}  \right) + g_T^s  \textrm{Im}[ C^{\rm TRR}_{ed}]_{1122} \right) \\
C_T^{(1)}  &=  -\frac{v^2}{2} \frac{g_T^u - g_T^d}{2}  \left(  \textrm{Im}[C^{\rm TRR}_{eu}]_{1111}  - \textrm{Im}[ C^{\rm TRR}_{ed}]_{1111}  \right).
\end{align}
The coefficients $\alpha^N_S$ are given by
\begin{equation}
\alpha_S^p = \mathcal R_{\rm SP} \frac{Z}{A}, \qquad  \alpha_S^n = \mathcal R_{\rm SP} \frac{N}{A}.
\end{equation}
$(Z, N, A)$= (80, 121, 201), (54, 77, 131), (87, 136,  223) are
the atomic number, number of neutrons, and mass number of Hg, Xe, and Ra, respectively.
$\mathcal R_{\rm SP}$ depends solely on atomic theory, and, for Mercury and Xexon, we use the 
values \cite{Fleig:2018bsf,Fleig:2020obr}.  
\begin{equation}
\left. \mathcal R_{\rm SP} \right|_{\rm Hg} = -\left(2.8 \pm 0.6\right) \cdot  10^{-22} e\, {\rm cm} \qquad 
\left. \mathcal R_{\rm SP} \right|_{\rm Xe} = -\left(0.71  \pm 0.18\right) \cdot  10^{-23} e\, {\rm cm}.
\end{equation}
These values are somewhat smaller than in other calculations, especially in the case of xenon \cite{Gaul:2023hdd}.
The tensor coefficients can be written as 
\begin{equation}
\alpha_T^{p}  = \mathcal R_T \langle \sigma_{p z} \rangle, \qquad  \alpha_T^n =  \mathcal R_T \langle \sigma_{n z} \rangle,
\end{equation}
where $\mathcal R_T$ is calculated in atomic theory, while   $\langle \sigma_{p z} \rangle$
and $\langle \sigma_{n z} \rangle$ are nuclear spin matrix elements.
We use \cite{Fleig:2018bsf,Fleig:2020obr}
\begin{equation}
\left. \mathcal R_{\rm T} \right|_{\rm Hg} = -\left( 2.8 \pm 0.6 \right) \cdot  10^{-20} e\, {\rm cm} \qquad 
\left. \mathcal R_{\rm T} \right|_{\rm Xe} =  \left(0.520  \pm 0.049\right) \cdot  10^{-20} e\, {\rm cm},
\end{equation}
and \cite{Yanase:2018qqq,Menendez:2012tm}
\begin{align}
\left. \langle \sigma_{n z} \rangle \right|_{\rm Hg}  &= -0.377, \qquad  \left. \langle \sigma_{p z} \rangle\right|_{\rm Hg}   =  0.009 \\
\left. \langle \sigma_{n z} \rangle \right|_{\rm Xe}  &= + 0.658, \qquad  \left. \langle \sigma_{p z} \rangle\right|_{\rm Xe}   =  0.020.
\end{align}
Finally, $\alpha^N_P$ are related to the tensor contributions by
\cite{Ginges:2003qt}
\begin{equation}
\alpha^N_P = \frac{Z \alpha_{\rm em}}{5 m_N R} \alpha^N_T 
\end{equation}
where $R = 1.2 A^{1/3}$ fm is the nuclear radius.

\section{Renormalization group running and threshold effects}
\label{sec:runmatch}
The WET operators containing heavier fermions (charm, bottom, muon, and tau) can contribute to the electron, neutron, atomic, and molecular EDMs via 
 renormalization group evolution and threshold effects arising when the heavy particles are integrated out. In this section, we discuss the following 
 three kinds of  contributions of WET operators with heavier generations to the 
 EDMs observables given in Sec.~\ref{sec:edms}, focusing only on the leading effects\footnote{Meaning if an 
 operator contributes at the tree-level its 1-loop effects are not discussed in this section, and so on. }:
 \begin{itemize}
 \item Leading-log QCD+QED RG running from EW scale to $\mu=2$ GeV.
 \item 1-loop threshold effects at $\mu=m_c$ and $\mu=m_b$. 
 \item Next-to-leading-log 2-loop QCD+QED RG running from EW scale to $\mu=2$ GeV.
 \end{itemize}
The logarithmic order at which each operator contributes is indicated in Table \ref{tab:wet4}.
In the following, we will dissect how these contributions arise.

 \subsection{\boldmath Leading-log 1-loop RG running}

\begin{figure}
\hspace{0.8cm}
\includegraphics[width=0.9\textwidth]{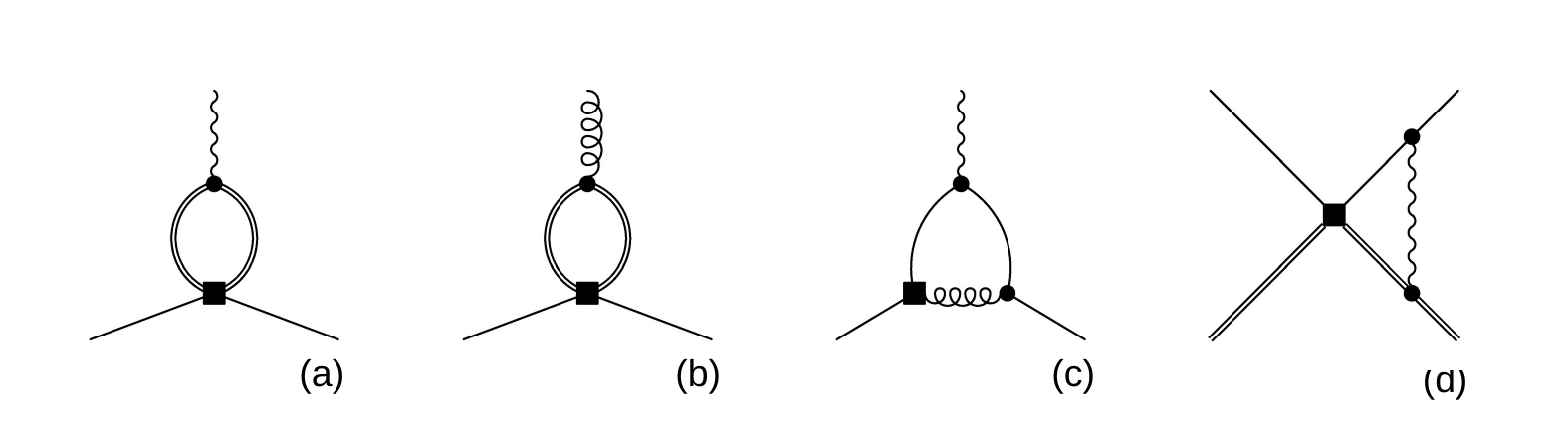}
\caption{Representative diagrams contributing to the 1-loop anomalous dimension, which induces LL contributions to EDMs.
Plain and double lines denote light ($e$ and $u$, $d$, $s$ quarks) and heavy fermions ($\mu$, $\tau$ leptons and $c$, $b$ quarks), respectively.
Wavy and wiggly lines denote photons and gluons. Insertions of WET operators are denoted by squares, while SM vertices by dots. 
Diagrams $(a)$ and $(b)$ contribute to the mixing of heavy-light four-fermion operators onto photon and gluon dipole operators.
Four-quark operators of the type $\mathcal O_{uu}^{S1, RR}$,  $\mathcal O_{uu}^{S8, RR}$, 
 $\mathcal O_{dd}^{S8, RR}$,  $\mathcal O_{dd}^{S8, RR}$,  $\mathcal O_{uddu}^{S1, RR}$ and  $\mathcal O_{uddu}^{S8, RR}$,
with two heavy and two light quarks,  
semileptonic tensor operators $\mathcal O^{T, RR}_{eu}$ 
$\mathcal O^{T, RR}_{ed}$ with heavy quarks or heavy leptons, and the leptonic scalar operator $\mathcal O^{S, RR}_{ee}$ with two heavy and two light leptons
generate LL EDMs via this path. Diagram $(c)$ denotes the mixing of dipole operators. 
Diagram $(d)$ denotes the mixing of the semileptonic scalar operators 
$\mathcal O^{S, RR}_{eu}$ and $\mathcal O^{S, RR}_{ed}$ onto tensor operators.
Combined with the running of the tensor into dipoles, these induce corrections to the electron EDM starting at $\mathcal O(L_\mu^2)$.
}\label{Fig:RGE1}
\end{figure}

The complete 1-loop ADMs  in WET including  $\Delta F=0$ 
operators have been calculated in \cite{Jenkins:2017dyc}.  In what follows, we 
identify all relevant terms in the ADMs and provide the naive solutions for the corresponding RGEs. 
Through diagrams such as diagram $(a)$ and $(c)$ in Figure \ref{Fig:RGE1}, 
the QCD dipole and 
four-quark scalar operators mix onto the photon dipole. Neglecting the running of the electromagnetic and strong couplings, and the running of the
Wilson coefficients, we can denote the solution of the RGEs as
\be \label{eq:agamma-sol}
\begin{aligned}
\wc[]{f\gamma}{ii} & =  {8  e g_s q_i C_F }  \wc[]{f G}{ii} L_\mu\,, \\
\wc[]{f\gamma}{ii}  & =  {2 e q_j  m_j  } \wc[S1,RR]{ff}{i jj i}   L_\mu\,, \\
\wc[]{f\gamma}{ii}  & =  {2 e  q_j  m_j C_F} \wc[S8,RR]{ff}{i jj i}  L_\mu\,, \\
%%%%%%%%%
\wc[]{f\gamma}{ii} & =  -8 e q_e  m_j \wc[T,RR]{ef}{jj ii}  L_\mu \,, \\
%%%%
\wc[]{u\gamma}{ii} & =   {e q_j  m_j  }  \wc[S1,RR]{uddu}{i jj i}   L_\mu\,, \\
\wc[]{u\gamma}{ii} & =   {e q_j   m_j C_F } \wc[S8,RR]{uddu}{i jj i}   L_\mu .
%\wc[]{b\gamma}{ii} & =   e q_i  C_F m_i \wc[S8,RR]{adda}{i jj i} \log \left ( \frac{\mu_{ew}}{\mu_{low}} \right ) .
\end{aligned}
\ee
where we define
\begin{equation}
L_\mu = \frac{1}{16 \pi^2} \log \frac{\mu_{low}}{\mu_{ew}}.
\end{equation}
Here $f=u$ or $d$, and $\mu_{low}$ denotes the scale to which we run, which depends on the flavor of the particle in the loop.
$e$ denotes the electric charge, while $q_j$ is the charge of the quark or lepton
$j$, with $q_e = -1$, $q_d = -1/3$ and $q_u = 2/3$.
 $C_F = (N_c^2 -1)/(2 N_c)$ and $C_A = N_c$, with $N_c$ the number of colors.
The mixing of $\wc[S1(S8),RR]{uddu}{ijji}$ onto $\wc[]{d\gamma}{}$ is governed the 
last two equations in  \eqref{eq:agamma-sol} by 
replacing $q_j, m_j$ by $q_i, m_i$, respectively. 
The QCD dipoles mix with the electromagnetic dipoles and 
four-quark scalar operators
\be \label{eq:aG-sol}
\begin{aligned}
\wc[]{fG}{ii} & =  {2 g_s  m_j }  \wc[S1,RR]{ff}{i jj i}  L_\mu\,, \\
\wc[]{fG}{ii} & =  {2 g_s m_j}  \left(C_F - {C_A \over 2 }\right)  \wc[S8,RR]{ff}{i jj i}  L_\mu\,, \\
%%%%
\wc[]{uG}{ii} & =   {g_s m_j } \wc[S1,RR]{uddu}{i jj i}  L_\mu\,, \\
\wc[]{uG}{ii} & =   {g_s m_j }  \left( C_F- {C_A \over 2}\right) \wc[S8,RR]{uddu}{i jj i} L_\mu.
\end{aligned}
\ee
The mixing of $\wc[]{f\gamma}{}$ onto  $\wc[]{f G}{}$ follows the exact same relation as the 
first equation in \eqref{eq:agamma-sol}. Also, the mixing of $\wc[S1(S8),RR]{uddu}{}$ onto $\wc[]{dG}{}$ 
follows the same relations as the last two equations in \eqref{eq:aG-sol}.
Eqs. \eqref{eq:agamma-sol} and \eqref{eq:aG-sol} show that all scalar operators with two heavy and two light quarks induce light quark EDMs
and CEDMs at one loop, leading, as we will see, to strong constraints from neutron and mercury EDM experiments.
%%%%%
%%%%%%
%
For the electron EDM, the mixing of the scalar and tensor operators onto the dipole operator 
is governed by pure QED
\be \label{eq:egamma-sol}
\begin{aligned}
\wc[]{e\gamma}{ii} & =  {-8 e   q_i m_j N_c}  \wc[T,RR]{ed}{iijj} L_\mu\,, \\
\wc[]{e\gamma}{ii} & =  {-8 e  q_j  m_j N_c} \wc[T,RR]{eu}{iijj} L_\mu\,, \\
\wc[]{e\gamma}{ii} & =  {2 e  q_j  m_j} \wc[S,RR]{ee}{i jj i} L_\mu. 
\end{aligned}
\ee
The first two relations in \eqref{eq:egamma-sol} indicate that semileptonic operators with 
 heavier generations of up and down quarks can mix with the electromagnetic dipole operator for the electron. 
Similarly, the muon and tau scalar operators can also mix with the dipole operator as indicated by the last relation 
in \eqref{eq:egamma-sol}.

Finally, through diagrams such as diagram $(d)$ in Fig. \ref{Fig:RGE1}, 
the leptonic operators 
 $\wc[S,RR]{ee}{1122}$ and  $\wc[S,RR]{ee}{1133}$ mix with 
$\wc[S,RR]{e\gamma}{11}$ via the $\wc[S,RR]{ee}{1221}$ and $\wc[S,RR]{ee}{1331}$ operators.
Similarly, semileptonic scalar operators mix into tensor operators,
and scalar operators of the form 
 $\wc[S1 (8),RR]{uu}{iijj}$,
  $\wc[S1 (8),RR]{dd}{iijj}$
 and  $\wc[S1 (8),RR]{ud}{iijj}$
 mix onto $\wc[S1 (8),RR]{uu}{ijji}$, $\wc[S1 (8),RR]{dd}{ijji}$
and $\wc[S1 (8),RR]{uddu}{ijji}$, respectively.
In this way, these operators induce  LL contribution to the electron and quark dipole operators, which however start at two loops
and go as  $L_\mu^2$. 
 \be \label{eq:egamma11_SRR}
 \begin{aligned}
 \wc[]{ e\gamma}{11} & =  { 32q_e^3 e^3 m_j}
    \wc[S,RR]{ee}{11jj}   \, \frac{  L^2_\mu}{2}
     \,, \\ 
         \wc[]{ e\gamma}{11} & =   { 16   }   {q_e q_u^2 e^3 m_j N_c }
    \wc[S,RR]{eu}{11jj}  \, { \frac{L^2_\mu}{2}}  \,,\\
      \wc[]{ e\gamma}{11}  & = { 16  }   {q_e q_d^2 e^3 m_j N_c }
    \wc[S,RR]{ed}{11jj} \, { \frac{L^2_\mu}{2}} .  
    %%%%
 \end{aligned}
 \ee
In the first relation in \eqref{eq:egamma11_SRR}, $jj=22,33 $, for the second $jj= 11,22$ and for third $jj=11,22$, or $33$.
Similarly, the semileptonic operators also mix with the quark dipoles
  \be \label{eq:egamma11_SRR_2}
 \begin{aligned}
 \wc[]{ u\gamma}{ii}  & =  { 16 }   {q_e^2 q_u e^3  m_j}
    \wc[S,RR]{eu}{jjii}  { \frac{L^2_\mu}{2}} 
     \,, \\ 
         \wc[]{ d\gamma}{ii}  & =  { 16  }   {q_e^2 q_d e^3  m_j}
    \wc[S,RR]{ed}{jjii}   { \frac{L^2_\mu}{2}}.
 \end{aligned}
 \ee
Here, $jj =11,22$ or $33$, however, only $33$ cases can induce significant effects. 
Finally, four-quark scalar operators also generate $L^2_\mu$ contributions. In the 
case of $uu$ and $dd$ operators, the mixing is given by
\be
\begin{aligned}
\wc[]{f\gamma}{ii} & = \frac{2e q_f m_j }{N_c}   (16e^2 q_u^2 + 16 g_s^2 C_F)  \wc[S1,RR]{ff}{iijj}   
\frac{L^2_\mu}{2} \,,\\
%%%%%%%%
\wc[]{f\gamma}{ii}  & = \frac {2e C_F q_f m_j }{N_c}  
\left ( 16e^2 q_u^2 C_F + 
2g_s^2 \left ( \frac{2}{N_c^2} +  N_c^2 -3 \right) \right )  \wc[S8,RR]{ff}{iijj}   
\frac{L^2_\mu}{2} \,, \\
%%%%%%%%%%%%
\wc[]{fG}{ii} & = \frac{2 g_s m_j }{N_c}    (16e^2 q_u^2 + 16 g_s^2 C_F)  \wc[S1,RR]{ff}{iijj}   
\frac{L^2_\mu}{2} \,,\\
%%%%%%%%
\wc[]{fG}{ii}  & =  -\frac{2 g_s}{N_c} m_j  \left(C_F -{C_A \over 2}\right)   
\left ( 16e^2 q_u^2 - 
4g_s^2 \left ( \frac{2}{N_c} +  N_c  \right) \right )  \wc[S8,RR]{ff}{iijj}   
\frac{L^2_\mu}{2} \,, 
\end{aligned}
\ee
for $f=u$ or $d$.  $\wc[S1(8), RR]{ud}{}$ operators induce both QED and QCD dipoles. For the QED dipoles, one finds
\be
\begin{aligned}\label{eq:scalar_to_dipole1}
\wc[]{u\gamma}{ii} & = \frac {e q_d m_j }{N_c}    (16e^2 q_u q_d + 16 g_s^2 C_F)  \wc[S1,RR]{ud}{iijj}   
\frac{L_\mu^2}{2}  \,,\\
%%%%%%%%
\wc[]{u\gamma}{ii}  & = \frac{2e C_F q_d m_j }{N_c} 
\left ( 16e^2 q_u q_d C_F + 
2g_s^2 \left ( \frac{2}{N_c^2} +  N_c^2 -3 \right) \right )  \wc[S8,RR]{ud}{iijj}   
\frac{L_\mu^2}{2} \,, \\
%%%%%%
\wc[]{d\gamma}{jj}  & = \frac{e q_d m_i }{N_c}    (16e^2 q_u q_d + 16 g_s^2 C_F)  \wc[S1,RR]{ud}{iijj}  
\frac{L_\mu^2}{2}   \,,\\
%%%%%%%%
\wc[]{d\gamma}{jj} & = \frac{2e C_F q_d m_i}{N_c}  
\left ( 16e^2 q_u q_d C_F + 
2g_s^2 \left ( \frac{2}{N_c^2} +  N_c^2 -3 \right) \right )  \wc[S8,RR]{ud}{iijj}   
\frac{L_\mu^2}{2} ,
\end{aligned}
\ee
and, for the QCD dipoles,
\be
\begin{aligned}\label{eq:scalar_to_dipole2}
\wc[]{u G}{ii} & = \frac {g_s m_j  }{N_c}   (16e^2 q_u q_d + 16 g_s^2 C_F)  \wc[S1,RR]{ud}{iijj}   
\frac{L_\mu^2}{2}  \,,\\
%%%%%%%%
\wc[]{u G}{ii}  & = \frac{g_s  }{N_c}     m_j \left(C_F - {C_A \over 2}\right)  
\left ( 16e^2 q_u q_d C_F + 
2g_s^2 \left ( \frac{2}{N_c^2} +  N_c^2 -3 \right) \right )  \wc[S8,RR]{ud}{iijj}   
\frac{L_\mu^2}{2} \,, \\
%%%%%%
\wc[]{d G}{jj} & = \frac{ g_s  }{N_c}      m_i  (16e^2 q_u q_d + 16 g_s^2 C_F)  \wc[S1,RR]{ud}{iijj}   
\frac{L_\mu^2}{2}  \,,\\
%%%%%%%%
\wc[]{d G}{jj}  & = \frac{ g_s  }{N_c}  m_j  \left(C_F - {C_A \over 2}\right)   
\left ( 16e^2 q_u q_d C_F + 
2g_s^2 \left ( \frac{2}{N_c^2} +  N_c^2 -3 \right) \right )  \wc[S8,RR]{ud}{iijj}   
\frac{L_\mu^2}{2}.
\end{aligned}
\ee
This exhausts all the operators denoted as LL in Table \ref{tab:wet4}.

 %%%%%%%
 \begin{center}
\begin{figure}[t]
\hspace{1.5cm}
\includegraphics[width=0.75\textwidth]{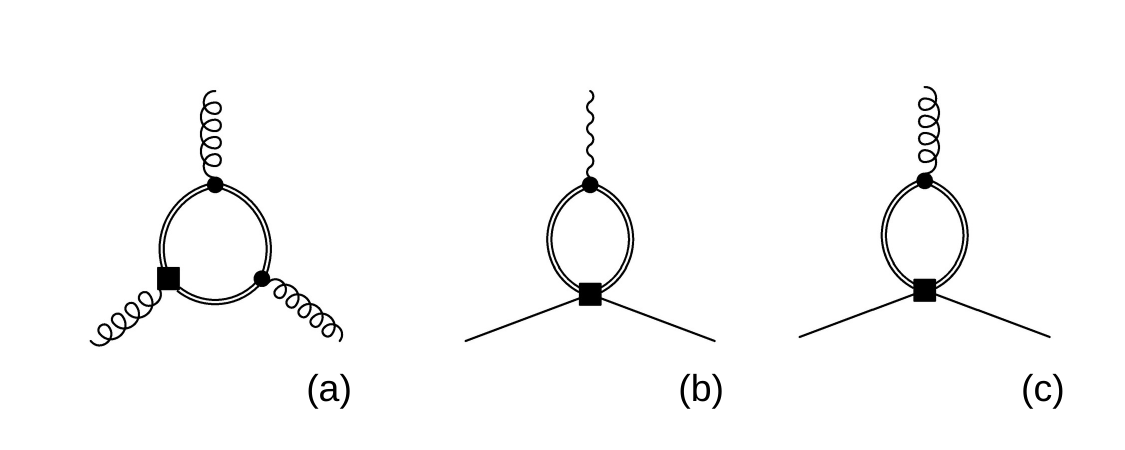}
\caption{Matching contributions arising from integrating out heavy particles. Diagram $(a)$ shows contributions to the Weinberg three-gluon operator 
from gluon dipole operators with heavy quarks.  Diagram $(b)$ and $(c)$ illustrate matching corrections from the four-quark operators 
$\mathcal O^{V1, LR}_{uddu}$, $\mathcal O^{V8, LR}_{uddu}$, $\mathcal O^{V1, LR}_{dd}$, $\mathcal O^{V8, LR}_{dd}$,
$\mathcal O^{V1, LR}_{uu}$ and $\mathcal O^{V8, LR}_{dd}$ and from the leptonic vector operators $\mathcal O^{V, LR}_{ee}$ onto dipole operators.}
\label{Fig:RGE3}
\end{figure}
\end{center}
 
\subsection{\boldmath 1-loop Matching}
In this subsection, we discuss the threshold effects at $m_c$, $m_b$, $m_\mu$, and $m_\tau$ at the 1-loop level.  
\subsubsection{\boldmath Matching to Weinberg operators }
$\wc[]{dG}{33}$ mixes with $\wc[]{\widetilde G}{}$ after integrating out $b$-quark \cite{Braaten:1990gq}.
After adjusting the normalization of the operators, in the JMS basis,
we find
\begin{equation} \label{eq:ggg1}
\wc[]{\widetilde G}{}(\mu_b^-) = \wc[]{\widetilde G}{}(\mu_b^+) +\frac{1}{3 m_b}   \frac{\alpha_s}{8 \pi} {\rm Im}\wc[]{dG}{33}(\mu_b^+).
\end{equation}
The coefficient of second term is $\approx 7.1 \cdot 10^{-4}$. 
The 1-loop Feynman diagram governing this mixing is given in   Fig.~\ref{Fig:RGE3}, diagram $(a)$. 
Similarly, we can integrate out the charm quark  leading to
\begin{equation} \label{eq:ggg2}
\wc[]{\widetilde G}{}(\mu_c^-) = \wc[]{\widetilde G}{}(\mu_c^+) + \frac{1}{3 m_c} \frac{\alpha_s}{8 \pi} {\rm Im}\wc[]{uG}{22}(\mu_c^+).
\end{equation}
The coefficient of second term is $\approx 5.2 \cdot 10^{-3}$. 
\subsubsection{Matching to dipole operators }\label{sec:vector}
The vector operators with left-right chiral currents match onto dipole operators at one loop  \cite{Alioli:2017ces}. For chromo-dipoles, in the JMS basis, we find
\be
\begin{aligned}
\wc[]{uG}{ii}(\mu^-) & = \wc[]{uG}{ii}(\mu^+)  - { g_s m_j \over 16\pi^2 } \left( \wc[V1,LR]{uddu}{ijji}(\mu^+) -{1\over 6}    \wc[V8,LR]{uddu}{ijji}(\mu^+) \right)   \,,\\ 
\wc[]{dG}{jj} (\mu^-)& = \wc[]{dG}{jj}(\mu^+)  - { g_s m_i \over 16\pi^2 } \left(\wc[V1,LR]{uddu}{ijji}(\mu^+)  - {1\over 6}  \wc[V8,LR]{uddu}{ijji}(\mu^+)     \right) \,,  \\
%%%%%%%%%%
\wc[]{uG}{ii}(\mu^-) & = \wc[]{uG}{ii}(\mu^+)  - { g_s m_j \over 16\pi^2 } \left( \wc[V1,LR]{uu}{ijji}(\mu^+)  -{1\over 6}    \wc[V8,LR]{uu}{ijji}(\mu^+) \right)    \,,\\ 
%%%%%%%%%%%%%
\wc[]{dG}{jj} (\mu^-)& = \wc[]{dG}{jj}(\mu^+)  - { g_s m_i \over 16\pi^2 } \left(\wc[V1,LR]{dd}{ijji}(\mu^+)  - {1\over 6}  \wc[V8,LR]{dd}{ijji} (\mu^+)    \right) \,, 
\end{aligned}
\ee
similarly, using the last two equations one can obtain the matching for $\wc[]{uG}{jj}$ and $\wc[]{dG}{ii}$ . For the photon dipoles, we have 
\be
\begin{aligned}
\wc[]{u\gamma}{ii}(\mu^-) & =  \wc[]{u\gamma}{ii}(\mu^+)  + { e Q_j m_j \over 16\pi^2 } \left( \wc[V1,LR]{uddu}{ijji}(\mu^+)  -{4\over 3}    \wc[V8,LR]{uddu}{ijji} (\mu^+) \right)  \,,\\ 
\wc[]{d\gamma}{jj}(\mu^-) & =  \wc[]{d\gamma}{jj}(\mu^+)  + { e Q_i m_i \over 16\pi^2 } \left( \wc[V1,LR]{uddu}{ijji}(\mu^+)  -{4\over 3}    \wc[V8,LR]{uddu}{ijji}(\mu^+) \right)  \,,\\
%%%%%%%%%%%
\wc[]{u\gamma}{ii}(\mu^-) & =  \wc[]{u\gamma}{ii}(\mu^+)  + { e Q_j m_j \over 16\pi^2 } \left( \wc[V1,LR]{uu}{ijji} (\mu^+) -{4\over 3}    \wc[V8,LR]{uu}{ijji}(\mu^+) \right)  \,,\\ 
\wc[]{d\gamma}{jj}(\mu^-) & =  \wc[]{d\gamma}{jj}(\mu^+)  + { e Q_i m_i \over 16\pi^2 } \left( \wc[V1,LR]{dd}{ijji}(\mu^+)  -{4\over 3}    \wc[V8,LR]{dd}{ijji}(\mu^+) \right) .
\end{aligned}
\ee
In the above relations, the scale $\mu$ can be set to the appropriate matching scale of either $m_c$ or $m_b$. Finally, for the leptonic
 dipoles, at $\mu=m_\tau$, we obtain \cite{Aebischer:2021uvt}
\be
\wc[]{e\gamma}{ii}(\mu^-) =  \wc[]{e\gamma}{ii}(\mu^+)  + { e Q_j m_j \over 16\pi^2 }  \wc[V,LR]{ee}{ijji} (\mu^+).
\ee
%%%%%

\subsection{\boldmath Two-step operator mixing}

\begin{figure}
\hspace{0.4cm}
\includegraphics[width=0.95\textwidth]{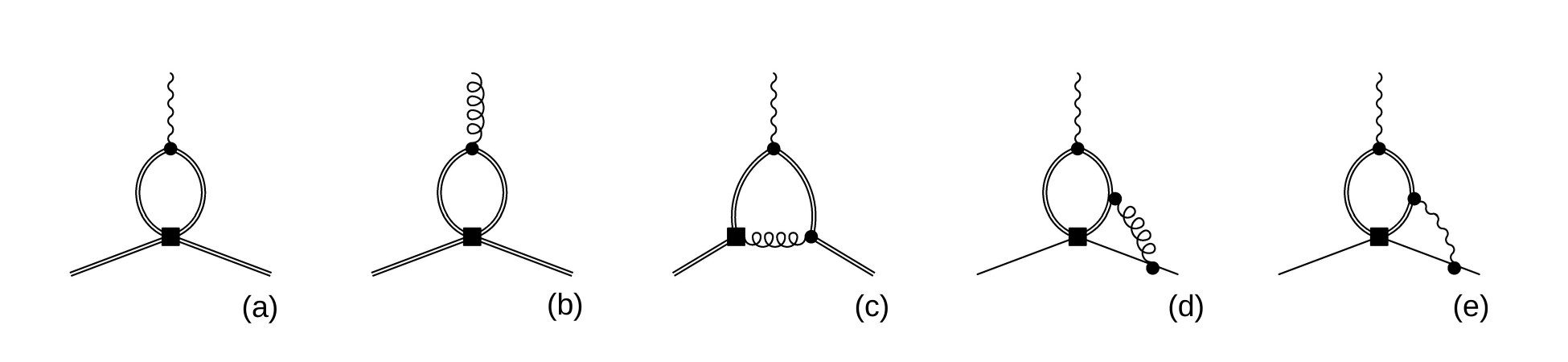}
\caption{Representative diagrams that induce NLL contributions to EDMs. 
Diagrams $(a)$ and $(b)$ denote the LL mixing of scalar four-fermion operators with four heavy quarks onto heavy flavor dipoles, while diagram $(c)$ 
the running of heavy flavor dipoles into themselves. At the heavy flavor threshold, the gluonic dipole is integrated out, inducing $\mathcal O( (4\pi)^{-2} L_\mu )$ contributions to $C_{\tilde G}$. Diagrams $(d)$ and $(e)$ contribute to the 2-loop mixing of heavy-light vector operators onto dipoles.   }\label{Fig:RGE4}
\end{figure}

We next consider possible NLL effects, which give rise to the dominant EDM contributions 
for those operators that do not contribute at LL.
As shown in \eqref{eq:ggg1} and \eqref{eq:ggg2} the QCD dipoles $\wc[]{dG}{33}$ and $\wc[]{uG}{22}$ 
can mix with the three-gluon operator $\wc[]{\widetilde G}{}$ due to 1-loop threshold at the bottom and 
charm thresholds, respectively. 
As illustrated in diagram $(b)$ in Fig. \ref{Fig:RGE4},
the  scalar operators $\wc[S1,RR]{uu}{2222}$ and $\wc[S8,RR]{uu}{2222}$
 mix with $\wc[]{uG}{22}$  at 1-loop level. 
 Therefore, $\wc[S1,RR]{uu}{2222}$ and $\wc[S8,RR]{uu}{2222}$ induce a correction to the  three gluon operator of the form
  \be
 \begin{aligned}
 \wc[]{\widetilde G}{}(m_c) & =  { 2    \over 3}   {g_s^3 }
    \wc[S1,RR]{uu}{2222 } (\mu_{ew})  {L_{\mu_c} \over 16 \pi^2}  \,, \\ 
     \wc[]{\widetilde G}{}(m_c) & =  { 2    \over 3 }   g_s^3 \left(C_F - {C_A \over 2}\right) 
    \wc[S8,RR]{uu}{2222}  (\mu_{ew})   {L_{\mu_c} \over 16 \pi^2} , \label{eq:heavy2}
 \end{aligned}
 \ee
where, again, the approximate solution in Eq. \eqref{eq:heavy2} is obtained by neglecting the   
running of the Wilson coefficients and of the strong coupling constant.
Similarly, for the operators containing bottom quarks, it reads
  \be
 \begin{aligned}
 \wc[]{\widetilde G}{} (m_b)& =  { 2  \over 3 }  g_s^3  
    \wc[S1,RR]{dd}{3333}  (\mu_{ew}) {L_{\mu_b} \over 16 \pi^2} \,, \\ 
     \wc[]{\widetilde G}{} (m_b)& =  { 2   \over 3 }  g_s^3 \left(C_F - {C_A \over 2}\right)
    \wc[S8,RR]{dd}{3333} (\mu_{ew})  {L_{\mu_b} \over 16 \pi^2} .\label{eq:heavy3}
 \end{aligned}
 \ee
Here we denoted
\begin{equation}
L_{\mu_c} = \frac{1}{16\pi^2} \log \left( \frac{m_c}{\mu_{ew}} \right), \qquad  L_{\mu_b} = \frac{1}{16\pi^2} \log  \left(\frac{m_b}{\mu_{ew}}\right), 
\end{equation}
and the appearance of one power of $L_\mu$ and an additional power of $(16\pi^2)^{-1}$ is typical of a NLL correction.
Eqs. \eqref{eq:heavy2} and \eqref{eq:heavy3}
are the leading contributions of scalar operators with charm and bottom quarks 
  $\wc[S1(S8),RR]{uu}{2222}$ and $\wc[S1(S8),RR]{dd}{3333}$ 
 to EDMs.
The operators $\wc[S1(S8)]{uddu}{2332}$ can also mix with the three-gluon operator
by first generating a $b$ or $c$ chromoelectric EDM
   \be
 \begin{aligned}
 \wc[]{\widetilde G}{} (m_c)& =  { m_b  \over 3 m_c }   {g_s^3} 
    \wc[S1,RR]{uddu}{2332}  (\mu_{ew}) { L_{\mu_c} \over 16 \pi^2} \,, \\ 
     \wc[]{\widetilde G}{} (m_c)& =  { m_b   \over 3  m_c}   {g_s^3} \left(C_F - {C_A \over 2}\right)
    \wc[S8,RR]{uddu}{2332} (\mu_{ew})  { L_{\mu_c} \over 16 \pi^2}  \,, \\
     \wc[]{\widetilde G}{} (m_b)& =  { m_c  \over 3 m_b }   {g_s^3} 
    \wc[S1,RR]{uddu}{2332}  (\mu_{ew}) { L_{\mu_b} \over 16 \pi^2} \,, \\ 
     \wc[]{\widetilde G}{} (m_b)& =  { m_c   \over 3  m_b } g_s^3  \left(C_F - {C_A \over 2}\right)
    \wc[S8,RR]{uddu}{2332} (\mu_{ew})  { L_{\mu_b} \over 16 \pi^2} .
 \end{aligned}
 \ee
 { The operators $\wc[S1(8), RR]{ud}{2233}$ have 1-loop QCD and QED mixing with $\wc[S1(8), RR]{uddu}{2332}$,
 and thus generate contributions to the $b$ and $c$ QED and QCD dipoles analogous to Eqs. \eqref{eq:scalar_to_dipole1}
 and \eqref{eq:scalar_to_dipole2}. These generate corrections to $\wc[]{\widetilde G}{}$ that scale as $L^2_\mu/(16\pi^2)$. 
 } 
 
\subsection{\boldmath 2-loop anomalous dimension}
We have seen that four-quark scalar operators in the $(LR)(LR)$ class 
contribute to EDM at LL, in the case of operators with four light quarks or two heavy and two light quarks, 
or NLL, for operators with four heavy quarks. We next analyze vector operators in the $(LL)(RR)$ class.
For heavy-light operators, the first contribution arises at NLL. At this order, we need to consider the 2-loop running between 
$\mu_{ew}$ and $m_b$ or $m_c$, and the 1-loop matching contribution discussed in Sec. \ref{sec:vector}. Both the anomalous dimension and the 
matching coefficient are separately scheme-dependent, and the scheme dependence cancels between the two \cite{Ciuchini:1993ks}. 
The 2-loop anomalous dimension can be obtained by generalizing the results of Misiak and Cho \cite{Cho:1993zb}.
They calculated the mixing between $\Delta F=1$ 
vector left-right and dipole operators in the ``standard'' basis.
Here we generalize their results to 
to the mixing of $\Delta F=0$ four-fermion and dipole operators in the JMS basis. A typical 2-loop Feynman diagram governing this mixing is given by diagram $(d)$ in Fig.~\ref{Fig:RGE4}. 
 The $\Delta F=1$ {effective Lagrangian} used in Ref. \cite{Cho:1993zb}
was 
\be
\mathcal{L}_{\rm eff}^{\Delta F=1}  = V_{tb} V_{ts}^* \frac{G_F}{\sqrt{2}}  \sum_{i \in 7,8,9,10}  C_i Q_i\,,
\ee
with 
\be \label{eq:standard}
\begin{aligned} 
Q_7 & =  \frac{ e}{16 \pi^2} m_b (\bar s_\alpha \sigma^{\mu\nu} P_R b_\alpha) F_{\mu \nu}\,, \quad 
Q_8  =  \frac{g_s}{16\pi^2} m_b (\bar s_\alpha \sigma^{\mu\nu} P_R  T^A_{\alpha \beta} b_\beta) G_{\mu \nu}^A\,,  \\
Q_9 & =  {m_b \over m_c}   (\bar s_\alpha \gamma_\mu P_L u_\beta)   (\bar u_\beta  \gamma^\mu P_R b_\alpha)\,,    \qquad
Q_{10}  = {m_b \over m_c} (\bar s_\alpha \gamma_\mu P_L u_\alpha)   (\bar u_\beta  \gamma^\mu P_R b_\beta).
\end{aligned}
\ee
The RGE for corresponding WCs can be written as 
\be
  2\mu^2 {d \over d \mu^2} \vec{ C_S}(\mu) = {\alpha_s \over 4\pi} (\hat \gamma_{S}^{u})^T \vec{ C}_S(\mu).
\ee
The $4\times 4$ ADM for the standard operator vector $\vec{Q}_S =  \{ Q_7, Q_8, Q_9, Q_{10}\}$ is given in 
equation (4.7) of Ref.~\cite{Cho:1993zb} for 
$u=c$. Keeping the explicit dependence on the electric charges of the $u$-type and $d$-type quarks, it reads\footnote{We thank  M. Misiak for providing us the expressions of the ADM for generic $q_u$ and $q_d$ charges.}
 \be \label{eq:misiak}
  {\hat \gamma_{S}^{u}  \over 2 }= 
 \begin{pmatrix}
16 \over 3 & 0 & 0 & 0 \\
-16 \over 9 & 14 \over 3 & 0 & 0 \\
40 q_u & -2 & -8 & 0\\
8 q_u + {16  \over 3} q_d & 4 \over 3 & -3 & 1
 \end{pmatrix}.
 \ee 
The superscript $u$ indicates that the $u$-quarks are closed in diagram $(d)$ in Fig.~\ref{Fig:RGE4}.
Since the JMS basis uses the unit normalization for the operators, using \eqref{eq:jms-rge}, first we go to an intermediate basis and obtain the 
ADM for it
\be
\vec{Q}^\prime =  \{ Q_7^\prime, Q_8^\prime, Q_9^\prime, Q_{10}^\prime\}.
\ee
The $Q_i^\prime$'s are the same as the operators given in \eqref{eq:standard} but with unit normalization like the JMS operators and 
 follows a JMS like RGEs for  the WCs
\be \label{eq:jms-rge}
16 \pi^2 \mu {d \vec{ C}^\prime (\mu) \over d \mu}  ={\hat \gamma}^\prime  \vec{C}^\prime (\mu).
\ee 
The ADM in this basis (when closing $u$-quark) comes out to be 
\be \label{eq:adm-itermediate}
 {\hat \gamma}^{\prime u} = 
 \begin{pmatrix}
{8 \over 3} g_s^2 & -32 eg_s \over 9 & {5eg_s^2 m_u  \over \pi^2} q_u & {eg_s^2 m_u \over 8\pi^2}  ({16 q_d \over 3} + 8 q_u)   \\
0 & -19 g_s^2 \over 3  & - g_s^3 m_u \over 4 \pi^2 &  g_s^3 m_u \over 6 \pi^2 \\
0 & 0 & - 16 g_s^2 & -6 g_s^2\\
0 & 0 & 0 & 2 g_s^2
 \end{pmatrix}.
 \ee
The relation between ${\hat \gamma}^{\prime u}$ and ${\hat \gamma}_S^{ u}$ is given by
\be \label{eq:jms-std} 
 {\hat \gamma}^{\prime u}
 = 16\pi^2 \mu {d R(\mu) \over d \mu} R(\mu)^{-1} +   R(\mu)  g_s^2 ({\hat \gamma}_{S}^{u})^T R(\mu)^{-1} .
\ee 
The matrix $R(\mu)$ is the transformation matrix defined by $C^\prime(\mu) = R(\mu) C_{S}(\mu)$.
The RGE for $R(\mu)$ is dictated by the RGEs of $g_s(\mu)$ and $m_b(\mu)$
 \be \label{eq:rge-Rmu}
16\pi^2 \mu {d R(\mu) \over d \mu}=
\begin{pmatrix}
-8e g_s^2 m_b   & 0& 0& 0 \\
0 & {-47 \over 3} g_s^3 m_b   & 0 & 0 \\
0 & 0 &  0 & 0 \\
0 & 0 &0  &  0 
\end{pmatrix}.
\ee
We have used $16\pi^2 \mu ({d m_b /d \mu}) = -6 C_F g_s^2 m_b  $ and 
 $16 \pi^ 2\mu ({d g_s / d \mu}) = -23  g_s^3/3$ for five flavors. 
The correspondence between the $Q^\prime$ and the JMS bases is given by
\be
\hat \gamma _{\text{JMS}}  = P  \hat \gamma^{\prime}  P^{-1} .
\ee
Here the matrix $P$ defines the transformation between the WCs in the two bases, namely 
$C_{\text{JMS}} = P C^\prime$. 
In what follows we will derive $\hat \gamma_{\text{JMS}}$ for specific sets of the  
JMS $\Delta F=0$ operators entering the EDMs.
{\boldmath
\subsubsection{$\wcL[V1,LR]{uddu}{prrp}, \wcL[V8,LR]{uddu}{prrp}$ onto dipoles}
}
In this case, the operators 
$\wcL[V1,LR]{uddu}{prrp}$ and  $\wcL[V8,LR]{uddu}{prrp}$  contribute to EDMs, here the index  $p$ 
can be equal or unequal to $r$.   
At 2-loop order, these two operators can mix with $\{ \wcL[]{u\gamma}{pp}, \wcL[]{uG}{pp} \}$,
if one closes the $d$ quark loop,
or 
$\{ \wcL[]{d\gamma}{rr}, \wcL[]{dG}{rr} \}$, by closing the $u$ quark loop.
Using the color identity $T^A_{ij} T^A_{kl} = {1\over 2} \left ( \delta_{il} \delta_{jk} - {1 \over 3} \delta_{ij}  \delta _{kl}       \right  )$, the tree-level transformation matrix between the intermediate basis
 with properly adjusted down-quark flavors  to $\Delta F=0$ and the JMS 
 basis $\vec{\mathcal{C}}_{u}= \{
 \wcL[]{d\gamma}{rr}, \wcL[]{dG}{rr}, $
 $ \wcL[V1,LR]{uddu}{prrp}, \wcL[V8,LR]{uddu}{prrp} \}$  is found to be 
\be \label{eq:tree}
P=
\begin{pmatrix}
1  & 0& 0& 0 \\
0 & 1 & 0 & 0 \\
0 & 0 &  1/3 & 1 \\
0 & 0 &2 &  0 
\end{pmatrix}.
\ee
Using \eqref{eq:misiak},    \eqref{eq:rge-Rmu}  and  \eqref{eq:tree} in \eqref{eq:jms-std}, we 
obtain \footnote{Note that one needs to properly adjust the value of electric charge for the case of up-type dipoles in contrast to the original result which is valid for the down-quark dipoles.} the ADMs for $\vec{\mathcal{C}}_{u}$   when the $u$-quark loop is closed\footnote{Note that in the original 
result of Misiak and Cho the charm-quark loop was closed.}
\be \label{eq:2loop1}
\gamma_{\text{JMS}}^u  = 
\begin{pmatrix}
8 g_s^2 \over 3 & -32 e g_s \over 9 & 4eg_s^2 m_u \over 9\pi^2 & 43 e g_s^2 m_u \over 27 \pi^2 \\
-8 e g_s \over 3 & -19 g_s^2 \over 3 & g_s^3 m_u \over 6 \pi^2 & -11 g_s^3 m_u \over 72 \pi^2\\
0 & 0& 0& -8 g_s^2 \over 3\\
0 & 0 & -12 g_s^2 & -14 g_s^2
\end{pmatrix}.
\ee
Here $u = u$ or $c$.

 For down-quark loops we have a four-vector  $\vec{\mathcal{C}}_{d}= \{
 \wcL[]{u\gamma}{pp}, \wcL[]{uG}{pp}, $
 $ \wcL[V1,LR]{uddu}{prrp}, \wcL[V8,LR]{uddu}{prrp} \}$.
 The transformation matrix is the same as given in \eqref{eq:tree}.
However, in this case, first, we need to   
make the replacements $q_u \to q_d$ and $q_d \to q_u$ in the original ADM \eqref{eq:misiak}. 
This brings us 
\be \label{eq:2loop2}
\gamma_{\text{JMS}}^d  = 
\begin{pmatrix}
8 g_s^2 \over 3 & 64 e g_s \over 9 & eg_s^2 m_d \over 9\pi^2 & -23 e g_s^2 m_d \over 27 \pi^2 \\
16eg_s \over 3 & -19 g_s^2 \over 3 & g_s^3 m_d \over 6 \pi^2 & -11 g_s^3 m_d \over 72 \pi^2\\
0 & 0& 0& -8 g_s^2 \over 3\\
0 & 0 & -12 g_s^2 & -14 g_s^2
\end{pmatrix}.
\ee
Here $d$ represents $d,s$ or $b$. 
Obviously, only the  $1-3$ and $1-4$ entries corresponding to the mixing of 
 vector left-right operators onto QED dipole operators are modified as 
 compared to $\gamma_{\text{JMS}}^u$.

{\boldmath
\subsubsection{$\wcL[V1,LR]{uu}{prrp}, \wcL[V8,LR]{uu}{prrp}$ onto dipoles  }
}
The operators $\wcL[V1,LR]{uu}{prrp}$ and  $\wcL[V8,LR]{uu}{prrp}$  
can contribute to EDMs if $p \neq r$.   These mix with 
$\{ \wcL[]{u\gamma}{pp}, \wcL[]{uG}{pp} \}$  and $\{ \wcL[]{u\gamma}{rr}, \opL[]{uG}{rr} \}$ depending on
 whether the quarks with  $rr$ or $pp$ flavor indices are  
 closed. In this case, we have two four vectors, namely $\vec{\mathcal{C}}_{ur}= \{
 \wcL[]{u\gamma}{pp}, \wcL[]{uG}{pp}, $
 $ \wcL[V1,LR]{uu}{prrp}, \wcL[V8,LR]{uu}{prrp} \}$  
and $\vec{\mathcal{C}}_{up}= \{
 \wcL[]{u\gamma}{rr}, \wcL[]{uG}{rr}, $
 $ \wcL[V1,LR]{uu}{prrp}, \wcL[V8,LR]{uu}{prrp} \}$  . Both cases involve 
 an up-quark loop with flavor indices $rr$ and $pp$ respectively.  The transformation matrix 
 between the standard (after properly adjusting quark flavors in \eqref{eq:standard}) and JMS 
 vectors $\vec{\mathcal{C}}_{ur}$ and $\vec{\mathcal{C}}_{up}$  is given by \eqref{eq:tree}.
In this case, in the original 2-loop ADM \eqref{eq:misiak} one has to make replacements 
$q_u \to q_u$ and $q_d \to q_u$ which lead us to 
\be \label{eq:2loop3}
\gamma_{ \text{JMS}}^{ur}  = 
\begin{pmatrix}
8 g_s^2 \over 3 & {64} e g_s \over 9 & 10eg_s^2 m_u \over 9\pi^2 & 40 e g_s^2 m_u \over 27 \pi^2 \\
16eg_s\over 3 & -19 g_s^2 \over 3 & g_s^3 m_u \over 6 \pi^2 & -11 g_s^3 m_u \over 72 \pi^2\\
0 & 0& 0& -8 g_s^2 \over 3\\
0 & 0 & -12 g_s^2 & -14 g_s^2
\end{pmatrix}.
\ee
The $\gamma_{ \text{JMS}}^{up}$ is exactly same as $\gamma_{ \text{JMS}}^{ur}$.
\\

{\boldmath
\subsubsection{$\wcL[V1,LR]{dd}{prrp}, \wcL[V8,LR]{dd}{prrp}$ onto dipoles}
}

In this case, we have two four vectors, namely $\vec{\mathcal{C}}_{dr}= \{
 \wcL[]{d\gamma}{pp}, \wcL[]{dG}{pp}, $
 $ \wcL[V1,LR]{dd}{prrp}, \wcL[V8,LR]{dd}{prrp} \}$  
and $\vec{\mathcal{C}}_{dp}= \{
 \wcL[]{d\gamma}{rr}, \wcL[]{dG}{rr}, $
 $ \wcL[V1,LR]{dd}{prrp}, \wcL[V8,LR]{dd}{prrp} \}$ . 
 In the original 2-loop ADM \eqref{eq:misiak} one has to make replacements 
$q_u \to q_d$ and $q_d \to q_d$. This leads us to
 \be \label{eq:2loop4}
\gamma_{\text{JMS}}^{dr}  = 
\begin{pmatrix}
8 g_s^2 \over 3 & -32 e g_s \over 9 & -5eg_s^2 m_d \over 9\pi^2 & -20 e g_s^2 m_d \over 27 \pi^2 \\
-8 eg_s \over 3 & -19 g_s^2 \over 3 & g_s^3 m_d \over 6 \pi^2 & -11 g_s^3 m_d \over 72 \pi^2\\
0 & 0& 0& -8 g_s^2 \over 3\\
0 & 0 & -12 g_s^2 & -14 g_s^2
\end{pmatrix}.
\ee
The $\gamma_{ \text{JMS}}^{dp}$ is exactly same as $\gamma_{ \text{JMS}}^{dr}$.

%'uG_11', 'uG_22', 'ugamma_11', 'ugamma_22', 'dG_11', 'dG_22', 'dG_33',   'dgamma_11', 'dgamma_22', 'dgamma_33', 'G', 'Gtilde'
\begin{table}[H]
\begin{center}
%\hspace{-2.cm}
 \renewcommand*{\arraystretch}{1.5}
 \resizebox{1.0\textwidth}{!}{
\begin{tabular}{ |c|cccccc||c||c|cccc|}
\hline
  \multicolumn{13}{|c|} {\boldmath All four-quark and dipole operators with bottom-quark in 5+3 WET } \\
\hline
&\multicolumn{8}{|c|}{\boldmath $\eta_i^j (\mu_{\rm low}, \mu_{\rm ew})$  }  & 
\multicolumn{4}{|c|}{\boldmath $\eta_i^j (\mu_{b}, \mu_{\rm ew})$  } \\
%& \multicolumn{7}{|c|}{\boldmath $N_f = 3 $ WET at $\mu_{\rm low} $} & \multicolumn{4}{|c|}{\boldmath $N_f = 5$ WET  at $\mu_{\rm low}$}   \\
\hline
\boldmath $i \downarrow$ & $ \underset{11}{C_{uG}}$  & $\underset{11}{C_{u\gamma}}$ & $\underset{11}{C_{dG}}$ &  $\underset{22}{C_{dG}}$  & $\underset{11}{C_{d\gamma}}$ & $\underset{22}{C_{d\gamma}}$ & $C_{\tilde G}$ &  $\underset{11}{C_{e\gamma}}$  & $\underset{33}{C_{dG}}$ & $\underset{33}{C_{d\gamma}}$ &  $\underset{22}{C_{uG}}$  & $\underset{22}{C_{u\gamma}}$   \\ \hline \hline
%%%%%%%%%%%%%%%

$[C^{S1,RR}_{dd}]_{3333}$&$4.6\cdot 10^{-8}$&-&$9.8\cdot 10^{-8}$&$2.0\cdot 10^{-6}$&-&-&$-1.4\cdot 10^{-4}$&-&$-2.6\cdot 10^{-1}$&$1.2\cdot 10^{-2}$&-&-\\
$[C^{S8,RR}_{dd}]_{3333}$&-&-&$-2.0\cdot 10^{-8}$&$-4.1\cdot 10^{-7}$&-&-&$3.0\cdot 10^{-5}$&-&$5.5\cdot 10^{-2}$&$1.4\cdot 10^{-2}$&-&-\\
$[C^{S1,RR}_{dd}]_{1133}$&-&-&$2.6\cdot 10^{-2}$&-&$3.6\cdot 10^{-4}$&-&$1.3\cdot 10^{-8}$&-&$2.4\cdot 10^{-5}$&$1.4\cdot 10^{-7}$&-&-\\
$[C^{S8,RR}_{dd}]_{1133}$&-&-&$8.9\cdot 10^{-3}$&-&$-1.1\cdot 10^{-3}$&-&-&-&$8.1\cdot 10^{-6}$&$-1.2\cdot 10^{-6}$&-&-\\
$[C^{S1,RR}_{dd}]_{2233}$&-&-&-&$2.6\cdot 10^{-2}$&-&$3.6\cdot 10^{-4}$&$2.6\cdot 10^{-7}$&-&$4.8\cdot 10^{-4}$&$2.7\cdot 10^{-6}$&-&-\\
$[C^{S8,RR}_{dd}]_{2233}$&-&-&-&$8.9\cdot 10^{-3}$&-&$-1.1\cdot 10^{-3}$&$8.9\cdot 10^{-8}$&-&$1.6\cdot 10^{-4}$&$-2.4\cdot 10^{-5}$&-&-\\
$[C^{S1,RR}_{dd}]_{1331}$&-&-&$-1.7\cdot 10^{-1}$&-&$4.2\cdot 10^{-3}$&-&$-8.5\cdot 10^{-8}$&-&$-1.6\cdot 10^{-4}$&$5.9\cdot 10^{-6}$&-&-\\
$[C^{S8,RR}_{dd}]_{1331}$&-&-&$2.2\cdot 10^{-2}$&-&$8.2\cdot 10^{-3}$&-&$1.1\cdot 10^{-8}$&-&$2.0\cdot 10^{-5}$&$8.4\cdot 10^{-6}$&-&-\\
$[C^{S1,RR}_{dd}]_{2332}$&-&-&-&$-1.7\cdot 10^{-1}$&-&$4.2\cdot 10^{-3}$&$-1.7\cdot 10^{-6}$&-&$-3.2\cdot 10^{-3}$&$1.2\cdot 10^{-4}$&-&-\\
$[C^{S8,RR}_{dd}]_{2332}$&-&-&-&$2.2\cdot 10^{-2}$&-&$8.2\cdot 10^{-3}$&$2.2\cdot 10^{-7}$&-&$4.0\cdot 10^{-4}$&$1.7\cdot 10^{-4}$&-&-\\
$[C^{S1,RR}_{ud}]_{1133}$&$2.6\cdot 10^{-2}$&$-7.6\cdot 10^{-4}$&-&-&-&-&-&-&$1.1\cdot 10^{-5}$&$2.5\cdot 10^{-8}$&-&-\\
$[C^{S8,RR}_{dd}]_{1133}$&-&-&$8.9\cdot 10^{-3}$&-&$-1.1\cdot 10^{-3}$&-&-&-&$8.1\cdot 10^{-6}$&$-1.2\cdot 10^{-6}$&-&-\\
$[C^{S1,RR}_{uddu}]_{1331}$&$-1.7\cdot 10^{-1}$&$1.4\cdot 10^{-2}$&-&-&-&-&$-4.0\cdot 10^{-8}$&-&$-7.3\cdot 10^{-5}$&$-8.1\cdot 10^{-6}$&-&-\\
$[C^{S8,RR}_{uddu}]_{1331}$&$2.1\cdot 10^{-2}$&$6.9\cdot 10^{-3}$&-&-&-&-&-&-&$9.0\cdot 10^{-6}$&$-7.5\cdot 10^{-6}$&-&-\\
$[C^{S1,RR}_{ud}]_{2233}$&-&-&-&$-4.5\cdot 10^{-8}$&-&-&$9.9\cdot 10^{-5}$&-&$6.0\cdot 10^{-3}$&$1.3\cdot 10^{-5}$&$2.4\cdot 10^{-2}$&$-3.2\cdot 10^{-4}$\\
$[C^{S8,RR}_{ud}]_{2233}$&-&-&-&$-1.5\cdot 10^{-8}$&-&-&$3.2\cdot 10^{-5}$&-&$1.9\cdot 10^{-3}$&$6.5\cdot 10^{-4}$&$7.7\cdot 10^{-3}$&$-1.4\cdot 10^{-3}$\\
$[C^{S1,RR}_{uddu}]_{2332}$&-&-&$1.4\cdot 10^{-8}$&$2.9\cdot 10^{-7}$&-&-&$-6.5\cdot 10^{-4}$&-&$-3.9\cdot 10^{-2}$&$-4.3\cdot 10^{-3}$&$-1.6\cdot 10^{-1}$&$1.1\cdot 10^{-2}$\\
$[C^{S8,RR}_{uddu}]_{2332}$&-&-&-&$-3.6\cdot 10^{-8}$&-&-&$7.9\cdot 10^{-5}$&-&$4.8\cdot 10^{-3}$&$-4.0\cdot 10^{-3}$&$1.9\cdot 10^{-2}$&$7.5\cdot 10^{-3}$\\
$[C^{V1,LR}_{uddu}]_{1331}$&$-5.0\cdot 10^{-2}$&$1.1\cdot 10^{-3}$&-&-&-&-&-&-&$-1.7\cdot 10^{-6}$&$-2.3\cdot 10^{-6}$&-&-\\
$[C^{V8,LR}_{uddu}]_{1331}$&$1.3\cdot 10^{-2}$&$1.2\cdot 10^{-2}$&-&-&-&-&-&-&$2.6\cdot 10^{-6}$&$-5.5\cdot 10^{-6}$&-&-\\
$[C^{V1,LR}_{dd}]_{1331}$&-&-&$-4.9\cdot 10^{-2}$&-&$3.2\cdot 10^{-3}$&-&-&-&$-3.6\cdot 10^{-6}$&$4.1\cdot 10^{-6}$&-&-\\
$[C^{V8,LR}_{dd}]_{1331}$&-&-&$1.4\cdot 10^{-2}$&-&$1.2\cdot 10^{-2}$&-&-&-&$5.7\cdot 10^{-6}$&$5.6\cdot 10^{-6}$&-&-\\
$[C^{V1,LR}_{dd}]_{2332}$&-&-&-&$-4.9\cdot 10^{-2}$&-&$3.2\cdot 10^{-3}$&$-3.9\cdot 10^{-8}$&-&$-7.1\cdot 10^{-5}$&$8.1\cdot 10^{-5}$&-&-\\
$[C^{V8,LR}_{dd}]_{2332}$&-&-&-&$1.4\cdot 10^{-2}$&-&$1.2\cdot 10^{-2}$&$6.1\cdot 10^{-8}$&-&$1.1\cdot 10^{-4}$&$1.1\cdot 10^{-4}$&-&-\\
$[C^{V1,LR}_{uddu}]_{2332}$&-&-&-&-&-&-&$-1.8\cdot 10^{-4}$&-&$-8.6\cdot 10^{-4}$&$-1.2\cdot 10^{-3}$&$-4.5\cdot 10^{-2}$&$2.9\cdot 10^{-4}$\\
$[C^{V8,LR}_{uddu}]_{2332}$&-&-&-&-&-&-&$5.0\cdot 10^{-5}$&-&$1.3\cdot 10^{-3}$&$-2.7\cdot 10^{-3}$&$1.3\cdot 10^{-2}$&$1.2\cdot 10^{-2}$\\
$[C_{dG}]_{33}$&$-2.2\cdot 10^{-7}$&-&$-4.7\cdot 10^{-7}$&$-9.5\cdot 10^{-6}$&-&$-4.2\cdot 10^{-8}$&$6.9\cdot 10^{-4}$&-&$1.3\cdot 10^{0}$&$3.2\cdot 10^{-2}$&-&-\\
$[C_{d\gamma}]_{33}$&-&-&-&$-1.9\cdot 10^{-7}$&-&-&$1.3\cdot 10^{-5}$&-&$2.5\cdot 10^{-2}$&$9.0\cdot 10^{-1}$&-&-\\

%%%%%%%%%%%%%
\hline  \hline
& $ \underset{11}{C_{uG}}$  & $\underset{11}{C_{u\gamma}}$ & $\underset{11}{C_{dG}}$ &  $\underset{22}{C_{dG}}$  & $\underset{11}{C_{d\gamma}}$ & $\underset{22}{C_{d\gamma}}$ & $C_{\tilde G}$   &$\underset{11}{C_{e\gamma}}$      & $\underset{33}{C_{dG}}$ & $\underset{33}{C_{d\gamma}}$
&  $\underset{22}{C_{uG}}$  & $\underset{22}{C_{u\gamma}}$   \\ \hline
\end{tabular}
}
\caption{Operator mixing of the bottom quark operators in $5+3$ WET (1st column) onto dipole operators (2nd row) in $3 +1$  and $5+3$ flavor WET is shown.  Here the entries represent the quantity $\eta_i^j (\mu_{\rm low}, \mu_{\rm ew})$, $\mu_{\rm ew}=91.1876$ GeV, $\mu_{\rm low}=2.0$ GeV .  Entries below $10^{-8}$ are dropped. Note that for the charm and bottom dipoles, we stop running at $\mu=m_b=4.18$ GeV.}\label{tab:mixing2}
\end{center}
\end{table}

{\boldmath
\subsubsection{$\wcL[V,LR]{ee}{prrp}$ onto dipole}
}

The leptonic $\wcL[V,LR]{ee}{prrp}$ operator can break CP is $p \neq r$.
In this case, we have two vectors, namely $\vec{\mathcal{C}}_{er}= \{
 \wcL[]{e\gamma}{pp},  \wcL[V,LR]{ee}{prrp} \}$  
and $\vec{\mathcal{C}}_{ep}= \{
 \wcL[]{e\gamma}{rr}, \wcL[V,LR]{ee}{prrp} \}$ . 
 One example of the 2-loop Feynman diagrams governing this mixing is given in diagram (e) of Fig.~\ref{Fig:RGE4}. 
 In the 2-loop ADM \eqref{eq:adm-itermediate} one has to make replacements 
$q_u \to q_e$ and $q_d \to q_e$.  However, since these are purely leptonic operators which exhibit only 
QED mixing we need to replace gluons in these diagrams with 
photons. So we need to take out the color factors in the 
original result.  We found that the 2-loop $P$-type diagrams (which are 
the only ones allowed for the pure leptonic  operators)
shown in Fig. 4 of \cite{Ciuchini:1993ks} has a universal color factor $C_F= (T^A T^A)_{ij}   =\delta_{ij}  (N^2-1)/2N $ .

Therefore, we need to divide the elements of ADM  by this overall color factor $C_F$.
With all these adjustments, we arrived at the RG equation for $\wcL[]{e\gamma}{pp}$
{
\be \label{eq:2loop5}
{16 \pi^2 \mu} ({ d \wcL[]{e\gamma}{pp} / d \mu}) = 
 {15e^3 m_r Q_e \over 12 \pi^2  } \wcL[V,LR]{ee}{prrp}.  
\ee}
Trivially, the same RGE holds for $  \wcL[]{e\gamma}{rr} $.

%%%%%%%%%%%%
\definecolor{Grayd}{gray}{0.8}
\definecolor{Gray}{gray}{0.95} 
% \rowcolor{Grayd}
\begin{table}[H] 
\begin{center}
 \renewcommand*{\arraystretch}{1.5}
 \resizebox{1.0\textwidth}{!}{
\begin{tabular}{|c|cccccc||c||c|cc|} 
\hline
  \multicolumn{11}{|c|} {\boldmath All operators with charm-quark in 4+1 WET } \\
\hline
  \multicolumn{11}{|c|} {\boldmath $\eta_i^j (\mu_{\rm low}, \mu_{\rm ew})$  } \\
\hline
%&\multicolumn{7}{|c|}{\boldmath  $N_f=3$ WET  } & \multicolumn{2}{|c|}{\boldmath $N_f=4$ WET  }   \\
%\hline
\boldmath $i \downarrow$ & $[C_{uG}]_{11}$  & $[C_{u\gamma}]_{11}$  & $[C_{dG}]_{11}$ &  $[C_{dG}]_{22}$ & $[C_{d\gamma}]_{11}$ & $[C_{d\gamma}]_{22}$ & $C_{\tilde G}$ &     $[C_{e\gamma}]_{11}$    & $[C_{uG}]_{22}$  & $[C_{u\gamma}]_{22}$    \\ \hline \hline
%%%%%%%%%%%%%%%%%%%%%%%%%%%%%%%%%%%%%%%%%%%%%%%%%%

$[C^{S1,RR}_{uu}]_{2222}$&-&-&-&-&-&-&$-3.7\cdot 10^{-4}$&-&$-1.0\cdot 10^{-1}$&$-7.0\cdot 10^{-3}$\\
$[C^{S8,RR}_{uu}]_{2222}$&-&-&-&-&-&-&$8.5\cdot 10^{-5}$&-&$2.3\cdot 10^{-2}$&$-8.5\cdot 10^{-3}$\\
$[C^{S1,RR}_{uu}]_{1122}$&$1.3\cdot 10^{-2}$&$-8.8\cdot 10^{-5}$&-&-&-&-&$9.2\cdot 10^{-8}$&-&$2.5\cdot 10^{-5}$&$-1.6\cdot 10^{-7}$\\
$[C^{S8,RR}_{uu}]_{1122}$&$4.4\cdot 10^{-3}$&$9.7\cdot 10^{-4}$&-&-&-&-&$3.0\cdot 10^{-8}$&-&$8.2\cdot 10^{-6}$&$1.8\cdot 10^{-6}$\\
$[C^{S1,RR}_{ud}]_{2211}$&-&-&$1.3\cdot 10^{-2}$&-&$8.4\cdot 10^{-5}$&-&$2.0\cdot 10^{-7}$&-&$5.4\cdot 10^{-5}$&$-9.5\cdot 10^{-7}$\\
$[C^{S8,RR}_{ud}]_{2211}$&-&-&$4.0\cdot 10^{-3}$&-&$1.1\cdot 10^{-3}$&-&$5.9\cdot 10^{-8}$&-&$1.6\cdot 10^{-5}$&$-2.4\cdot 10^{-6}$\\
$[C^{S1,RR}_{ud}]_{2222}$&-&-&-&$1.3\cdot 10^{-2}$&-&$8.4\cdot 10^{-5}$&$4.0\cdot 10^{-6}$&-&$1.1\cdot 10^{-3}$&$-1.9\cdot 10^{-5}$\\
$[C^{S8,RR}_{ud}]_{2222}$&-&-&-&$4.0\cdot 10^{-3}$&-&$1.1\cdot 10^{-3}$&$1.2\cdot 10^{-6}$&-&$3.3\cdot 10^{-4}$&$-4.9\cdot 10^{-5}$\\
$[C^{S1,RR}_{uddu}]_{2112}$&-&-&$-6.4\cdot 10^{-2}$&-&$-6.1\cdot 10^{-3}$&-&$-9.4\cdot 10^{-7}$&-&$-2.6\cdot 10^{-4}$&$1.7\cdot 10^{-5}$\\
$[C^{S8,RR}_{uddu}]_{2112}$&-&-&$6.9\cdot 10^{-3}$&-&$-4.9\cdot 10^{-3}$&-&$1.0\cdot 10^{-7}$&-&$2.7\cdot 10^{-5}$&$9.0\cdot 10^{-6}$\\
$[C^{S1,RR}_{uddu}]_{2222}$&-&-&-&$-6.4\cdot 10^{-2}$&-&$-6.1\cdot 10^{-3}$&$-1.9\cdot 10^{-5}$&-&$-5.2\cdot 10^{-3}$&$3.5\cdot 10^{-4}$\\
$[C^{S8,RR}_{uddu}]_{2222}$&-&-&-&$6.9\cdot 10^{-3}$&-&$-4.9\cdot 10^{-3}$&$2.0\cdot 10^{-6}$&-&$5.5\cdot 10^{-4}$&$1.8\cdot 10^{-4}$\\
$[C^{S1,RR}_{uu}]_{1221}$&$-6.4\cdot 10^{-2}$&$-3.4\cdot 10^{-3}$&-&-&-&-&$-4.4\cdot 10^{-7}$&-&$-1.2\cdot 10^{-4}$&$-6.3\cdot 10^{-6}$\\
$[C^{S8,RR}_{uu}]_{1221}$&$7.2\cdot 10^{-3}$&$-5.2\cdot 10^{-3}$&-&-&-&-&$4.9\cdot 10^{-8}$&-&$1.4\cdot 10^{-5}$&$-9.8\cdot 10^{-6}$\\
$[C^{V1,LR}_{uddu}]_{2112}$&-&-&$-1.3\cdot 10^{-2}$&-&$-2.3\cdot 10^{-3}$&-&$-1.7\cdot 10^{-8}$&-&$-4.8\cdot 10^{-6}$&$2.0\cdot 10^{-6}$\\
$[C^{V8,LR}_{uddu}]_{2112}$&-&-&$4.3\cdot 10^{-3}$&-&$-8.1\cdot 10^{-3}$&-&$3.4\cdot 10^{-8}$&-&$9.2\cdot 10^{-6}$&$9.2\cdot 10^{-6}$\\
$[C^{V1,LR}_{uddu}]_{2222}$&-&-&-&$-1.3\cdot 10^{-2}$&-&$-2.3\cdot 10^{-3}$&$-3.4\cdot 10^{-7}$&-&$-9.4\cdot 10^{-5}$&$4.0\cdot 10^{-5}$\\
$[C^{V8,LR}_{uddu}]_{2222}$&-&-&-&$4.3\cdot 10^{-3}$&-&$-8.1\cdot 10^{-3}$&$6.7\cdot 10^{-7}$&-&$1.8\cdot 10^{-4}$&$1.8\cdot 10^{-4}$\\
$[C^{V1,LR}_{uu}]_{1221}$&$-1.3\cdot 10^{-2}$&$-3.3\cdot 10^{-3}$&-&-&-&-&-&-&$-1.9\cdot 10^{-6}$&$-6.1\cdot 10^{-6}$\\
$[C^{V8,LR}_{uu}]_{1221}$&$4.5\cdot 10^{-3}$&$-8.0\cdot 10^{-3}$&-&-&-&-&$1.8\cdot 10^{-8}$&-&$5.0\cdot 10^{-6}$&$-8.3\cdot 10^{-6}$\\
$[C_{u\gamma}]_{22}$&-&-&-&-&-&-&$-4.9\cdot 10^{-4}$&-&$-1.4\cdot 10^{-1}$&$8.7\cdot 10^{-1}$\\
$[C_{uG}]_{22}$&-&-&-&-&-&-&$5.1\cdot 10^{-3}$&-&$1.4\cdot 10^{0}$&$-8.7\cdot 10^{-2}$\\
$[C^{S,RR}_{eu}]_{1122}$&-&-&-&-&-&-&-&$-1.5\cdot 10^{-4}$&-&$5.8\cdot 10^{-8}$\\
$[C^{T,RR}_{eu}]_{1122}$&-&-&-&-&-&-&-&$7.4\cdot 10^{-2}$&$1.5\cdot 10^{-6}$&$-3.0\cdot 10^{-5}$\\

%%%%%%%%%%%%%%%%%%%%%
\hline
  & $[C_{uG}]_{11}$  & $[C_{u\gamma}]_{11}$  & $[C_{dG}]_{11}$ &  $[C_{dG}]_{22}$ & $[C_{d\gamma}]_{11}$ & $[C_{d\gamma}]_{22}$ & $C_{\tilde G}$ &     $[C_{e\gamma}]_{11}$    & $[C_{uG}]_{22}$  & $[C_{u\gamma}]_{22}$    \\ \hline \hline
%%%%%%%%%%%%%%%%%%%%%%%%%%%%%%%%%%%%%%%%%%%%%%%%%%
\end{tabular}
}
\caption{Operator mixing of the charm quark operators in $4+1$ WET (1st column) onto dipole operators (2nd row) in $ 3+1 $  and $4+1$ flavor WET is shown.  Here the entries represent the quantity $\eta_i^j (\mu_{\rm low}, \mu_{\rm ew})$, $\mu_{\rm ew}=91.1876$ GeV, $\mu_{\rm low}=2.0$ GeV.  Entries below $10^{-8}$ are dropped.}
\label{tab:mixing1}\end{center}
\end{table}

\subsection{Numerical solution for the RGEs}
In this section, we provide the numerical solution to the RGEs, which allows us to quantify the
impact of such effects on the low-energy operators that contribute to EDMs at the tree level.
As we have learned in the previous subsections, 1-loop and 2-loop RG running effects and
threshold effects cause operator mixing leading to the expansion of 
operator basis for the EDMs.

\begin{table}[H]
\begin{center}
%\hspace{-2.cm}
 \renewcommand*{\arraystretch}{1.3}
 \resizebox{1.0\textwidth}{!}{
\begin{tabular}{ |c|cccccc||c||c|cccc|}
\hline
  \multicolumn{13}{|c|} {\boldmath All semileptonic operators in 5+3 WET } \\
\hline

&\multicolumn{8}{|c|}{\boldmath $\eta_i^j (\mu_{\rm low}, \mu_{\rm ew})$ } & 
\multicolumn{4}{|c|}{\boldmath $\eta_i^j (\mu_{b}, \mu_{\rm ew})$ }  \\
%& \multicolumn{7}{|c|}{\boldmath $N_f = 3 $ WET at $\mu_{\rm low} $} & \multicolumn{4}{|c|}{\boldmath $N_f = 5$ WET  at $\mu_{\rm low}$}   \\
\hline
\boldmath $i \downarrow$ & $ \underset{11}{C_{uG}}$  & $\underset{11}{C_{u\gamma}}$ & $\underset{11}{C_{dG}}$ &  $\underset{22}{C_{dG}}$  & $\underset{11}{C_{d\gamma}}$ & $\underset{22}{C_{d\gamma}}$ & $C_{\tilde G}$ &  $\underset{11}{C_{e\gamma}}$  & $\underset{33}{C_{dG}}$ & $\underset{33}{C_{d\gamma}}$ &  $\underset{22}{C_{uG}}$  & $\underset{22}{C_{u\gamma}}$   \\ \hline \hline
%%%%%%%%%%%%%%%

$[C^{S,RR}_{eu}]_{1111}$&-&$5.8\cdot 10^{-8}$&-&-&-&-&-&$-2.9\cdot 10^{-7}$&-&-&-&-\\
$[C^{S,RR}_{ed}]_{1111}$&-&-&-&-&$-2.9\cdot 10^{-8}$&-&-&$-1.5\cdot 10^{-7}$&-&-&-&-\\
$[C^{S,RR}_{ed}]_{1122}$&-&-&-&-&-&$-2.9\cdot 10^{-8}$&-&$-3.1\cdot 10^{-6}$&-&-&-&-\\
$[C^{S,RL}_{eu}]_{1111}$&-&-&-&-&-&-&-&-&-&-&-&-\\
$[C^{S,RL}_{ed}]_{1111}$&-&-&-&-&-&-&-&-&-&-&-&-\\
$[C^{S,RL}_{ed}]_{1122}$&-&-&-&-&-&-&-&-&-&-&-&-\\
$[C^{T,RR}_{eu}]_{1111}$&$1.5\cdot 10^{-6}$&$-3.0\cdot 10^{-5}$&-&-&-&-&-&$1.4\cdot 10^{-4}$&-&-&-&-\\
$[C^{T,RR}_{ed}]_{1111}$&-&-&$-5.8\cdot 10^{-7}$&-&$-3.0\cdot 10^{-5}$&-&-&$-1.5\cdot 10^{-4}$&-&-&-&-\\
$[C^{T,RR}_{ed}]_{1122}$&-&-&-&$-5.8\cdot 10^{-7}$&-&$-3.0\cdot 10^{-5}$&-&$-3.0\cdot 10^{-3}$&-&-&-&-\\
$[C^{T,RR}_{eu}]_{1122}$&-&-&-&-&-&-&-&$7.4\cdot 10^{-2}$&-&-&$6.7\cdot 10^{-7}$&$-2.6\cdot 10^{-5}$\\
$[C^{S,RR}_{eu}]_{1122}$&-&-&-&-&-&-&-&$-1.5\cdot 10^{-4}$&-&-&-&$4.0\cdot 10^{-8}$\\
$[C^{T,RR}_{ed}]_{1133}$&-&-&-&-&-&-&-&$-1.1\cdot 10^{-1}$&$-3.4\cdot 10^{-7}$&$-2.6\cdot 10^{-5}$&-&-\\
$[C^{T,RR}_{ed}]_{2211}$&-&-&$-1.2\cdot 10^{-4}$&-&$-6.3\cdot 10^{-3}$&-&-&-&-&-&-&-\\
$[C^{T,RR}_{ed}]_{2222}$&-&-&-&$-1.2\cdot 10^{-4}$&-&$-6.3\cdot 10^{-3}$&-&-&-&-&-&-\\
$[C^{T,RR}_{ed}]_{2233}$&-&-&-&-&-&-&$-3.8\cdot 10^{-8}$&-&$-7.0\cdot 10^{-5}$&$-5.4\cdot 10^{-3}$&-&-\\
$[C^{T,RR}_{ed}]_{3311}$&-&-&$-2.0\cdot 10^{-3}$&-&$-1.1\cdot 10^{-1}$&-&-&-&-&-&-&-\\
$[C^{T,RR}_{ed}]_{3322}$&-&-&-&$-2.0\cdot 10^{-3}$&-&$-1.1\cdot 10^{-1}$&-&-&-&-&-&-\\
$[C^{T,RR}_{ed}]_{3333}$&-&-&-&-&-&-&$-6.4\cdot 10^{-7}$&-&$-1.2\cdot 10^{-3}$&$-9.0\cdot 10^{-2}$&-&-\\
$[C^{T,RR}_{eu}]_{2211}$&$3.1\cdot 10^{-4}$&$-6.3\cdot 10^{-3}$&-&-&-&-&-&-&-&-&-&-\\
$[C^{T,RR}_{eu}]_{2222}$&-&-&-&-&-&-&$1.1\cdot 10^{-6}$&-&-&-&$1.4\cdot 10^{-4}$&$-5.3\cdot 10^{-3}$\\
$[C^{T,RR}_{eu}]_{3311}$&$5.3\cdot 10^{-3}$&$-1.1\cdot 10^{-1}$&-&-&-&-&-&-&-&-&-&-\\
$[C^{T,RR}_{eu}]_{3322}$&-&-&-&-&-&-&$1.9\cdot 10^{-5}$&-&-&-&$2.3\cdot 10^{-3}$&$-9.0\cdot 10^{-2}$\\
$[C^{S,RR}_{ed}]_{1133}$&-&-&-&-&-&-&-&$-8.1\cdot 10^{-5}$&-&$-2.0\cdot 10^{-8}$&-&-\\
$[C^{S,RR}_{ed}]_{2211}$&-&-&$-3.6\cdot 10^{-8}$&-&$-6.0\cdot 10^{-6}$&-&-&-&-&-&-&-\\
$[C^{S,RR}_{ed}]_{2222}$&-&-&-&$-3.6\cdot 10^{-8}$&-&$-6.0\cdot 10^{-6}$&-&-&-&-&-&-\\
$[C^{S,RR}_{ed}]_{2233}$&-&-&-&-&-&-&-&-&-&$-4.1\cdot 10^{-6}$&-&-\\
$[C^{S,RR}_{ed}]_{3311}$&-&-&$-6.0\cdot 10^{-7}$&-&$-1.0\cdot 10^{-4}$&-&-&-&-&-&-&-\\
$[C^{S,RR}_{ed}]_{3322}$&-&-&-&$-6.0\cdot 10^{-7}$&-&$-1.0\cdot 10^{-4}$&-&-&-&-&-&-\\
$[C^{S,RR}_{ed}]_{3333}$&-&-&-&-&-&-&-&-&-&$-6.9\cdot 10^{-5}$&-&-\\
$[C^{S,RR}_{eu}]_{2211}$&$-2.6\cdot 10^{-7}$&$1.2\cdot 10^{-5}$&-&-&-&-&-&-&-&-&-&-\\
$[C^{S,RR}_{eu}]_{2222}$&-&-&-&-&-&-&-&-&-&-&-&$8.2\cdot 10^{-6}$\\
$[C^{S,RR}_{eu}]_{3311}$&$-4.4\cdot 10^{-6}$&$2.0\cdot 10^{-4}$&-&-&-&-&-&-&-&-&-&-\\
$[C^{S,RR}_{eu}]_{3322}$&-&-&-&-&-&-&$-1.6\cdot 10^{-8}$&-&-&-&-&$1.4\cdot 10^{-4}$\\
$[C^{V,LR}_{ee}]_{1221}$&-&-&-&-&-&-&-&$-2.0\cdot 10^{-4}$&-&-&-&-\\
$[C^{V,LR}_{ee}]_{1331}$&-&-&-&-&-&-&-&$-3.2\cdot 10^{-3}$&-&-&-&-\\
$[C^{S,RR}_{ee}]_{1221}$&-&-&-&-&-&-&-&$6.3\cdot 10^{-4}$&-&-&-&-\\
$[C^{S,RR}_{ee}]_{1331}$&-&-&-&-&-&-&-&$1.1\cdot 10^{-2}$&-&-&-&-\\
$[C^{S,RR}_{ee}]_{1122}$&-&-&-&-&-&-&-&$-1.3\cdot 10^{-5}$&-&-&-&-\\
$[C^{S,RR}_{ee}]_{1133}$&-&-&-&-&-&-&-&$-2.2\cdot 10^{-4}$&-&-&-&-\\

%%%%%%%%%%%%%
\hline  \hline
& $ \underset{11}{C_{uG}}$  & $\underset{11}{C_{u\gamma}}$ & $\underset{11}{C_{dG}}$ &  $\underset{22}{C_{dG}}$  & $\underset{11}{C_{d\gamma}}$ & $\underset{22}{C_{d\gamma}}$ & $C_{\tilde G}$   &$\underset{11}{C_{e\gamma}}$      & $\underset{33}{C_{dG}}$ & $\underset{33}{C_{d\gamma}}$
&  $\underset{22}{C_{uG}}$  & $\underset{22}{C_{u\gamma}}$   \\ \hline
\end{tabular}
}
\caption{Operator mixing of the semileptonic operators in $5+3$ WET (1st column) onto dipole operators (2nd row) in $3 +1$  and $5+3$ flavor WET is shown. Here the entries represent the quantity $\eta_i^j (\mu_{\rm low}, \mu_{\rm ew})$, $\mu_{\rm ew}=91.1876$ GeV, $\mu_{\rm low}=2.0$ GeV . Entries below $10^{-8}$ are dropped. Note that for the charm and bottom dipoles, we stop running at $\mu=m_b=4.18$ GeV.}\label{tab:mixing3}
\end{center}
\end{table}

In Tabs.~\ref{tab:mixing2}, \ref{tab:mixing1} and \ref{tab:mixing3}, we show the numerical values of the quantity
\be 
\eta_i^j (\mu_{\rm low}, \mu_{\rm ew}) =   {\text{Im} C_j(\mu_{low}) \over \text{Im}C_i(\mu_{ew}) }
\ee 
for $\mu_{low} = 2$ GeV and $\mu_{ew}=91$ GeV. 
In these tables, we focus mainly on heavy flavor operators, that do not generate EDMs at tree level.
In this case, $C_j(\mu_{low})$ stands for the WCs of the dipole and Weinberg operators that enter at tree-level to the EDM expressions. 
The $C_i(\mu_{ew})$ represent the all other WET operators which mix with 
$C_j(\mu_{low})$ via any of the mechanism discussed above. In the 4+1 and 5+3 theories we only show the operators that contain at least one 
heavy flavor fermion current.  Running effects, in particular QCD running, are also important for light flavor operators, and they are fully accounted for in the master formulae that we provide in Section \ref{sec:master}.

\section{EDM Master Formulae in WET}\label{sec:master}
In this section, we derive master formulae for the EDMs for various species such as electron (via the HfF, ThO and YbF precession frequencies), neutron, and proton as well as 
diamagnetic atoms such as Hg, Xe, and  Ra in terms of WET WCs at the EW scale in the JMS basis.  By matching the WET to SMEFT 
and HEFT these formulae can be used to get the EDMs predictions in 
 a large class of UV scenarios with linear and non-linear realizations of EW symmetry breaking, respectively. 
 The main strengths of these formulae are the systematic inclusion of short-distance effects due to RG running and matching 
 and the inclusion of the heavy flavor for the quark and leptons which are novel. 

Without loss of generality, the EDMs can be expressed as a linear function of WET WCs as 
\be \label{eq:master}
d_X =  \sum_{I \in \rm dipoles} \alpha_I^X (\mu_{ew})  { C_{I}^{(5)}(\mu_{ew})  \left[{\rm TeV^{-1}}\right] } + \sum_{J \in \rm 4f} \alpha_J^X (\mu_{ew}){  C_{J}^{(6)}(\mu_{ew})  \left[{\rm {TeV^{-2}}}\right] } \,,
\ee
here $X=n,p$ or Hg, Xe, Ra and the indices $I$ and $J$ run over all contributing dipole and four-fermion operators in the 
$n_f=5+3$ $\Delta F=0$ WET  at the 
EW scale. The complete list of relevant operators can be found in Sec.~\ref{sec:wet}. The coefficients $\alpha_I$ encode the 
long-distance effects due to matrix elements and RG running 
at the 2-loop level as described in the previous section. These coefficients have units of $e\, \rm cm\,  {TeV}^2$ or $e\, \rm cm\,  TeV$ for the 
dim-6 and dim-5 WCs, respectively.    The WCs in \eqref{eq:master} have $\rm {TeV}^{-2}$ and TeV$^{-1}$ units for dim-6 and dim-5 operators. {This is indicated by the units in square brackets in Eq. \eqref{eq:master}}.
In this way, the overall units of  $d_X$  are $e$ cm.   To obtain the EDMs in the $\rm {GeV}^{-1}$   
units we need to divide the $\alpha_I$  coefficients by a conversion factor according to ${\rm GeV}^{-1}= 6.52 \times 10^{-14} e\,\textrm{cm}$.

We can write very similar expressions for the frequencies for the electron EDM. In this case, the $\alpha_I^X$ coefficients  will 
have (mrad/s)$\rm {TeV}^2$ or (mrad/s)TeV units.   
 The values for $\alpha_I$ coefficients at the EW scale and their theoretical uncertainties for the HfF precession frequency, the Hg and neutron  EDMs
 are given in Tab.~\ref{tab:masterHfF},  \ref{tab:masterdHg1}, \ref{tab:masterdHg2} and \ref{tab:masterdn}.
 The master formulae for $d_p$, $\omega_{\rm ThO}$ and $\omega_{\rm YbF}$ can be found in appendix \ref{app:additional}.
 We do not provide explicit master formulae for $d_{\rm Xe}$ and $d_{\rm Ra}$, but they  have been implemented in \texttt{flavio}.
 
 %%%%%%%%%%%%%
 {\boldmath
 \subsection{Theoretical uncertainties}
 }
 The theoretical uncertainties in the $\alpha_I^X$ coefficients can stem from various sources such as 
 the SM parameters and the matrix elements. 
 The latter arise from both nuclear theory and hadronization uncertainties and, for most species, they are the dominant source.
 The uncertainties from atomic theory are in most cases better understood.
 As discussed in Section \ref{sec:edms}, the quoted hadronic and nuclear uncertainties provide at best 
 a very rough estimate of the real theoretical uncertainty. In this respect, the field is less advanced compared to
flavor physics, where, for several observables, lattice QCD matrix elements have been evaluated, or collider physics. 
Here we assume theoretical uncertainties to be Gaussian distributed. 
 The estimate of the theoretical uncertainties is based on the random sampling of the input parameters and 
the evaluation of the standard deviation of the corresponding EDM observables. 
 
 {\boldmath
 \subsection{$\omega_{\text{HfF}}$}
 }
 In this subsection, we provide a master formula for  $\omega_{\text{HfF}}$ (mrad/s), which is sensitive to the electron 
 EDM, see Tab.~\ref{tab:masterHfF}. Similar formulae for the $\omega_{\text{ThO}}$ and $\omega_{\text{YbF}}$ 
 are given in App.~\ref{app:additional}.
 %%%%%%%%%
\begin{table}[t]
\begin{center}
 \renewcommand*{\arraystretch}{0.9}
 \resizebox{0.8\textwidth}{!}{
 {
\begin{tabular}{ |cc|cc|}
\hline
\hline
\multicolumn{4}{|c|}{{\bf $\alpha_I^{\rm HfF} (\mu_{ew})$ for the $\omega_{\text{HfF}}$}}  \\
\hline
\hline
\multicolumn{2}{|c|}{$n_f=3+1$}     & \multicolumn{2}{|c|}{$n_f=5+3$}  \\ \hline
%%%%%%%%%%

$[C_{e\gamma}]_{11}$&$(-2.9\pm0.1) \cdot 10^{12}$&$[C^{S,RR}_{eu}]_{1122}$&$(9.2\pm0.3) \cdot 10^{5}$\\
$[C^{S,RR}_{ee}]_{1111}$&$(-2.7\pm0.1) \cdot 10^{4}$&$[C^{T,RR}_{eu}]_{1122}$&$(-3.8\pm0.2) \cdot 10^{8}$\\
$[C^{S,RR}_{eu}]_{1111}$&$(-2.6\pm0.2) \cdot 10^{8}$&$[C^{S,RR}_{ed}]_{1133}$&$(3.7\pm0.1) \cdot 10^{5}$\\
$[C^{S,RR}_{ed}]_{1111}$&$(-2.6\pm0.3) \cdot 10^{8}$&$[C^{T,RR}_{ed}]_{1133}$&$(4.8\pm0.2) \cdot 10^{8}$\\
$[C^{S,RR}_{ed}]_{1122}$&$(-1.9\pm0.1) \cdot 10^{7}$&$[C^{S,RR}_{ee}]_{1122}$&$(5.8\pm0.2) \cdot 10^{4}$\\
$[C^{S,RL}_{eu}]_{1111}$&$(-2.6\pm0.3) \cdot 10^{8}$&$[C^{S,RR}_{ee}]_{1133}$&$(1.3\pm0.0) \cdot 10^{6}$\\
$[C^{S,RL}_{ed}]_{1111}$&$(-2.6\pm0.2) \cdot 10^{8}$&$[C^{S,RR}_{ee}]_{1221}$&$(-2.8\pm0.1) \cdot 10^{6}$\\
$[C^{S,RL}_{ed}]_{1122}$&$(-1.9\pm0.1) \cdot 10^{7}$&$[C^{S,RR}_{ee}]_{1331}$&$(-5.5\pm0.2) \cdot 10^{7}$\\
$[C^{T,RR}_{eu}]_{1111}$&$(2.3\pm0.1) \cdot 10^{8}$&$[C^{V,LR}_{ee}]_{1221}$&$(9.1\pm0.4) \cdot 10^{5}$\\
$[C^{T,RR}_{ed}]_{1111}$&$(-1.2\pm0.0) \cdot 10^{8}$&$[C^{V,LR}_{ee}]_{1331}$&$(1.5\pm0.1) \cdot 10^{7}$\\
$[C^{T,RR}_{ed}]_{1122}$&$(-9.0\pm0.4) \cdot 10^{7}$&~&-\\

%%%%%%%%%%%%
\hline
\end{tabular}}
}
\caption{Master formula for $\omega_{\rm HfF}$ in WET in the JMS basis at the EW scale.
The $\alpha_I^{\rm HfF}$ coefficients have units of (mrad/s)$\rm TeV^{2}$ and
(mrad/s)$\rm TeV$ for the four-fermion and dipole operators, respectively.
The $90\%$ CL upper bound on $\omega_{\rm HfF}$ is $0.17$ mrad/s.}
\label{tab:masterHfF}
\end{center}
\end{table}

Recalling that the $90\%$ CL upper bound on $\omega_{\rm HfF}$ is $0.17$ mrad/s,
we can immediately see that the scale of the operators that contribute at tree level, which comprises all the operators in the 
$n_f = 3 + 1$ column of Table \ref{tab:masterHfF} with the exception of $\wc[S, RR]{ee}{1111}$, has to be very high,  $\Lambda \sim 10^4$ TeV.
Even for $\wc[S, RR]{ee}{1111}$, which mixes onto $C_{e\gamma}$ at one loop, $\Lambda$ has to be close to $400$ TeV.
The $n_f = 5+3$ column in Table \ref{tab:masterHfF} contains operators with electrons and heavy quarks 
or electrons and muons or taus. All these operators are very strongly constrained.

 {\boldmath
 \subsection{ $d_{\rm Hg}$}
}
The master formulae for $d_{\rm Hg}$ are presented in Tab.~\ref{tab:masterdHg1} and Tab.~\ref{tab:masterdHg2} for
 the semileptonic and four-quark operators, respectively.
From Table \ref{tab:masterdHg2}
we notice that for all the four-quark operators in Table \ref{tab:wet4},
the central value of $d_{\rm Hg}$ is larger than the experimental bound, 
implying that, even in the worst case scenario, Hg EDM experiments probe four-quark scales of $1$ TeV or larger. 
This is true even for four-quark operators with four heavy quarks, such as $\wc[V1, LR]{uddu}{2332}$ (with two $b$ and two $c$ quarks) or $\wc[S8, RR]{dd}{3333}$
(with four $b$ quarks).
The scale of light flavor operators needs in general to be much larger than the TeV.
This picture is complicated by theoretical uncertainties, which are generally very large.

\begin{table}[H]
\begin{center}
 \renewcommand*{\arraystretch}{1.0}
 \resizebox{\textwidth}{!}{
\begin{tabular}{ |cc|cc|cc|}
\hline
\hline
\multicolumn{6}{|c|}{{\bf $\alpha_I^{\rm Hg} (\mu_{ew})$ for the Hg EDM: SL operators}}  \\
\hline 
\hline
\multicolumn{2}{|c|}{$n_f=3+1$} & \multicolumn{2}{|c|}{$n_f=4+1$} & \multicolumn{2}{|c|}{$n_f=5+3$}  \\ \hline
%%%%%%%%

$[C^{S,RR}_{eu}]_{1111}$&$(8.3\pm1.5) \cdot 10^{-22}$&$[C^{T,RR}_{eu}]_{1122}$&-&$[C^{T,RR}_{ed}]_{1133}$&-\\
$[C^{S,RR}_{ed}]_{1111}$&$(-3.8\pm1.3) \cdot 10^{-22}$&$[C^{S,RR}_{eu}]_{1122}$&-&$[C^{T,RR}_{ed}]_{2211}$&$(-2.6\pm0.7) \cdot 10^{-25}$\\
$[C^{S,RR}_{ed}]_{1122}$&$(1.7\pm0.4) \cdot 10^{-23}$&~&-&$[C^{T,RR}_{ed}]_{2222}$&$(1.0\pm0.7) \cdot 10^{-27}$\\
$[C^{S,RL}_{eu}]_{1111}$&$(-3.8\pm1.4) \cdot 10^{-22}$&~&-&$[C^{T,RR}_{ed}]_{2233}$&-\\
$[C^{S,RL}_{ed}]_{1111}$&$(8.3\pm1.6) \cdot 10^{-22}$&~&-&$[C^{T,RR}_{ed}]_{3311}$&$(-4.8\pm1.4) \cdot 10^{-24}$\\
$[C^{S,RL}_{ed}]_{1122}$&$(1.7\pm0.4) \cdot 10^{-23}$&~&-&$[C^{T,RR}_{ed}]_{3322}$&$(1.9\pm1.3) \cdot 10^{-26}$\\
$[C^{T,RR}_{eu}]_{1111}$&$(-3.7\pm0.9) \cdot 10^{-23}$&~&-&$[C^{T,RR}_{ed}]_{3333}$&$(3.5\pm2.1) \cdot 10^{-31}$\\
$[C^{T,RR}_{ed}]_{1111}$&$(-2.4\pm0.6) \cdot 10^{-22}$&~&-&$[C^{T,RR}_{eu}]_{2211}$&$(5.0\pm8.0) \cdot 10^{-26}$\\
$[C^{T,RR}_{ed}]_{1122}$&$(1.7\pm0.6) \cdot 10^{-24}$&~&-&$[C^{T,RR}_{eu}]_{2222}$&$(-5.9\pm3.8) \cdot 10^{-31}$\\
~&-&~&-&$[C^{T,RR}_{eu}]_{3311}$&$(9.1\pm12.1) \cdot 10^{-25}$\\
~&-&~&-&$[C^{T,RR}_{eu}]_{3322}$&$(-9.9\pm5.7) \cdot 10^{-30}$\\
~&-&~&-&$[C^{S,RR}_{ed}]_{1133}$&-\\
~&-&~&-&$[C^{S,RR}_{ed}]_{2211}$&$(-2.6\pm0.6) \cdot 10^{-28}$\\
~&-&~&-&$[C^{S,RR}_{ed}]_{2222}$&$(1.0\pm0.6) \cdot 10^{-30}$\\
~&-&~&-&$[C^{S,RR}_{ed}]_{2233}$&-\\
~&-&~&-&$[C^{S,RR}_{ed}]_{3311}$&$(-5.2\pm1.4) \cdot 10^{-27}$\\
~&-&~&-&$[C^{S,RR}_{ed}]_{3322}$&$(2.0\pm1.3) \cdot 10^{-29}$\\
~&-&~&-&$[C^{S,RR}_{ed}]_{3333}$&-\\
~&-&~&-&$[C^{S,RR}_{eu}]_{2211}$&$(-8.9\pm5.2) \cdot 10^{-29}$\\
~&-&~&-&$[C^{S,RR}_{eu}]_{2222}$&-\\
~&-&~&-&$[C^{S,RR}_{eu}]_{3311}$&$(-1.8\pm1.0) \cdot 10^{-27}$\\
~&-&~&-&$[C^{S,RR}_{eu}]_{3322}$&-\\

%%%%%%%%%%%%%%%%%%
\hline
\end{tabular}
}

\caption{Master formula for $d_{\rm Hg}$ in terms of semileptonic WET operators in the JMS basis at the EW scale.
The $\alpha_I^{\rm Hg}$ have units of $e\,\rm cm\, \rm TeV^{2}$.
The $90\%$ CL upper bound on $d_{\rm Hg}$ is $6.2\times 10^{-30} $ $e$ cm. {The entries below $10^{-33}$ have been dropped.}
}
\label{tab:masterdHg1}
\end{center}
\end{table}

\begin{table}[H]
\begin{center}
 \renewcommand*{\arraystretch}{1.0}
 \resizebox{\textwidth}{!}{
\begin{tabular}{ |cc|cc|cc|}
\hline
\hline
\multicolumn{6}{|c|}{{\bf $\alpha_I^{\rm Hg} (\mu_{ew})$ for the Hg EDM}}  \\
\hline 
\hline
\multicolumn{2}{|c|}{$n_f=3+1$} & \multicolumn{2}{|c|}{$n_f=4+1$} & \multicolumn{2}{|c|}{$n_f=5+3$}  \\ \hline
%%%%%%%%%%%%%

$C_{ \widetilde G}$&$(-2.4\pm3.9) \cdot 10^{-25}$&$[C_{uG}]_{22}$&$(-2.6\pm1.5) \cdot 10^{-24}$&$[C_{dG}]_{33}$&$(-3.8\pm2.6) \cdot 10^{-25}$\\
$[C_{dG}]_{11}$&$(5.0\pm28.3) \cdot 10^{-20}$&$[C_{u\gamma}]_{22}$&$(2.5\pm1.6) \cdot 10^{-25}$&$[C_{d\gamma}]_{33}$&$(-7.3\pm5.3) \cdot 10^{-27}$\\
$[C_{dG}]_{22}$&$(-6.7\pm4.9) \cdot 10^{-24}$&$[C^{S1,RR}_{uu}]_{2222}$&$(1.9\pm1.0) \cdot 10^{-28}$&$[C^{V1,LR}_{uddu}]_{1331}$&$(-1.4\pm7.4) \cdot 10^{-24}$\\
$[C_{uG}]_{11}$&$(4.1\pm32.9) \cdot 10^{-20}$&$[C^{S8,RR}_{uu}]_{2222}$&$(-4.3\pm2.6) \cdot 10^{-29}$&$[C^{V8,LR}_{uddu}]_{1331}$&$(3.2\pm21.5) \cdot 10^{-25}$\\
$[C_{d\gamma}]_{11}$&$(3.5\pm1.3) \cdot 10^{-20}$&$[C^{S1,RR}_{uu}]_{1122}$&$(5.9\pm46.4) \cdot 10^{-25}$&$[C^{V1,LR}_{dd}]_{1331}$&$(-1.6\pm7.4) \cdot 10^{-24}$\\
$[C_{d\gamma}]_{22}$&$(-1.3\pm0.8) \cdot 10^{-22}$&$[C^{S8,RR}_{uu}]_{1122}$&$(1.7\pm15.9) \cdot 10^{-25}$&$[C^{V8,LR}_{dd}]_{1331}$&$(9.5\pm20.4) \cdot 10^{-25}$\\
$[C_{u\gamma}]_{11}$&$(-9.4\pm21.8) \cdot 10^{-21}$&$[C^{S1,RR}_{ud}]_{2211}$&$(7.0\pm35.1) \cdot 10^{-25}$&$[C^{V1,LR}_{dd}]_{2332}$&$(-5.0\pm3.3) \cdot 10^{-28}$\\
$[C^{V1,LR}_{uddu}]_{1111}$&$(8.9\pm30.3) \cdot 10^{-24}$&$[C^{S8,RR}_{ud}]_{2211}$&$(2.6\pm12.5) \cdot 10^{-25}$&$[C^{V8,LR}_{dd}]_{2332}$&$(-1.8\pm1.1) \cdot 10^{-27}$\\
$[C^{V8,LR}_{uddu}]_{1111}$&$(1.4\pm5.3) \cdot 10^{-23}$&$[C^{S1,RR}_{ud}]_{2222}$&$(-1.2\pm0.6) \cdot 10^{-29}$&$[C^{S1,RR}_{dd}]_{3333}$&$(7.8\pm5.0) \cdot 10^{-29}$\\
$[C^{V1,LR}_{uddu}]_{1221}$&$(4.8\pm14.4) \cdot 10^{-24}$&$[C^{S8,RR}_{ud}]_{2222}$&$(-2.3\pm1.4) \cdot 10^{-28}$&$[C^{S8,RR}_{dd}]_{3333}$&$(-1.6\pm1.2) \cdot 10^{-29}$\\
$[C^{V8,LR}_{uddu}]_{1221}$&$(7.4\pm23.5) \cdot 10^{-24}$&$[C^{S1,RR}_{uddu}]_{2112}$&$(-3.1\pm15.0) \cdot 10^{-24}$&$[C^{S1,RR}_{dd}]_{1133}$&$(9.1\pm52.2) \cdot 10^{-25}$\\
$[C^{V1,LR}_{dd}]_{1221}$&$(-4.1\pm13.9) \cdot 10^{-24}$&$[C^{S8,RR}_{uddu}]_{2112}$&$(5.1\pm201.5) \cdot 10^{-26}$&$[C^{S8,RR}_{dd}]_{1133}$&$(2.6\pm14.4) \cdot 10^{-25}$\\
$[C^{V8,LR}_{dd}]_{1221}$&$(-6.2\pm20.0) \cdot 10^{-24}$&$[C^{S1,RR}_{uddu}]_{2222}$&$(1.1\pm0.6) \cdot 10^{-27}$&$[C^{S1,RR}_{dd}]_{2233}$&$(-5.6\pm3.5) \cdot 10^{-29}$\\
$[C^{S1,RR}_{uu}]_{1111}$&$(-5.9\pm21.7) \cdot 10^{-24}$&$[C^{S8,RR}_{uddu}]_{2222}$&$(8.5\pm6.3) \cdot 10^{-28}$&$[C^{S8,RR}_{dd}]_{2233}$&$(1.6\pm1.1) \cdot 10^{-28}$\\
$[C^{S8,RR}_{uu}]_{1111}$&$(1.8\pm5.6) \cdot 10^{-24}$&$[C^{S1,RR}_{uu}]_{1221}$&$(-2.4\pm15.3) \cdot 10^{-24}$&$[C^{S1,RR}_{dd}]_{1331}$&$(-5.7\pm34.0) \cdot 10^{-24}$\\
$[C^{S1,RR}_{dd}]_{1111}$&$(4.9\pm24.4) \cdot 10^{-24}$&$[C^{S8,RR}_{uu}]_{1221}$&$(2.7\pm14.9) \cdot 10^{-25}$&$[C^{S8,RR}_{dd}]_{1331}$&$(1.1\pm4.2) \cdot 10^{-24}$\\
$[C^{S8,RR}_{dd}]_{1111}$&$(-1.5\pm4.6) \cdot 10^{-24}$&$[C^{V1,LR}_{uddu}]_{2112}$&$(-5.5\pm22.3) \cdot 10^{-25}$&$[C^{S1,RR}_{dd}]_{2332}$&$(-6.4\pm4.7) \cdot 10^{-28}$\\
$[C^{S1,RR}_{dd}]_{2222}$&$(-4.4\pm2.8) \cdot 10^{-29}$&$[C^{V8,LR}_{uddu}]_{2112}$&$(-1.7\pm10.2) \cdot 10^{-25}$&$[C^{S8,RR}_{dd}]_{2332}$&$(-1.3\pm0.8) \cdot 10^{-27}$\\
$[C^{S8,RR}_{dd}]_{2222}$&$(-5.3\pm3.2) \cdot 10^{-29}$&$[C^{V1,LR}_{uddu}]_{2222}$&$(3.6\pm2.4) \cdot 10^{-28}$&$[C^{S1,RR}_{ud}]_{1133}$&$(7.7\pm61.1) \cdot 10^{-25}$\\
$[C^{S1,RR}_{dd}]_{1122}$&$(1.3\pm5.7) \cdot 10^{-24}$&$[C^{V8,LR}_{uddu}]_{2222}$&$(1.2\pm0.8) \cdot 10^{-27}$&$[C^{S8,RR}_{ud}]_{1133}$&$(2.6\pm13.2) \cdot 10^{-25}$\\
$[C^{S8,RR}_{dd}]_{1122}$&$(-3.7\pm14.2) \cdot 10^{-25}$&$[C^{V1,LR}_{uu}]_{1221}$&$(-3.6\pm21.9) \cdot 10^{-25}$&$[C^{S1,RR}_{uddu}]_{1331}$&$(-5.1\pm24.2) \cdot 10^{-24}$\\
$[C^{S1,RR}_{dd}]_{1221}$&$(1.1\pm4.1) \cdot 10^{-24}$&$[C^{V8,LR}_{uu}]_{1221}$&$(1.8\pm6.7) \cdot 10^{-25}$&$[C^{S8,RR}_{uddu}]_{1331}$&$(5.6\pm33.0) \cdot 10^{-25}$\\
$[C^{S8,RR}_{dd}]_{1221}$&$(-3.5\pm13.1) \cdot 10^{-25}$&~&-&$[C^{S1,RR}_{ud}]_{2233}$&$(-5.1\pm2.8) \cdot 10^{-29}$\\
$[C^{S1,RR}_{ud}]_{1111}$&$(-2.3\pm3.6) \cdot 10^{-25}$&~&-&$[C^{S8,RR}_{ud}]_{2233}$&$(-1.7\pm0.9) \cdot 10^{-29}$\\
$[C^{S8,RR}_{ud}]_{1111}$&$(7.0\pm10.8) \cdot 10^{-26}$&~&-&$[C^{S1,RR}_{uddu}]_{2332}$&$(3.3\pm2.0) \cdot 10^{-28}$\\
$[C^{S1,RR}_{ud}]_{1122}$&$(-1.4\pm5.1) \cdot 10^{-24}$&~&-&$[C^{S8,RR}_{uddu}]_{2332}$&$(-4.0\pm1.8) \cdot 10^{-29}$\\
$[C^{S8,RR}_{ud}]_{1122}$&$(4.5\pm18.1) \cdot 10^{-25}$&~&-&$[C^{V1,LR}_{uddu}]_{2332}$&$(9.3\pm4.7) \cdot 10^{-29}$\\
$[C^{S1,RR}_{uddu}]_{1111}$&$(-2.4\pm3.8) \cdot 10^{-25}$&~&-&$[C^{V8,LR}_{uddu}]_{2332}$&$(-2.6\pm1.4) \cdot 10^{-29}$\\
$[C^{S8,RR}_{uddu}]_{1111}$&$(7.0\pm10.6) \cdot 10^{-26}$&~&-&~&-\\
$[C^{S1,RR}_{uddu}]_{1221}$&$(-1.6\pm6.0) \cdot 10^{-24}$&~&-&~&-\\
$[C^{S8,RR}_{uddu}]_{1221}$&$(4.6\pm14.6) \cdot 10^{-25}$&~&-&~&-\\

%%%%%%%%%%%%%
\hline
\end{tabular}
}
\caption{Master formula for $d_{\rm Hg}$ in WET in the JMS basis at the EW scale. The $\alpha_I^{\rm Hg}$ have units of  $e$ cm $\rm TeV^{2}$ and 
$e$ cm $\rm TeV$ for the four-fermion and dipole operators, respectively. The $90\%$ CL upper bound on $d_{\rm Hg}$ is $6.2\times 10^{-30}$ $e$ cm.}
\label{tab:masterdHg2}
\end{center}
\end{table}
%%%%%%%%%%%%%

Table \ref{tab:masterdHg1} provides a similar message for semileptonic operators,
where one can extract strong constraints on operators such as $\wc[T,RR]{eu}{3311}$ (two $\tau$ leptons and two $u$ quarks)
or $\wc[T,RR]{ed}{3311}$ (two $\tau$ leptons and two $d$ quarks). In this case, however,
operators with both heavy leptons and heavy quarks are less constrained and their scale can drop below the TeV.
This is for example the case of  $\wc[T,RR]{eu}{2222}$ (with a naive scale of $\Lambda \sim 0.3$ TeV, neglecting theoretical uncertainties)
or $\wc[S,RR]{eu}{3322}$, with naive scale below the electroweak.

 {\boldmath
\subsection{$d_n$}
}
The master formula for the neutron EDM is presented in Tab.~\ref{tab:masterdn}.  
As in the case of $d_{\rm Hg}$, we notice that, if we neglect theoretical uncertainties, all four-quark operators
in Table \ref{tab:wet4} need to have scales larger than the TeV.

\begin{table}[H]
\begin{center}
 \renewcommand*{\arraystretch}{1.0}
 \resizebox{\textwidth}{!}{
\begin{tabular}{ |cc|cc|cc|}
\hline
\hline
\multicolumn{6}{|c|}{{\bf $\alpha_I^n (\mu_{ew})$ for neutron EDM}}  \\
\hline 
\hline
\multicolumn{2}{|c|}{$n_f=3+1$} & \multicolumn{2}{|c|}{$n_f=4+1$} & \multicolumn{2}{|c|}{$n_f=5+3$}  \\ \hline
%%%%%%%%%%%%%

$C_{ \widetilde G}$&$(5.5\pm2.8) \cdot 10^{-22}$&$[C_{uG}]_{22}$&$(7.7\pm3.9) \cdot 10^{-21}$&$[C_{dG}]_{33}$&$(1.1\pm0.5) \cdot 10^{-21}$\\
$[C_{dG}]_{11}$&$(-3.7\pm1.8) \cdot 10^{-17}$&$[C_{u\gamma}]_{22}$&$(-7.4\pm3.4) \cdot 10^{-22}$&$[C_{d\gamma}]_{33}$&$(2.0\pm1.1) \cdot 10^{-23}$\\
$[C_{dG}]_{22}$&$(1.5\pm0.9) \cdot 10^{-20}$&$[C^{S1,RR}_{uu}]_{2222}$&$(-5.6\pm3.0) \cdot 10^{-25}$&$[C^{V1,LR}_{uddu}]_{1331}$&$(6.0\pm3.1) \cdot 10^{-22}$\\
$[C_{uG}]_{11}$&$(-1.9\pm0.9) \cdot 10^{-17}$&$[C^{S8,RR}_{uu}]_{2222}$&$(1.3\pm0.5) \cdot 10^{-25}$&$[C^{V8,LR}_{uddu}]_{1331}$&$(1.5\pm0.9) \cdot 10^{-22}$\\
$[C_{d\gamma}]_{11}$&$(-8.9\pm0.4) \cdot 10^{-17}$&$[C^{S1,RR}_{uu}]_{1122}$&$(-2.3\pm1.1) \cdot 10^{-22}$&$[C^{V1,LR}_{dd}]_{1331}$&$(8.1\pm6.4) \cdot 10^{-22}$\\
$[C_{d\gamma}]_{22}$&$(3.0\pm1.7) \cdot 10^{-19}$&$[C^{S8,RR}_{uu}]_{1122}$&$(-3.6\pm4.1) \cdot 10^{-23}$&$[C^{V8,LR}_{dd}]_{1331}$&$(-1.5\pm0.2) \cdot 10^{-21}$\\
$[C_{u\gamma}]_{11}$&$(2.5\pm0.2) \cdot 10^{-17}$&$[C^{S1,RR}_{ud}]_{2211}$&$(-4.8\pm2.6) \cdot 10^{-22}$&$[C^{V1,LR}_{dd}]_{2332}$&$(1.1\pm0.6) \cdot 10^{-24}$\\
$[C^{V1,LR}_{uddu}]_{1111}$&$(-4.3\pm2.4) \cdot 10^{-22}$&$[C^{S8,RR}_{ud}]_{2211}$&$(-2.8\pm0.7) \cdot 10^{-22}$&$[C^{V8,LR}_{dd}]_{2332}$&$(4.1\pm2.5) \cdot 10^{-24}$\\
$[C^{V8,LR}_{uddu}]_{1111}$&$(-6.6\pm3.7) \cdot 10^{-22}$&$[C^{S1,RR}_{ud}]_{2222}$&$(2.9\pm1.4) \cdot 10^{-26}$&$[C^{S1,RR}_{dd}]_{3333}$&$(-2.2\pm1.1) \cdot 10^{-25}$\\
$[C^{V1,LR}_{uddu}]_{1221}$&$(-9.0\pm4.7) \cdot 10^{-22}$&$[C^{S8,RR}_{ud}]_{2222}$&$(5.2\pm3.3) \cdot 10^{-25}$&$[C^{S8,RR}_{dd}]_{3333}$&$(4.6\pm2.5) \cdot 10^{-26}$\\
$[C^{V8,LR}_{uddu}]_{1221}$&$(-1.4\pm0.7) \cdot 10^{-21}$&$[C^{S1,RR}_{uddu}]_{2112}$&$(2.6\pm1.0) \cdot 10^{-21}$&$[C^{S1,RR}_{dd}]_{1133}$&$(-6.4\pm3.2) \cdot 10^{-22}$\\
$[C^{V1,LR}_{dd}]_{1221}$&$(-4.8\pm2.6) \cdot 10^{-22}$&$[C^{S8,RR}_{uddu}]_{2112}$&$(3.8\pm1.0) \cdot 10^{-22}$&$[C^{S8,RR}_{dd}]_{1133}$&$(-9.7\pm9.5) \cdot 10^{-23}$\\
$[C^{V8,LR}_{dd}]_{1221}$&$(-7.4\pm3.1) \cdot 10^{-22}$&$[C^{S1,RR}_{uddu}]_{2222}$&$(-2.5\pm1.5) \cdot 10^{-24}$&$[C^{S1,RR}_{dd}]_{2233}$&$(1.3\pm0.7) \cdot 10^{-25}$\\
$[C^{S1,RR}_{uu}]_{1111}$&$(1.1\pm0.5) \cdot 10^{-21}$&$[C^{S8,RR}_{uddu}]_{2222}$&$(-1.9\pm1.3) \cdot 10^{-24}$&$[C^{S8,RR}_{dd}]_{2233}$&$(-3.7\pm2.1) \cdot 10^{-25}$\\
$[C^{S8,RR}_{uu}]_{1111}$&$(-3.4\pm1.5) \cdot 10^{-22}$&$[C^{S1,RR}_{uu}]_{1221}$&$(8.3\pm4.2) \cdot 10^{-22}$&$[C^{S1,RR}_{dd}]_{1331}$&$(3.5\pm2.0) \cdot 10^{-21}$\\
$[C^{S1,RR}_{dd}]_{1111}$&$(5.7\pm3.1) \cdot 10^{-22}$&$[C^{S8,RR}_{uu}]_{1221}$&$(-2.5\pm0.4) \cdot 10^{-22}$&$[C^{S8,RR}_{dd}]_{1331}$&$(-1.3\pm0.3) \cdot 10^{-21}$\\
$[C^{S8,RR}_{dd}]_{1111}$&$(-1.8\pm0.9) \cdot 10^{-22}$&$[C^{V1,LR}_{uddu}]_{2112}$&$(5.4\pm1.4) \cdot 10^{-22}$&$[C^{S1,RR}_{dd}]_{2332}$&$(1.4\pm0.8) \cdot 10^{-24}$\\
$[C^{S1,RR}_{dd}]_{2222}$&$(1.0\pm0.6) \cdot 10^{-25}$&$[C^{V8,LR}_{uddu}]_{2112}$&$(7.2\pm0.5) \cdot 10^{-22}$&$[C^{S8,RR}_{dd}]_{2332}$&$(2.9\pm1.6) \cdot 10^{-24}$\\
$[C^{S8,RR}_{dd}]_{2222}$&$(1.2\pm0.6) \cdot 10^{-25}$&$[C^{V1,LR}_{uddu}]_{2222}$&$(-8.1\pm4.5) \cdot 10^{-25}$&$[C^{S1,RR}_{ud}]_{1133}$&$(-3.2\pm1.4) \cdot 10^{-22}$\\
$[C^{S1,RR}_{dd}]_{1122}$&$(1.2\pm0.7) \cdot 10^{-22}$&$[C^{V8,LR}_{uddu}]_{2222}$&$(-2.8\pm1.8) \cdot 10^{-24}$&$[C^{S8,RR}_{ud}]_{1133}$&$(-1.4\pm0.5) \cdot 10^{-22}$\\
$[C^{S8,RR}_{dd}]_{1122}$&$(-4.7\pm2.4) \cdot 10^{-23}$&$[C^{V1,LR}_{uu}]_{1221}$&$(6.3\pm7.9) \cdot 10^{-23}$&$[C^{S1,RR}_{uddu}]_{1331}$&$(2.3\pm1.0) \cdot 10^{-21}$\\
$[C^{S1,RR}_{dd}]_{1221}$&$(2.5\pm1.0) \cdot 10^{-22}$&$[C^{V8,LR}_{uu}]_{1221}$&$(-2.6\pm0.3) \cdot 10^{-22}$&$[C^{S8,RR}_{uddu}]_{1331}$&$(-5.8\pm11.6) \cdot 10^{-23}$\\
$[C^{S8,RR}_{dd}]_{1221}$&$(-7.8\pm2.0) \cdot 10^{-23}$&~&-&$[C^{S1,RR}_{ud}]_{2233}$&$(1.5\pm0.7) \cdot 10^{-25}$\\
$[C^{S1,RR}_{ud}]_{1111}$&$(4.2\pm2.1) \cdot 10^{-22}$&~&-&$[C^{S8,RR}_{ud}]_{2233}$&$(4.9\pm2.0) \cdot 10^{-26}$\\
$[C^{S8,RR}_{ud}]_{1111}$&$(-1.3\pm0.7) \cdot 10^{-22}$&~&-&$[C^{S1,RR}_{uddu}]_{2332}$&$(-9.7\pm4.9) \cdot 10^{-25}$\\
$[C^{S1,RR}_{ud}]_{1122}$&$(2.6\pm1.4) \cdot 10^{-22}$&~&-&$[C^{S8,RR}_{uddu}]_{2332}$&$(1.2\pm0.6) \cdot 10^{-25}$\\
$[C^{S8,RR}_{ud}]_{1122}$&$(-8.8\pm4.4) \cdot 10^{-23}$&~&-&$[C^{V1,LR}_{uddu}]_{2332}$&$(-2.7\pm1.5) \cdot 10^{-25}$\\
$[C^{S1,RR}_{uddu}]_{1111}$&$(4.2\pm2.2) \cdot 10^{-22}$&~&-&$[C^{V8,LR}_{uddu}]_{2332}$&$(7.5\pm3.6) \cdot 10^{-26}$\\
$[C^{S8,RR}_{uddu}]_{1111}$&$(-1.3\pm0.7) \cdot 10^{-22}$&~&-&~&-\\
$[C^{S1,RR}_{uddu}]_{1221}$&$(3.4\pm1.4) \cdot 10^{-22}$&~&-&~&-\\
$[C^{S8,RR}_{uddu}]_{1221}$&$(-8.4\pm3.9) \cdot 10^{-23}$&~&-&~&-\\

%%%%%%%%%%%%%
\hline
\end{tabular}
}
\caption{Master formula for $d_n$ in WET in the JMS basis at the EW scale. The $\alpha_I^n$ have units of $e$ cm $\rm TeV^{2}$ and 
$e$ cm $\rm TeV$ for the four-fermion and dipole operators, respectively. The $90\%$ CL upper bound on $d_n$ is $1.8\times 10^{-26} $ $e$ cm.}
\label{tab:masterdn}
\end{center}
\end{table}

\section{Non-standard Higgs Couplings}\label{sec:Higgs}
In this section, we apply the master formulae derived so far to
constrain possible non-standard Yukawa couplings of the Higgs boson.
We consider the following Yukawa Lagrangian
\be
\mathcal{L}_{\rm Yuk}^{\rm eff} = - \left ( \sum_{i,j = d,s,b} y_{ij}^d ~\bar d_{L,i}  d_{R,j}  + \sum_{i,j=u,c,t}  y_{ij}^u ~\bar u_{L,i} u_{R,j} 
+ \sum_{i,j=e,\mu,\tau}  y_{ij}^e ~\bar e_{L,i} e_{R,j} \right ) h  + h.c.
\ee
This Lagrangian can arise in two theoretical frameworks that can be used for describing non-standard couplings of the Higgs to quarks, lepton, and gauge bosons.
In SMEFT, gauge symmetry is linearly realized, the Higgs boson belongs to a $SU_L(2)$
doublet and operators are organized according to their canonical dimension. 
In this framework, the SM Yukawa couplings arise at dim-4, they can always be chosen to be real and diagonal, and are determined by the quark and lepton masses.
Non-standard Yukawa couplings arise at dim-6 and are given by the Lagrangian 
\cite{Buchmuller:1986zs,Grzadkowski:2010es,Hamoudou:2022tdn}
 \be \label{eq:higgssmeft}
 \begin{aligned}
 \mathcal{L}_{h} &=
  (D_\mu H)^\dagger (D^\mu H) - \lambda\left(H^\dagger H -{1\over 2}v^2\right)^2 - 
  \left[
  \bar Q \hat Y_u u_R \widetilde H +  \bar Q \hat Y_d d_R H + \bar L \hat Y_u e_R  H \right.    \\
  & \left.  + (\bar Q  C_{uH} u_R \widetilde H
  + \bar Q  C_{dH} d_R  H
  +\bar L  C_{eH} e_R  H  ) (H^\dagger H) + h.c. \right].
 \end{aligned}
 \ee
In unitary gauge
\begin{equation}
H=\frac{1}{\sqrt{2}} (0,  v+h)^T,
\end{equation}
where $h$ is the physical Higgs boson field.
$C_{fH}$, for  $f= u,d ,e$, are the WC of  dim-6 SMEFT operators. We can rewrite Eq. \eqref{eq:higgssmeft} in terms of mass matrices, $M_f$, and 
Yukawa interactions
 \begin{align}
 M_u &=
 \frac{v}{\sqrt{2}} \left( \hat Y_u  + \frac{v^2}{2} C_{uH}\right), \qquad   y^u  =  \frac{1}{\sqrt{2}} \left( \hat Y_u  + \frac{3}{2} v^2 C_{uH} \right)  = \frac{M_u}{v}  + \frac{v^2}{\sqrt{2}} C_{uH}\,,  \\
 M_d &=
 \frac{v}{\sqrt{2}} \left( \hat Y_d  + \frac{v^2}{2} C_{dH}\right), \qquad   y^d  =  \frac{1}{\sqrt{2}} \left( \hat Y_d  + \frac{3}{2} v^2 C_{dH} \right)  = \frac{M_d}{v}  + \frac{v^2}{\sqrt{2}} C_{dH}\,, \\
 M_e &=
 \frac{v}{\sqrt{2}} \left( \hat Y_e  + \frac{v^2}{2} C_{eH}\right), \qquad   y^e  =  \frac{1}{\sqrt{2}} \left( \hat Y_e  + \frac{3}{2} v^2 C_{eH} \right)  = \frac{M_e}{v}  + \frac{v^2}{\sqrt{2}} C_{eH}. 
 \end{align}
Since at dim-6 the fermion masses and Yukawa couplings 
are proportional to different linear combinations of $\hat{Y}_f$ and $C_{fH}$,  once the mass matrices are diagonalized, the Yukawas are in general non-diagonal and complex.

The other possible theoretical framework is the Higgs Effective Field Theory (HEFT), in which gauge symmetry is realized non-linearly.
In this case, the Higgs field $h$ is a singlet under $SU(2)_L$, and the EFT provides a momentum expansion in $Q/\Lambda$. 
The LO HEFT Lagrangian is given by
\cite{Buchalla:2013rka}
\be\label{eq:higgsheft}
\begin{aligned}
\mathcal{L}_{Uh} &= \frac{v^2}{4}  \langle D_\mu U^\dagger D^\mu U   \rangle
(1+ F_U(h)) + \frac{1}{2} \partial_\mu h \partial^\mu h  -V(h) \\
& - v \left[    \bar  Q \left  (\hat Y_{u }+ \sum_{n=1}^{\infty}   \hat Y_{u}^{(n)} 
 \left  ({h \over v} \right )^n \right )  U P_{+} r   +    \bar  Q \left  (\hat Y_{d}+ \sum_{n=1}^{\infty}   \hat Y_{d}^{(n)} 
 \left  ({h \over v} \right )^n \right )  U P_{-} r      \right. \\
 & \left. +    \bar  L \left  (\hat Y_{e }+ \sum_{n=1}^{\infty}   \hat Y_{e}^{(n)} 
 \left  ({h \over v} \right )^n \right )  U P_{-} \eta          + h.c.  \right].
\end{aligned}
\ee
Here, $F_U(h)$ and $V(h)$ are arbitrary functions of the Higgs field
\be
F_U(h) =    \sum_{n=1}^{\infty}   f_{U,n}  \left  ({h \over v} \right )^n \ \,, \quad
V(h) =  v^4 \sum_{n=2}^{\infty}   f_{V,n}  \left  ({h \over v} \right )^n   .
\ee
 Also, the left and right-chiral fields are defined by $Q=(u_L,d_L)^T,  L = (\nu_L, e_L)^T$,  
 $r=(u_R,d_R)^T$ and $\eta=(\nu_R, e_R)^T$ and $P_{\pm} = (1 \pm \tau_3)/2$, with $\boldsymbol{\tau}$ Pauli matrices. 
 $U$ is a matrix containing the Goldstone fields
 \begin{equation}
 U = \textrm{exp}(2i\Phi/v), \qquad \Phi = \phi^a \frac{\tau^a}{2} = \frac{1}{\sqrt{2}} 
\begin{pmatrix}
\frac{\phi^0}{\sqrt{2}} & \phi^+\\
\phi^- & {-\frac{\phi^0}{\sqrt{2}}}
\end{pmatrix},
\end{equation}
 and  the action of covariant derivative on $U$ is given by
\be
D_\mu U =\partial_\mu + ig W_\mu U - i g' B_\mu U \frac{\tau_3}{2}.
\ee
The implications of  \eqref{eq:higgsheft} are that the $h W W$ and $h Z Z$ couplings are decoupled from the $W$ and $Z$ masses.
They are parameterized by a free coupling $\kappa_V$,  with $\kappa_V = 1$ in the SM.
Similarly, the fermion masses and Yukawas are characterized by independent matrices, so that, as in the SMEFT, the Yukawa couplings are in general complex and non-diagonal. In this case, we have
\begin{equation}
M_f = v \hat{Y}_f, \qquad  y^f = \hat{Y}_f^{(1)}.
\end{equation}
The main difference between the HEFT and SMEFT is that in HEFT the leading operator (as well as subleading operators) contain an arbitrary number of Higgs fields, while in SMEFT insertions of $H^\dagger H$ are suppressed by $\Lambda^2$ \cite{Buchalla:2013rka}. Thus, in HEFT,  Yukawa couplings and fermion masses differ already at leading order. 

\subsection{Matching at the EW scale}

Integrating out the Higgs boson at tree level leads to several four-fermion 
scalar operators with CPV WCs, if the Yukawa couplings are assumed to be complex \cite{Hamoudou:2022tdn}. 
\be \label{eq:tree-higgsexchange}
\begin{aligned}
{[C^{S1,RR}_{dd}]}_{ijkl}(\mu_{ ew})&= \frac{1}{2m_h^2} y_{ij}^d y^d_{kl}\,, \quad 
{[C^{S1,RR}_{uu}]}_{ijkl}(\mu_{ ew}) = \frac{1}{2m_h^2} y _{ij}^u y_{kl}^u\,,\\
{[C^{S1,RR}_{ud}]}_{ijkl}(\mu_{ ew}) &= \frac{1}{ m_h^2} y_{ij}^u y_{kl}^d\,, \quad
{[C^{S,RR}_{ee}]}_{ijkl}(\mu_{ ew})  = \frac{1}{2m_h^2} y_{ij}^e y_{kl}^e\,, \\ 
%%%%%%%%%%%%%%%%%
{[C^{S,RR}_{ed}]}_{ijkl}(\mu_{ ew})&= \frac{1}{m_h^2} y_{ij}^e y^{d}_{kl}\,, \quad 
{[C^{S,RR}_{eu}]}_{ijkl}(\mu_{ ew})= \frac{1}{m_h^2} y_{ij}^e y^{u}_{kl}\,, \\ 
%%%%%%%%%
{[C^{S,RL}_{ed}]}_{ijkl}(\mu_{ ew})&= \frac{1}{m_h^2} y_{ij}^e y^{d*}_{lk}\,, \quad 
{[C^{S,RL}_{eu}]}_{ijkl}(\mu_{ ew})= \frac{1}{m_h^2} y_{ij}^e y^{u*}_{lk}\,, \\ 
%%%%%%%%% DF=2
{[C^{V1,LR}_{dd}]}_{ijkl}(\mu_{ ew})&= -\frac{1}{6m_h^2} y_{il}^d y^{d*}_{jk}\,, \quad 
{[C^{V8,LR}_{dd}]}_{ijkl}(\mu_{ ew})= -\frac{1}{m_h^2} y_{il}^d y^{d*}_{jk}\,, \\
%{[C^{S1,RR}_{uu}]}_{ijkl}(\mu_{\rm ew}) = \frac{1}{2m_h^2} y _{ij}^u y_{kl}^u\,,\\
{[C^{V1,LR}_{uu}]}_{ijkl}(\mu_{ ew})&= -\frac{1}{6m_h^2} y_{il}^u y^{u*}_{jk}\,, \quad 
{[C^{V8,LR}_{uu}]}_{ijkl}(\mu_{ ew})= -\frac{1}{m_h^2} y_{il}^u y^{u*}_{jk}\,, \\
%%%%%%%%%%
{[C^{V1,LR}_{uddu}]}_{ijkl}(\mu_{ ew})&= -\frac{1}{6m_h^2} y_{il}^u y^{d*}_{jk}\,, \quad 
{[C^{V8,LR}_{uddu}]}_{ijkl}(\mu_{ ew})= -\frac{1}{m_h^2} y_{il}^u y^{d*}_{jk}\,,\\
%%%%%%%%%%
{[C^{V,LR}_{ee}]}_{ijkl}(\mu_{ ew})&= -\frac{1}{2m_h^2} y_{il}^e y^{e*}_{jk}.
\end{aligned}
\ee
Here the matching scale $\mu_{ew}$ can be set to $m_h$. Since EDMs can only constrain the imaginary parts of the WCs,  
they probe combinations of couplings such as
\be
\quad {\rm Im}(y^f_{ij} y^{f'}_{kl}) = \textrm{Re} (y^f_{ij} )  \textrm{Im} (y^{f'}_{kl} ) + \textrm{Im} (y^{f}_{ij} )  \textrm{Re} (y^{f'}_{kl} )   \,, \quad f,f' = u,d, e\,.
\ee
In the SMEFT, the imaginary part of the Yukawa coupling can be non-zero only at dim-6, so that four-fermion operators become proportional to 
\be
\left. \quad {\rm Im}(y^f_{ij} y^{f'}_{kl}) \right|_{\rm SMEFT} =  \frac{m_f}{v}  \delta_{ij} \textrm{Im} (y^{f'}_{kl} ) +  \frac{m_{f^\prime}}{v} \delta_{k l} \textrm{Im} (y^{f}_{ij} )   + \mathcal O\left(\frac{v^4}{\Lambda^4}\right).
\ee
In this case, flavor-violating couplings only contribute to EDMs at dim-8 and can be neglected.
Both flavor-conserving and flavor-violating couplings contribute at leading order in the HEFT.
In the following, we analyze the flavor-conserving and flavor-violating couplings 
separately. 
Eq. \eqref{eq:tree-higgsexchange} is the initial condition for the renormalization group evolution. As discussed in Section \ref{sec:runmatch},
the renormalization group running of these operators below the EW scale
results in contributions to four-fermion operators, and, in particular in the case of operators with heavy quarks and leptons,
to quark and gluon dipole operators. All these contributions are included in the master formulae.

Beyond tree level, non-standard Yukawa couplings generate dipole contributions at one loop \cite{Hisano:2012cc,Chien:2015xha}.
These are proportional to the same combination of couplings in  Eq. \eqref{eq:tree-higgsexchange}
and provide a next-to-leading logarithmic correction, which we can neglect.

\begin{figure}
\hspace{0.8cm}
\includegraphics[width=0.9\textwidth]{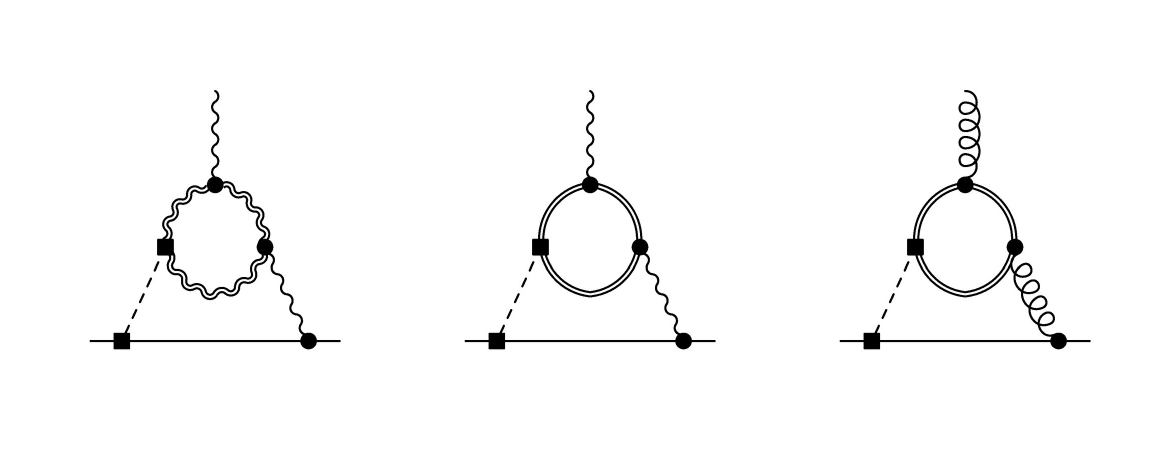}
\caption{Two loop Barr-Zee diagrams contributing to $\Delta F=0$ dipole operators. Here plain lines denote quarks and leptons lighter than the electroweak scale, double lines the top quark and double wiggly lines the $W$ boson. Insertions of the fermion Yukawa and of the $h W W$ couplings are denoted by squares, while gauge couplings are denoted by dots. }\label{fig:BZ}
\end{figure}
At two loops,  Barr-Zee diagrams \cite{Barr:1990vd} provide sizable contributions, 
and are sensitive to different combinations of Yukawa couplings. 
The diagrams are shown in Fig. \ref{fig:BZ}.
Integrating out the top-quark, Higgs boson, and $W$-boson at $\mu_{ew}$ leads to contributions to the quark and lepton dipole operators $C_{f \gamma}$
and to the quark chromo-dipole operator $C_{q G}$
\begin{align} \label{eq:BZ}
\textrm{Im}\left[C_{f \gamma}\right]_{i i}(\mu_{ ew}) &=  -12 e \frac{\alpha}{(4\pi)^3} q_f   q^2_{t} \frac{1}{m_{t}} \left[
f(x_{t}) \textrm{Re} (y^f_{i i})  \textrm{Im} (y^{u}_{33})
+ g(x_{t}) \textrm{Im}  (y^f_{i i})  \textrm{Re} (y^{u}_{33}) \right]  \nonumber \\
& + 2 e q_f \frac{\alpha}{(4\pi)^3} \frac{1}{v} \kappa_V (3 f(x_W) + 5 g (x_W)) \, \textrm{Im} (y^f _{ii})\,, \\
\textrm{Im}\left[C_{q G}\right]_{i i} (\mu_{ ew}) & =  2 g_s  \frac{\alpha_s}{(4\pi)^3}  \frac{1}{m_{t}} \left[
f(x_{t}) \textrm{Re}  (y^q_{i i})  \textrm{Im} (y^{u}_{33})
+ g(x_{t}) \textrm{Im} (y^q_{i i})  \textrm{Re} (y^{u}_{33}) \right],
\end{align}
where $x_i =  m_i^2/m_h^2$, and $q_i$ is the electric charge. The loop functions are given by
\begin{align}
f(z) = \frac{z}{2} \int_0^1 d x \frac{1 -2 x (1-x)}{ x (1-x) - z} \ln \frac{x (1-x)}{z}, \qquad g(z) = \frac{z}{2} \int_0^1 d x \frac{1}{x (1-x) -z} \ln \frac{x(1-x)}{z} .
\end{align}
In the SMEFT, $\kappa_V = 1$, and we can set the real part of the Yukawa couplings to its dim-4 value. In the HEFT, $\kappa_V$ is a free, real parameter, that needs to be extracted from data. 
In the HEFT, the loop and momentum expansions are tied by $\Lambda = 4\pi v$
so that the contributions in Eq. \eqref{eq:BZ} arise at NNLO. At this order, one would also expect contributions from
dipole operators in the HEFT  \cite{Buchalla:2013rka}. Since the Barr-Zee diagrams are finite, we can however consider them in isolation.

\subsection{Constraints on flavor-violating Higgs couplings}
First, we discuss the constraints on flavor-violating Higgs couplings. 
In the SMEFT interpretation, flavor-violating Higgs couplings do not contribute to EDM at dim-6. The bounds in this subsection are thus valid in the HEFT picture.
It is instructive to have linearized formulae of the EDMs directly in terms of Higgs 
couplings.  Since $d_{\rm Hg}$ and electron EDM impose strongest limits,  
we employ our EFT master formulae to  obtain expressions for  $d_{\rm Hg}$  and $\omega_{\rm HfF}$. 
For $d_{\rm Hg}$, we find
\be \label{eq:HgFV}
\begin{aligned} 
d_{\rm Hg} \over [{\rm ecm}]  & \supset
{(1.1 \pm 4.1) \cdot 10^{-24}   \over (2 \times 0.125^2)}  {\rm Im}(y^d_{12} y^d_{21}) 
+  {(-5.7 \pm 34) \cdot 10^{-24}   \over (2 \times 0.125^2)}  {\rm Im}(y^d_{13} y^d_{31})  \\
&+  {(-6.4 \pm 4.7) \cdot 10^{-28}   \over (2 \times 0.125^2)}     {\rm Im}(y^d_{23} y^d_{32})
+  {(-2.4 \pm 15.3) \cdot 10^{-24}   \over (2 \times 0.125^2)}     {\rm Im}(y^u_{12} y^u_{21}) \,,
\end {aligned} 
\ee
while, for $\omega_{\textrm{HfF}}$, we have
{
\be \label{eq:HfFFV}
{\omega_{\textrm{HfF}} \over [\textrm{mrad/s}] } \supset 
{(9.1 \pm 0.4) \cdot 10^{5}   \over (2 \times 0.125^2)}  {\rm Im}(y^e_{12} y^e_{21}) 
+ {(1.5 \pm 0.1) \cdot 10^{7}   \over (2 \times 0.125^2)}  {\rm Im}(y^e_{13} y^e_{31}).
\ee
}
%%%%%%%
Similarly, one can find expressions for the other EDM observables. Only a specific combination of the Higgs coupling can be probed with EDMs. 
To get an estimate of the upper bounds on these couplings one can compare 
 the above expressions with the experimental limits, e.g. for $\left|  d_{\rm Hg} \right| <  6.2 \times 10^{-30}$ $e$ cm and 
$|\omega_{\textrm{HfF}}|  < 0.17$ mrad/s.

\begin{figure}[htb]
\begin{center}
\includegraphics[width=0.7\textwidth]{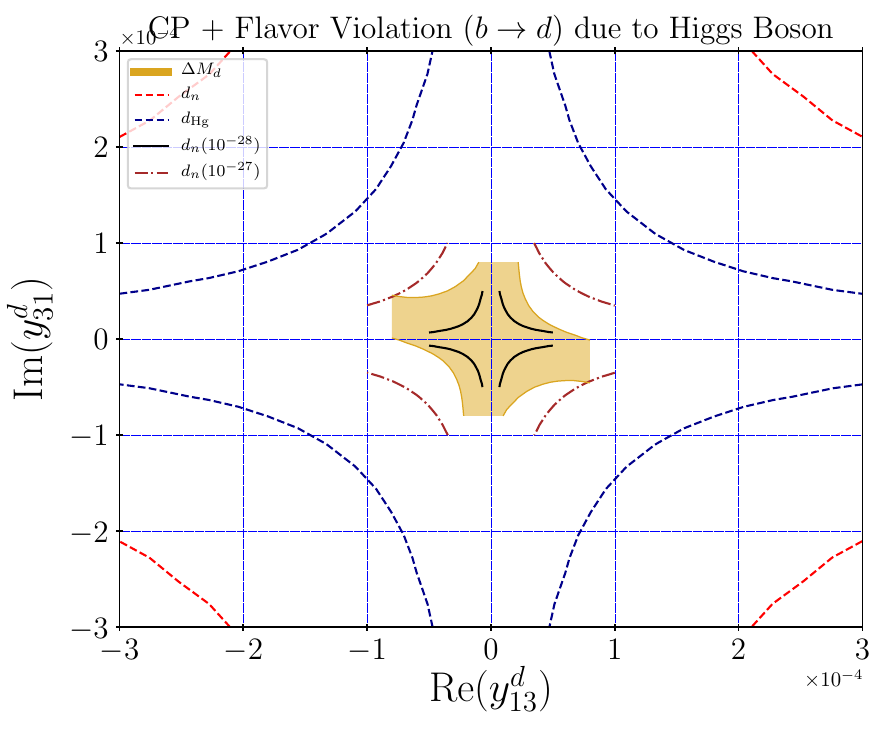}
\end{center}
%\end{minipage}
\caption{EDM and flavor constraints on the CP and flavor violating Higgs couplings. {Note that the black line, brown  and 
golden lines are truncated to increase the statistics.} }
\label{Fig:fvhiggs}
\end{figure}
 We also perform a combined fit of FV Higgs  couplings to all EDM observables ($d_n$, $d_{\rm Hg}$, $d_{\rm Xe}$, $d_{\rm Ra}$, $\omega_{\rm HfF}, \omega_{\rm YbF}, \omega_{\rm ThO}$) using 
 the current and future upper limits given in Tab.~\ref{tab:EDMexps}. 
 For this purpose, we assume only one combination of the couplings to be non-zero at a time. 
 The results are presented in Tab.~\ref{tab:FVHiggsCouplings} (column 2). 

In column 3, we indicate the origin of the EDMs in terms of low-energy WET operators 
at $\mu= 2$ GeV. At the EW scale, the scalar and vector operators are generated by integrating out the Higgs 
as shown in the matching relations \eqref{eq:tree-higgsexchange}. Some of these give a leading contribution to the EDMs through 
direct four-fermion matrix elements (see \eqref{eq:C1066}- \eqref{eq:C1133}) for the light flavors $f,f'= u,d,s$. 
1-loop QCD and QED effects through the operator mixing onto dipoles are also present for the light flavors. 
However, for the heavier generations, $c$ and $b$, 
the contribution to the EDMs arises solely due to operator mixing at 
1-loop level (see \eqref{eq:agamma-sol}-\eqref{eq:aG-sol}). Apart from that, as shown in column 3, the Weinberg operator also contributes 
for the case of heavier generations due to threshold corrections of the $b$ and $c$ dipoles onto it at 
the $m_b$ and $m_c$ scales (see Eqs. \eqref{eq:ggg1} and \eqref{eq:ggg2}). 

{In Fig.~\ref{Fig:fvhiggs} we show a comparison of the constraints from the EDMs and $\Delta M_d$
on the CP + flavor violating down quark Higgs couplings in the $bd$ sector. 
We see that current $d_n$ and $d_{\rm Hg}$ bounds are weaker than the constraints from $\Delta M_d$.
However, a neutron EDM bound at the level of $10^{-27}$ $e$ cm will be highly competitive with $B$-$\bar B$ oscillations.
For other sectors ($bs, sd$, and $uc$) the 
flavor bounds \cite{Blankenburg:2012ex} are found to be stronger.}

\begin{table}[htb!]
\begin{center}
 \renewcommand*{\arraystretch}{1.0}
 \resizebox{\textwidth}{!}{
\begin{tabular}{|c|cc|cc|}
\hline
\multicolumn{5}{|c|}{Generic FV Higgs couplings}  \\ \hline
couplings   & current limit& future limit  & \multicolumn{2}{|c|}{operators at hadronic scale}  \\
\hline
Im($y^d_{12} y_{21}^d$) & $ 1.2 \cdot 10^{-7}$ &  $ 7.0  \cdot 10^{-9}$  & $\wc[]{d\gamma(G)}{11, 22}, \wc[S1(8),RR]{dd}{1221}, \wc[S1(8),RR]{dd}{1122}$   & \\
%\hline
Im($y_{13}^d y_{31}^d$)   & $2.1 \cdot 10^{-8}$ &$ 6.0 \cdot 10^{-10}$  &  $\wc[]{\widetilde G}{}, \wc[]{d\gamma(G)}{11}$   & \\
%\hline
Im($y_{23}^d y_{32}^d$) & $ 1.5 \cdot 10^{-4}$& $1.4 \cdot  10^{-6}$  &  $\wc[]{\widetilde G}{}, \wc[]{d\gamma(G)}{22}$   & \\
%\hline
Im($y_{12}^u y_{21}^u$) & $ 5.0 \cdot 10^{-8}$& $2.3 \cdot 10^{-9}$  & $\wc[]{\widetilde G}{}, \wc[]{u\gamma(G)}{11}$   & \\
%\hline
Im($y_{12}^e y_{21}^e$) & $ 1.1 \cdot  10^{-9}$& $1.1 \cdot 10^{-10}$  & $\wc[]{e\gamma}{11}$   & \\
%\hline
Im($y_{13}^e y_{31}^e$) & $ 6.0 \cdot  10^{-11}$& $6.0 \cdot 10^{-12}$        & $\wc[]{e\gamma}{11}$   & \\
\hline
\end{tabular}}
\caption{Current and future EDM constraints on the combinations of the generic flavor 
violating Higgs couplings (column 2) {at $1\sigma$ level} . In the column 3, we also  
indicate the origin of EDM contributions at $\mu=2$ GeV. The matching scale is set to the Higgs mass.}
\label{tab:FVHiggsCouplings}
\end{center}
\end{table}

\subsection{EDM constraints on flavor-conserving Higgs couplings}

For the flavor-conserving Yukawas, we 
get contributions from the  tree-level matching onto four-fermion, given in Eq. \eqref{eq:tree-higgsexchange}, 
and from the 2-loop Barr-Zee diagrams, in Eq. \eqref{eq:BZ}.
Using these one can directly use the master formulae to obtain the EDMs in terms of the 
FC Yukawas, similar to the expressions in \eqref{eq:HgFV} and \eqref{eq:HfFFV}.  
%Together with the tree-level matching contributions presented in \eqref{eq:Hg-FCHiggs}-\eqref{eq:HfF-FCHiggs}, we have 
%now complete result for the EDMs in terms of the FC Higgs couplings that 
%can be used to obtain the estimate of constraints. 

We note some unique features of BZ contributions, which are in contrast to the tree-level matching, 
associated with the particular combination 
of the Higgs couplings appearing in the matching relations.   
First, the BZ diagrams  involve a 
top-quark Yukawa coupling which is correlated to the other quarks as 
well as leptonic Yukawas. 
Secondly,  the BZ effects can also correlate fermion Yukawas to the 
$W$-Higgs couplings. These features are absent in the tree-level matching and therefore 
allow additional constraints on the FC Higgs couplings.   Recall that, 
the EDMs through tree-level Higgs exchange \eqref{eq:tree-higgsexchange} probe following combination of FC couplings 
\be
\quad {\rm Im}(y^f_{ii} y^{f'}_{jj}) = \textrm{Re} (y^f_{ii} )  \textrm{Im} (y^{f'}_{jj} ) + \textrm{Im} (y^f_{jj} )  \textrm{Re} (y^{f'}_{ii} )   \,, \quad f, f' = u,d, e\,,
\ee
here $f$ can be equal to $f'$. Therefore, for $f=f'$ and $i=j$, one can also probe individual couplings (more precisely a combination of its Re and Im parts: $\textrm{Re}(y^f_{ii}) \textrm{Im}(y^f_{ii})$) in an uncorrelated manner to other couplings, under the assumption that only a single coupling is turned on.
Overall, the tree-level and 2-loop BZ diagram effects can test very different parts of the low-energy Higgs effective coupling parameter space.
In Tab.~\ref{tab:FCHiggsCouplings} we show upper bounds at $1\sigma$ level on the various combinations of the generic  FC Higgs couplings. 
%which needs to be compared to its experimental value $0.17$mrad/s.
\begin{table}[htb]
\begin{center}
 \renewcommand*{\arraystretch}{1.0}
 \resizebox{1.0\textwidth}{!}{
\begin{tabular}{ |c|cc|cc|}
\hline
\multicolumn{5}{|c|}{Generic FC Higgs couplings}  \\ \hline
couplings   & current limit & future limit  & \multicolumn{2}{|c|}{operators at hadronic scale}  \\
\hline
Im$(y_{11}^d y_{11}^d)$ & $ 2.4 \cdot 10^{-8}$&    $2.6 \cdot 10^{-9}$  &  & $\wc[S1(8),RR]{dd}{1111}, \wc[]{d\gamma(G)}{11}$  \\
Im$(y_{22}^d y_{22}^d)$ & $2.0 \cdot 10^{-3}$  &   $ 1.9 \cdot 10^{-5}$  & & $\wc[S1(8),RR]{dd}{2222},  \wc[]{d\gamma(G)}{22}$ \\
Im$(y_{33}^d y_{33}^d)$ &  $1.1 \cdot 10^{-3}$ &  $9.0\cdot 10^{-6}$  &  & $ \wc[]{\widetilde G}{}$  \\
%Im$(y_{11}^d y_{22}^d)$ & $(37\pm 27) \times 10^{-9}$  &  $(24\pm 12) \times 10^{-10}$    & & $\wc[S1(8),RR]{dd}{1122},  \wc[]{d\gamma(G)}{11,22}$    \\
%Im$(y_{11}^d y_{33}^d)$ &  $(31\pm 32) \times 10^{-9}$       &  $(21\pm 21) \times 10^{-10}$    & & $\wc[]{\widetilde G}{}, \wc[]{d\gamma(G)}{11}$  \\
%Im$(y_{22}^d y_{33}^d)$ & $(-1\pm 4) \times 10^{-4}$      &  $(6\pm 4) \times 10^{-6}$     & & $\wc[]{\widetilde G}{}, \wc[]{d\gamma(G)}{22}$  \\
\hline
%%%%%%%%%%%%
Im$(y_{11}^u y_{11}^u)$ & $2.6 \cdot 10^{-8}$  &  $1.6 \cdot 10^{-9}$    & & $\wc[S1(8),RR]{uu}{1111}, \wc[]{u\gamma(G)}{11}$  \\
Im$(y_{22}^u y_{22}^u)$ &  $4.0 \cdot 10^{-4}$ &   $3.5 \cdot 10^{-6}$   & & $ \wc[]{\widetilde G}{}$  \\
%%
%Im$(y_{11}^u y_{22}^u)$ & $(-15\pm 12) \times 10^{-9}$  &  $(4\pm 11) \times 10^{-10}$     & & $\wc[]{\widetilde G}{}, \wc[]{u\gamma(G)}{11}$  \\
\hline
%%%%%%%%
%Im$(y_{11}^u y_{11}^d)$ & $(-18\pm 17) \times 10^{-9}$  & $(6\pm 5) \times 10^{-10}$     & & $\wc[S1(8),RR]{ud}{1111},  \wc[]{u\gamma(G)}{11}, \wc[]{d\gamma(G)}{11}$ \\
%Im$(y_{11}^u y_{22}^d)$ & $(-10\pm 8) \times 10^{-9}$   & $(3\pm 6) \times 10^{-10}$     & &  $\wc[S1(8),RR]{ud}{1122},  \wc[]{u\gamma(G)}{11}, \wc[]{d\gamma(G)}{22} $  \\
%Im$(y_{11}^u y_{33}^d)$ & $(-16\pm 12) \times 10^{-9}$    & $(-9\pm 18) \times 10^{-10}$     & & $\wc[]{\widetilde G}{}, \wc[]{u\gamma(G)}{11}$ \\
%Im$(y_{22}^u y_{11}^d)$ & $(17\pm 13) \times 10^{-9}$  & $(4\pm 16) \times 10^{-10}$     & &  $\wc[]{\widetilde G}{}, \wc[]{d\gamma(G)}{11}$   \\
%Im$(y_{22}^u y_{22}^d)$ & $(-2\pm 8) \times 10^{-4}$    & $(9\pm 7) \times 10^{-6}$      & & $\wc[]{\widetilde G}{}, \wc[]{d\gamma(G)}{22}$   \\
%Im$(y_{22}^u y_{33}^d)$ &  --& --   & & $\wc[]{\widetilde G}{}$ \\
%\hline
Im$(y_{11}^e y_{11}^e)$ & $1.2  \cdot 10^{-7}$    & $1.2 \cdot 10^{-8}$      & & $\wc[S,RR]{ee}{1111}, \wc[]{e\gamma}{11}$   \\
\hline
\end{tabular}
}
\caption{Current and future EDM constraints on the combinations of the generic flavor 
conserving Higgs couplings {at $1\sigma$ level} (to obtain $90\%$CL these numbers have to be multiplied by factor 1.6 ). In column 3, we also  
indicate the origin of leading contributions to the EDMs at $\mu=2$GeV. It is worth mentioning that 
the EDMs can also probe other combinations of the Yukawa couplings which are not shown here.
{In that case, the scalar or vector operators with left-right chiralities are generated at the EW scale (see e.g. \eqref{eq:tree-higgsexchange}). The electron EDM dominates the constraints on the leptonic and semileptonic operators. The matching scale is set to the Higgs mass. }}
\label{tab:FCHiggsCouplings}
\end{center}
\end{table}

%\newpage

\subsection{LHC constraints on flavor-conserving Higgs couplings}\label{sec:HiggsLHC}
Finally, we want to compare the bounds in Table \ref{tab:FCHiggsCouplings} with complementary constraints from the LHC.
Non-standard Yukawa couplings affect Higgs production and decay processes and are thus constrained by measurements of the Higgs properties at the LHC  \cite{ATLAS:2022vkf,CMS:2022dwd} .
For a given Higgs production mechanism, $i \rightarrow H$, followed by the decay of the Higgs to the final state $f$,  
the signal strength in the presence of non-standard interactions is defined as 
\begin{equation}
\mu^{}_{i \rightarrow H \rightarrow f} = \mu^{}_i\, \mu^{}_f =   \left(1+ \frac{ \sigma^{}_{i \rightarrow H} }{\sigma^{\rm SM}_{i \rightarrow H}} \right) \frac{  1 + \frac{\Gamma^{}_{H \rightarrow f}}{\Gamma^{\rm SM}_{H \rightarrow f} }  }{
1 + \frac{\Gamma^{}_{\textrm{tot}}}{\Gamma^{\rm SM}_{\textrm{tot}}}
},
\end{equation}
where $\sigma^{\rm SM}$ and $\sigma^{}$ are, respectively, the production cross sections in the SM and the correction induced by non-standard interactions. 
$\Gamma^{\rm SM}_{H \rightarrow f}$ are the decay widths in the channel $f$ and $\Gamma^{\rm SM}_{\textrm{tot}}$
the Higgs total width.
In the SM, the Yukawas of the light $u$, $d$, and $s$ quarks are too small to significantly affect Higgs production and decay.
Direct constraints on $H \rightarrow c \bar c$ are sensitive to charm Yukawas about ten times larger than the SM expectation \cite{ATLAS:2022ers,CMS:2022fxs},
while the $b$ and $t$ Yukawas are compatible with the SM \cite{ATLAS:2022vkf,CMS:2022dwd}.
In the HEFT, the quark masses and Yukawas are decoupled from each other. Light quark Yukawas can thus affect the Higgs production via the partonic processes $\bar q q \rightarrow H$, which falls in the same category at gluon fusion, and $q \bar q^{(\prime)} \rightarrow H V$, with $V  = W,Z$. They also affect the decay signal strength by modifying the Higgs total width. 
The top Yukawa can be probed by the loop processes $g g\rightarrow H$, $H \rightarrow \gamma \gamma$, $H \rightarrow \gamma Z$,
and, at tree level, via $t \bar t H$ production.

At $\sqrt{S} = 13$ TeV, the gluon fusion SM cross section is \cite{LHCHiggsCrossSectionWorkingGroup:2016ypw}
\begin{equation*}
\sigma^{\rm SM}_{ggH} = 48.58^{+2.22}_{-3.27} \pm 1.56 \, {\rm pb}.
\end{equation*}
where the cross-section is evaluated at the factorization and renormalization scales $\mu_F = \mu_R  = m_h/2$.
The first error includes the contribution of missing orders and EW corrections, while the second error denotes the PDF uncertainties.

Non-standard Yukawas affect the production cross-section in several ways. 
A nonstandard Yukawa coupling of the top quark affects gluon fusion production. The real part of the coupling simply rescales the SM contribution.
In the large quark mass limit, the imaginary part of the top Yukawa coupling contributes to gluon fusion via
the operator $h G \tilde G$. Neglecting $\mathcal O(m_h^2/m_t^2)$ corrections and at leading order in QCD,  the contribution of the imaginary part 
of the top Yukawa to the gluon fusion cross section is a factor of $9/4$ larger than that of the real part.
Higher order QCD corrections are different for scalar and pseudoscalar couplings, and the expression of the difference at NLO and NNLO  can be found in Ref. \cite{Anastasiou:2002wq,Harlander:2002vv}.

The contributions of light quarks with scalar and pseudoscalar Yukawas at NNLO in QCD have been computed in Ref. \cite{Harlander:2003ai}.
In this case, the cross-section is proportional to the absolute value of the Yukawa coupling, up to corrections proportional to $m_q/m_h$, which we neglect.
We write the production cross-section as
\begin{align}
\sigma_{ggH} &= \left(({\rm Re}\, y^u_{33})^2 + \frac{9}{4}({\rm Im}\, y^u_{33})^2\right)\frac{1}{\left| (y^{u}_{33})_{\rm SM}\right|^2} \sigma^{\rm SM}_{ggH} 
+ ({\rm Im}\, y^{u}_{33})^2 \Delta \sigma_{ggH}  \nonumber \\
&
+ \sum_{i = 1}^3  \sigma^d_{ii}  \left| y^d_{ii} \right|^2 + \sum_{i = 1}^2  \sigma^u_{ii}  \left| y^u_{ii} \right|^2, 
\end{align}
with the Yukawa couplings evaluated at the scale $\mu = m_h$.
The results for $\sigma_{q \bar q}$ at $\sqrt{S} =13$ TeV are summarized in Tab.~\ref{tab:higgspro}. 
The central value of the cross-section is given at the scale  $\mu_F = \mu_R = m_h/2$. The scale uncertainties are obtained by varying $\mu= \mu_F = \mu_R$ between $m_h/4$ and $m_h$, while the pdf and $\alpha_s$ error was evaluated following the PDF4LHC21 recommendation \cite{PDF4LHCWorkingGroup:2022cjn}, using 100 replicas in the \texttt{PDF4LHC21$\_$mc$\_$pdfas} PDF set.

\begin{table}[H]
\center
\begin{tabular}{||c||ccccc ||}
\hline
 $\sigma$ ($10^{3}$ pb) &  $\sigma^{u}_{11}$  & $\sigma^{d}_{11}$  &  $\sigma^{d}_{22}$ &  $\sigma^{u}_{22}$ & $\sigma^{d}_{33}$ \\
  \hline
central         & $120.0  $               &   $83.3$               &  $31.7$               & $12.6$                &  $3.9$\\
scale           & $^{+2.7}_{-4.8}$    &  $^{+1.9}_{-3.3}$   & $^{+0.8}_{-1.1}$ & $^{+0.5}_{-0.5}$& $^{+0.2}_{-0.1}$\\
pdf               & $3.6$                     & $3.1$                      & $5.3$                  & $0.5$                  & $0.1$\\ 
\hline 
  $\mu_{ggH}$  & $[2.2, 2.8] \cdot 10^3$ & $[1.5,1.9] \cdot 10^3$ & $[5.1,8.2] \cdot 10^2$ & $[2.3,3.0] \cdot 10^2$ & $[72,91]$  \\
  \hline \hline
$\sigma_{HV}$ (pb)  &  $\sigma^u_{11}$  & $\sigma^d_{11}$  &  $\sigma^{d}_{22}$ &  $\sigma^{u}_{22}$ & $\sigma^{d}_{33}$  \\
  \hline
$\sigma_{HW+}$  & 16.0  &  15.3  & 1.99 & 1.44  &0.09 \\
$\sigma_{HW-}$  &  8.35  &  8.50 &  1.53 & 1.69  & 0.06 \\
$\sigma_{HZ}$  &   3.18 &  2.24 & 0.50  & 0.19 & 0.09  \\
\hline 
$\mu_{HW}$  & 158  &  154  & 22.8 & 20.3  &0.97 \\
$\mu_{HZ}$  &   106 &  75 & 17  & 6.37 & 3.02  \\
\hline \hline
\end{tabular}
\caption{Contribution of non-standard light quark Yukawa couplings to the Higgs production cross section in the gluon fusion and HV channels. }\label{tab:higgspro}
\end{table}

The corrections to associated $HW$ and $HZ$ production induced by non-standard Yukawas were computed at NLO in QCD in Ref. \cite{Alioli:2018ljm}
\begin{equation}
\sigma_{HV} =  \kappa_V \sigma^{\rm SM}_{HV}  +   \sum_{i = 1}^3  \sigma^d_{ii,\, HV}  \left| y^d_{ii} \right|^2  +   \sum_{i = 1}^2  \sigma^u_{ii,\, HV}  \left| y^u_{ii} \right|^2,
\end{equation}
with \cite{LHCHiggsCrossSectionWorkingGroup:2016ypw}
\begin{align}
\sigma^{SM}_{HW^+(\ell^+ \nu)} & =  94.26^{+ 0.5\%}_{-0.7\%} \pm 1.8\% \, {\rm fb}, \qquad 
\sigma^{SM}_{HW^-(\ell^- \bar{\nu})}  =  59.83^{+ 0.4\%}_{-0.7\%} \pm 2.0\% \, {\rm fb}, \nonumber \\
\sigma^{SM}_{HZ^-(\ell^+ \ell^-)} & =  29.82^{+ 3.8\%}_{-3.1\%} \pm 1.6\% \, {\rm fb}.
\end{align}
The theoretical uncertainties on the non-standard contributions to 
$\sigma_{HW+}$, $\sigma_{HW-}$ and $\sigma_{HZ}$ were shown to be 
at the 10\% level \cite{Alioli:2018ljm}. Since the HV channel gives weaker bounds and the theoretical uncertainties are  
small, we did not re-evaluate them in this work.
For the corrections to the $t \bar t H$ cross-section, we use 
\begin{equation}
\mu_{t\bar t h} = \frac{ \left(\textrm{Re}\, y^u_{33}\right)^2  + 0.4  \left(\textrm{Im}\,  y^u_{33}\right)^2  }{\left| (y^{u}_{33})_{\rm SM}\right|^2}.
\end{equation}

%%%
The Yukawa also affect the Higgs decay into fermions. At NLO in QCD, the decay of a Higgs boson to two quarks is given by  \cite{Spira:2016ztx}
\begin{equation}
\Gamma(H \rightarrow q_j \bar{q}_j) = \frac{3  m_h }{8 \pi} |y^q_{jj}|^2 \left(1 + 5.67 \frac{\alpha_s}{\pi}\right)  = (18.0 \; {\rm GeV}) |y^q_{jj}|^2.
\end{equation}
The decay into two leptons has a similar expression, without the $\mathcal O(\alpha_s)$ correction.
For illustration, we use the signal strength measurements of the ATLAS experiment \cite{ATLAS:2022vkf}, set $\kappa_V = 1$ and $\textrm{Re}\, y^u_{33}$ to its SM value. The latter assumption is justified by analyses of the CP properties of the couplings of the Higgs boson to the top, which prefer a dominantly real coupling
\cite{CMS:2020cga,ATLAS:2020ior}.
With these assumptions, a combined fit to the Yukawas
of the electron and of the  $u$, $d$, $s$, $c$ and $b$ quarks
to ATLAS data yields
\begin{align}\label{eq:HiggsBounds}
| y^e_{11} | &< 1.2 \cdot 10^{-2}, \qquad
| y^u_{11} | < 6.4 \cdot 10^{-3}, \qquad  | y^d_{11} | < 6.9 \cdot 10^{-3}, \nonumber\\
\quad | y^d_{22} | &< 7.1 \cdot 10^{-3}, \qquad
| y^u_{22} | < 7.1 \cdot 10^{-3},
\end{align}
at the 90\% confidence level, while, for the $b$ quark, we have 
\begin{align}\label{eq:HiggsBoundsB}
0.90 \cdot 10^{-2}  < |y^d_{33}| <  1.2 \cdot 10^{-2}.
\end{align}
The bounds are given on couplings at the renormalization scale $\mu = m_h$.
For the $u$ and $d$ quarks, corrections to the production cross section and Higgs total width provide similar effects,
while for $s$, $c$ and $b$ quarks the dominant effect are the corrections to the Higgs partial and total widths.
The data we use  are only sensitive to the absolute values of the couplings.

We next perform a joint fit to the Higgs and EDM data in the HEFT framework, in which fermion masses and Yukawa are independent. 
In our fit, we fix $\kappa_V$ and the real part of the top Yukawa coupling,
$\textrm{Re} y^u_{33}$, to their SM model values. We then perform a simultaneous fit to the real parts of the $e$, $u$, $d$, $s$, $c$ and $b$ Yukawas
and to one imaginary part at a time. We neglect theoretical errors on EDMs, so that the bounds give an indication of the full potential of EDMs experiments, but should be taken with some caution.
If we include both the Barr-Zee and four-fermion contributions, at the 90\% CL we find
\begin{align}
|\textrm{Im} y^e_{11}| < 1.2 \cdot 10^{-6}, \qquad |\textrm{Im} y^u_{11}| &< 1.2 \cdot 10^{-6}, \qquad 
|\textrm{Im} y^d_{11}| < 1 .7 \cdot 10^{-7},  \nonumber \\
|\textrm{Im} y^d_{22}| < 7.1 \cdot 10^{-3}, \qquad |\textrm{Im} y^u_{22}| &< 7.1 \cdot 10^{-3}, \qquad 
|\textrm{Im} y^d_{33}| < 1.2 \cdot 10^{-2}.
\end{align}
The 90\% CL for the real part of the couplings corresponds to the ranges given in Eqs. \eqref{eq:HiggsBounds} and
\eqref{eq:HiggsBoundsB}. We see that for the $e$,  $u$ and $d$  Yukawas, the EDM bound on the imaginary part is much stronger than the 
bounds from Higgs observables in Eq. \eqref{eq:HiggsBounds}.
For the $s$, $c$ and $b$ quark, however, the bounds go back to what we obtained from the signal strengths. 
This is partially due to cancellations between the Barr-Zee (Eq. \eqref{eq:BZ}) and four-fermion
(Eq. \eqref{eq:tree-higgsexchange}) contributions to EDMs, which can happen once we let the real part of the Yukawas deviate from its SM value.

\subsection{SMEFT Interpretation}

The Yukawa couplings in SMEFT are more constrained as 
compared to the HEFT. Since the leading non-standard 
effects to the Yukawa couplings are dim-6 \eqref{eq:higgssmeft}, we can not have the double 
insertion of the Yukawas at this level. Hence, the contributions to the EDMs through flavor-violating Higgs couplings in 
Tab.~\ref{tab:FVHiggsCouplings} is essentially 
a dim-8 effect in the SMEFT power counting,  
because for the matching of $\wc[S1,RR]{ff}{iijj}$  a double insertion of 
flavor-violating couplings is required. 
On the other hand, flavor-conserving CPV Yukawa couplings generate EDMs at dim-6 in SMEFT. In this case, one of the 
Yukawa has to be SM-like. Moreover, since the SM Yukawas are restricted to be real in the mass basis, one can directly constrain 
the individual CPV  couplings from EDM observables unlike the generic case presented in Tab.~\ref{tab:FCHiggsCouplings}.

Limiting to dim-6 effects in the SMEFT, the matching onto 4f WET operators can be obtained by setting  
one of the Yukawas in \eqref{eq:tree-higgsexchange} to its SM value 
\be
y^d_{ii, \textrm{SM}}= {m_{d_i} \over v}\,, \quad y^u_{ii, \textrm{SM}}= { m_{u_i} \over v}\,, 
\quad y^e_{ii, \textrm{SM}}=  {m_{e_i} \over v}.
\ee
Moreover, in the BZ contributions \eqref{eq:BZ}  we also set $\kappa_V=1$.
The current limits on the CPV FC Higgs couplings are
 given in Tab.~\ref{tab:smeft-higgs}. For each case, we present two sets of results using the 
 tree-level (4f) and 2-loop (BZ) matching.  Also, in this case, we neglect theoretical errors. 
 Once they are under better control, theoretical errors can straightforwardly be added to the analysis.

\begin{table}[H]
\begin{center}
 \renewcommand*{\arraystretch}{1.2}
 \resizebox{1\textwidth}{!}{
\begin{tabular}{ |c|cccccccc|}
\hline
\multicolumn{9}{|c|}{CPV FC Higgs couplings in SMEFT at {$1\sigma$ level}}  \\ \hline
\multicolumn{9}{|c|}{$d_n$, Hg, Xe, Ra}  \\ \hline
matching  &Im($y^d_{11}$) & Im($y^d_{22}$)  & Im($y^d_{33}$)   &    Im($y^u_{11}$)  &  Im($y^u_{22}$)  &  Im$(y^e_{11})$  &    Im$(y^e_{22})$    &  Im($y^e_{33})$   \\
 %%%%%%%%%
 4f   & $ 1.4 \cdot 10^{-5}$   & $ 2.6\cdot 10^{-3}$      & $2.6 \cdot 10^{-3}$   & $6.0  \cdot 10^{-6}$   & $4.0\cdot 10^{-3}$       & -- &--& --  \\
 BZ   &  $ 9.0 \cdot 10^{-8}$    & $ 7.0  \cdot 10^{-4}$     & $ 9.0 \cdot 10^{-3}$  & $ 1.1 \cdot 10^{-7}$   &  $1.3 \cdot 10^{-3}$ & -- &--&--  \\
 \hline
 %%%%%%%%%%%%%%%%%%%%%%%%%%%
 \multicolumn{9}{|c|}{ $\omega_{\rm HfF}$, $\omega_{\rm YbF}$,  $\omega_{\rm ThO}$}   \\ \hline
matching  &Im($y^d_{11}$) & Im($y^d_{22}$)  & Im($y^d_{33}$)   &    Im($y^u_{11}$)  &  Im($y^u_{22}$) &  Im$(y^e_{11})$  &    Im$(y^e_{22})$    &  Im($y^e_{33})$   \\
 %%%%%%%%%%%%
 4f   & $4.0\cdot 10^{-3}$   & $ 4.0 \cdot 10^{-3}$   & $ 2.0  \cdot 10^{-3}$    & $ 9.0  \cdot 10^{-4}$   & $ 8.0 \cdot 10^{-4}$ &  $ 2.6\cdot 10^{-8}$   & $ 1.6 \cdot 10^{-3}$   &   $ 1.0 \cdot 10^{-4}$   \\
 BZ   & --   &  --   & --    & --   & -- &  $1.5\cdot 10^{-9}$   &  --    &    --   \\
 %%%%%%%%%%%%%%%%%%%
 \hline
  \multicolumn{9}{|c|}{combined}  \\ \hline
matching  &Im($y^d_{11}$) & Im($y^d_{22}$)  & Im($y^d_{33}$)   &    Im($y^u_{11}$)  &  Im($y^u_{22}$) & Im$(y^e_{11})$  &  Im$(y^e_{22})$    &  Im($y^e_{33})$  \\
 %%%%%%%%%%%%
 4f +BZ  & $9.0\cdot 10^{-8}$   & $ 7.0 \cdot 10^{-4}$   & $ 1.8  \cdot 10^{-3}$    & $ 1.0  \cdot 10^{-7}$   & $ 7.0 \cdot 10^{-4}$  & $1.5 \cdot 10^{-9}$  &  $ 1.6 \cdot 10^{-3}$  &  $ 1.0 \cdot 10^{-4}$  \\
 \hline
 
\end{tabular}
}
\caption{The current limits on the CPV FC Yukawa couplings resulting due to tree-level and 2-loop Barr-Zee 
diagrams up to dim-6 level in SMEFT. The neutron, Hg, Xe, and Ra EDMs and 
the frequencies  $\omega_{\rm HfF}$, $\omega_{\rm YbF}$, $ \omega_{\rm ThO}$ are included in the fit. The real parts of the 
Yukawas are set to their SM values.  The matching scale is set to the Higgs mass.}
\label{tab:smeft-higgs}
\end{center}
\end{table}

The imaginary part of the  Yukawa couplings of the $u$ and $d$ quarks is predominantly constrained by the neutron and Hg EDMs, via the contributions arising from Barr-Zee diagrams. The constraints are very strong, and out of the reach of existing colliders.
The $s$ quark is also mostly constrained by hadronic EDMs. In this case, the bound is a factor of ten stronger than the constraints from Higgs observables discussed in Sec. \ref{sec:HiggsLHC}, so that, once theoretical errors are considered, the LHC and EDM experiments can probe similar scales. The $b$-quark Yukawa is mostly constrained by the electron EDM. These contributions arise from  first integrating out the Higgs and matching onto the scalar four-fermion operator
$\wc[S, RR]{e d}{}$, which consequently runs into $\wc[T, RR]{e d}{}$ and  $C_{e\gamma}$. Finally, the $c$-quark Yukawa receive similar constraints from $d_n$, $d_{\rm Hg}$ and $\omega_{\rm HfF}$. For the $c$ and $b$ Yukawas, the indirect bounds on the Higgs couplings are complementary to the constraints resulting from the direct 
Higgs searches as presented in the previous section.

The bounds in Table \ref{tab:smeft-higgs} are of the same order of magnitude as those presented in Ref. \cite{Chien:2015xha}. 
The main difference is in the treatment of the RGE between the EW and $\sim$ 2 GeV scale. Ref. \cite{Chien:2015xha} ignored the role of four-fermion operators,
which re-appeared as unresummed large logarithms of $m_q/m_h$ in the matching coefficients. Our analysis is more similar to Ref.
\cite{Brod:2023wsh}, which, for the $b$ and $c$ quarks, calculated the matching coefficients and the mixing between $C_{e\gamma}$,   
$\wc[S, RR]{e q}{}$ and $\wc[T, RR]{e q}{}$ at one order higher than in our study. The bounds 
on $y^{d}_{33}$ and $y^{u}_{22}$ obtained in Ref. \cite{Brod:2023wsh} are very close to those presented in Table \ref{tab:smeft-higgs}.

\section{Summary and Outlook}\label{sec:concl}
\label{sec:7}

The inability of the SM to explain the origin of the baryon asymmetry in the Universe provides a strong motivation to search for new sources of CP violation beyond the phase of the CKM quark mixing matrix. 
Electric dipole moments of leptons, nucleons, atoms, and molecules are extremely sensitive probes of flavor-diagonal CP violation. Upcoming experiments have the potential to improve the bounds on the neutron EDM and on EDMs of diamagnetic atoms by at least one order of magnitude, while molecular experiments will further constrain the electron EDM and start to probe hadronic EDMs. 
As they could offer the first hints of BSM physics, it is important to develop robust tools for 
the model-independent interpretation of EDM experiments. 
In addition, identifying the detailed features of BSM physics from one or multiple observations in EDM experiments requires understanding their correlations with observables in other systems and at other energy scales.
Effective Field Theories provide a natural framework to achieve these goals in a controlled and systematic way, with minimal reliance on specific BSM models.
In this paper, we discuss EDMs in the Weak EFT and leverage EFT tools such as renormalization group evolution and 
matching at the heavy fermion thresholds to derive master formulae for EDMs in terms of the complete set of $\Delta F=0$ WET operators at the electroweak scale. 
This set includes both operators with light $u$, $d$, and $s$ quarks, which have nonvanishing nucleon matrix elements and thus contribute to EDM at ``tree level'',
but also operators with heavy $b$ and $c$ quarks and leptons of the second and third generations. 
Indeed we find that the great majority of CP-violating $\Delta F=0$ operators with heavy fermions mix into tree-level operators at leading or next to leading-log.
EDM constraints are so strong that even in the case of operators that contribute at next-to-leading log (meaning two or even more loops)
the operator scale has to be larger than 1 TeV.
We now briefly summarize the main results of this paper.
\begin{itemize}
\item The main result of this paper is the master formulae for the HfF, ThO, and YbF precession frequencies, given in Tables
\ref{tab:masterHfF} and \ref{tab:masterThOYbF}, and for the neutron, proton, and Hg EDMs, in Tables \ref{tab:masterdHg1},
\ref{tab:masterdHg2}, \ref{tab:masterdn} and \ref{tab:masterdp}.
These formulae can be readily applied to EFTs at the electroweak scale (SMEFT or HEFT) or to BSM models
by calculating the WET matching coefficients at the electroweak scale. 
\end{itemize}
The coefficients $\alpha_I^X(\mu_{ew})$ provided in these tables encode both short-distance effects, from the running between 
the electroweak scale $\mu_{ew}$ and the low energy scales $\mu_{low} = 2$ GeV (or the heavy fermion thresholds),
and long-distance effects, from hadronic and nuclear matrix elements. Concerning the short-distance effects:
\begin{itemize}
\item 
 We consider the full WET 1-loop anomalous dimension, 
which is sufficient to capture the LL contributions from scalar four-quark operators  with two heavy and two light quarks ( 
$\wc[S1(8), RR]{u u}{}$, $\wc[S1(8), RR]{u d}{}$, and  $\wc[S1(8), RR]{uddu}{}$),
 scalar leptonic operators with two heavy and two light leptons ($\wc[S, RR]{ee}{}$)
and scalar and tensor semileptonic operators with light quarks and heavy leptons
or light leptons and heavy quarks ($\wc[T, RR]{eu(d)}{}$ and $\wc[S, RR]{eu(d)}{}$).
The 1-loop RGE and the 1-loop matching of heavy quark QCD dipoles onto the Weinberg three-gluon operator
provide NLL contributions from four-quark scalar operators with four heavy quarks, and semileptonic scalar and tensor operators with heavy quarks and heavy leptons.
\item The vector operators $\wc[V1(8), LR]{uu}{}$,  $\wc[V1(8), LR]{dd}{}$, $\wc[V1(8), LR]{uddu}{}$
and $\wc[V, LR]{ee}{}$, with two or four heavy fermions, do not generate LL contributions. 
For these operators, we identify NLL contributions by generalizing 2-loop anomalous dimensions originally derived 
for $B \rightarrow X_s \gamma$ \cite{Cho:1993zb}, and by considering the 1-loop matching at the heavy fermion threshold.
After 2-loop running and 1-loop matching onto the $b$ and $c$ QCD dipole, 
$\wc[V1(8), LR]{uddu}{2332}$ generates a NNLL contribution to EDMs.
\end{itemize}
For the neutron, proton and diamagnetic atoms, the long distance piece of  $\alpha_I^X(\mu_{ew})$ is affected by large hadronic and nuclear uncertainties.
Our master formulae use up-to-date available informations from Lattice QCD, QCD sum rules and chiral perturbation theory,
and the implementation in the package \texttt{flavio} allows for the prompt updating of long-distance matrix elements, as soon as they become available.

%{\color{blue} Mention here that in the future the theory errors can be added to the fits?}

As an example, we have applied the master formulae to study the constraints from EDMs on Higgs couplings.
For flavor-violating Higgs couplings, we showed that, at the moment, $\Delta F=2$ observables provide stronger bounds.
Future EDM experiments, however, will provide competitive constraints, especially in the $bd$ sector.
In the case of flavor conserving Yukawa couplings, we updated bounds on the imaginary part in two scenarios.
In the HEFT scenario, in which the real part of the Yukawa is also a free parameter, we found that EDMs give the strongest bounds
on the imaginary part of the $e$, $u$, and $d$  Yukawas, while Higgs observables dominate the bounds on the $s$, $c$ and $b$ Yukawas.
In the SMEFT scenario, in which, up to dimension-eight corrections, the real part of the Yukawa is fixed to its SM value,
EDMs give by far the strongest bounds.
In particular, the electron EDM gives the strongest bounds on the $b$ and $s$ Yukawas,
neutron and mercury dominate the $u$ and $d$ bounds, and for the charm quark both contributions are important.

This work can be extended in several directions.
First of all, we have in this paper relied on a logarithmic counting to organize the contributions to EDMs arising from the
solution of the WET RGEs at one and two loops. This counting does not take into account the presence of small couplings,
such as the QED coupling, for which the suppression from additional loops is not fully offset by the large logarithms.
It is thus possible that higher loop corrections to matching and running, while formally subleading,
will be numerically more important that the LL and NLL
terms identified here, especially in those cases in which these effectively arise at three loops
(e.g. $\wc[V1(8), LR]{uddu}{2332}$ or $\wc[S1(8), RR]{ud}{2233}$).
Especially for applications to very sensitive observables as EDMs, it is thus important to extend the calculations of WET anomalous dimensions and threshold corrections to higher order.

Secondly, higher order corrections are necessary to constrain the few $\Delta F =0$ CP-violating operators that are not listed in Table \ref{tab:wet4}.
These include  $\wc[S,RL]{eu}{11 22}$, $\wc[S,RL]{eu}{11 33}$
and all the entries with heavy leptons, for which it should be possible to generalize the 2-loop anomalous dimension for vector operators,
and  $\wc[S,RR]{ee}{ii jj}$ with $i, j > 1$.
In addition, we find very weak constraints on $\mu$-charm and $\tau$-charm semileptonic operators
($\wc[T (S), RR]{eu}{2222}$, $\wc[T (S), RR]{eu}{3322}$) and 
$\mu$-bottom and $\tau$-bottom semileptonic operators ($\wc[T (S), RR]{ed}{2233}$ and $\wc[T (S), RR]{ed}{3333}$).
For these classes, it will be important to find additional paths.

Finally, reducing the theoretical uncertainties is crucial to the success of the EDM program. This requires, in particular, to address the large hadronic and nuclear uncertainties that affect the master formulae provided in this paper.
In future the theoretical uncertaintities can be included in the fits using {\tt smelli} program \cite{Aebischer:2018iyb}.

%\newpage
\section*{Acknowledgements}
We acknowledge several useful discussions with W. Dekens and M. Misiak.
JK and EM are supported by the U.S. Department of Energy through the Los Alamos National Laboratory and by the Laboratory Directed Research and Development program of Los Alamos National Laboratory under project numbers 20220706PRD1 and  20240078DR. Los Alamos National Laboratory is operated by Triad National Security, LLC, for the National Nuclear Security Administration of U.S. Department of Energy (Contract No. 89233218CNA000001).
We acknowledge support from the DOE Topical Collaboration ``Nuclear Theory for New Physics,'' award No.\ DE-SC0023663.

\newpage
\appendix

\section{Additional Formulae}
\label{app:additional}
In this section, we provide some additional master formulae for the $\omega_{\rm ThO}$, $\omega_{\rm YbF}$ (\ref{tab:masterThOYbF}) and proton EDM (\ref{tab:masterdp}).
%%%%%%%%%
\begin{table}[H]
\begin{center}
 \renewcommand*{\arraystretch}{1.0}
 \resizebox{0.8\textwidth}{!}{
 {
\begin{tabular}{ |cc|cc|}
\hline
\hline
\multicolumn{4}{|c|}{{\bf $\alpha_I^{\rm ThO} (\mu_{ew})$ for the $\omega_{\text{ThO}}$}}  \\
\hline
\hline
\multicolumn{2}{|c|}{$n_f=3+1$}     & \multicolumn{2}{|c|}{$n_f=5+3$}  \\ \hline
%%%%%%%%%%

$[C_{e\gamma}]_{11}$&$(9.9\pm2.1) \cdot 10^{12}$&$[C^{S,RR}_{eu}]_{1122}$&$(-3.2\pm0.6) \cdot 10^{6}$\\
$[C^{S,RR}_{ee}]_{1111}$&$(9.4\pm2.1) \cdot 10^{4}$&$[C^{T,RR}_{eu}]_{1122}$&$(1.3\pm0.3) \cdot 10^{9}$\\
$[C^{S,RR}_{eu}]_{1111}$&$(1.5\pm0.3) \cdot 10^{9}$&$[C^{S,RR}_{ed}]_{1133}$&$(-1.3\pm0.2) \cdot 10^{6}$\\
$[C^{S,RR}_{ed}]_{1111}$&$(1.5\pm0.3) \cdot 10^{9}$&$[C^{T,RR}_{ed}]_{1133}$&$(-1.7\pm0.3) \cdot 10^{9}$\\
$[C^{S,RR}_{ed}]_{1122}$&$(1.1\pm0.2) \cdot 10^{8}$&$[C^{S,RR}_{ee}]_{1122}$&$(-2.0\pm0.4) \cdot 10^{5}$\\
$[C^{S,RL}_{eu}]_{1111}$&$(1.5\pm0.3) \cdot 10^{9}$&$[C^{S,RR}_{ee}]_{1133}$&$(-4.4\pm0.7) \cdot 10^{6}$\\
$[C^{S,RL}_{ed}]_{1111}$&$(1.5\pm0.3) \cdot 10^{9}$&$[C^{S,RR}_{ee}]_{1221}$&$(9.9\pm2.0) \cdot 10^{6}$\\
$[C^{S,RL}_{ed}]_{1122}$&$(1.1\pm0.3) \cdot 10^{8}$&$[C^{S,RR}_{ee}]_{1331}$&$(1.9\pm0.4) \cdot 10^{8}$\\
$[C^{T,RR}_{eu}]_{1111}$&$(-8.7\pm1.6) \cdot 10^{8}$&$[C^{V,LR}_{ee}]_{1221}$&$(-3.2\pm0.6) \cdot 10^{6}$\\
$[C^{T,RR}_{ed}]_{1111}$&$(4.4\pm0.8) \cdot 10^{8}$&$[C^{V,LR}_{ee}]_{1331}$&$(-5.1\pm1.0) \cdot 10^{7}$\\
$[C^{T,RR}_{ed}]_{1122}$&$(3.1\pm0.5) \cdot 10^{8}$&~&-\\

%%%%%%%%%%%%%%%%%%%%
\hline
\hline
\multicolumn{4}{|c|}{{\bf $\alpha_I^{\rm YbF} (\mu_{ew})$ for the $\omega_{\text{YbF}}$}}  \\
\hline
\hline
\multicolumn{2}{|c|}{$n_f=3+1$}     & \multicolumn{2}{|c|}{$n_f=5+3$}  \\ \hline
%%%%%%%%%%

$[C_{e\gamma}]_{11}$&$(1.6\pm0.1) \cdot 10^{12}$&$[C^{S,RR}_{eu}]_{1122}$&$(-5.2\pm0.4) \cdot 10^{5}$\\
$[C^{S,RR}_{ee}]_{1111}$&$(1.5\pm0.1) \cdot 10^{4}$&$[C^{T,RR}_{eu}]_{1122}$&$(2.1\pm0.2) \cdot 10^{8}$\\
$[C^{S,RR}_{eu}]_{1111}$&$(1.4\pm0.2) \cdot 10^{8}$&$[C^{S,RR}_{ed}]_{1133}$&$(-2.1\pm0.2) \cdot 10^{5}$\\
$[C^{S,RR}_{ed}]_{1111}$&$(1.4\pm0.2) \cdot 10^{8}$&$[C^{T,RR}_{ed}]_{1133}$&$(-2.7\pm0.2) \cdot 10^{8}$\\
$[C^{S,RR}_{ed}]_{1122}$&$(1.1\pm0.1) \cdot 10^{7}$&$[C^{S,RR}_{ee}]_{1122}$&$(-3.3\pm0.2) \cdot 10^{4}$\\
$[C^{S,RL}_{eu}]_{1111}$&$(1.4\pm0.2) \cdot 10^{8}$&$[C^{S,RR}_{ee}]_{1133}$&$(-7.1\pm0.5) \cdot 10^{5}$\\
$[C^{S,RL}_{ed}]_{1111}$&$(1.4\pm0.2) \cdot 10^{8}$&$[C^{S,RR}_{ee}]_{1221}$&$(1.6\pm0.1) \cdot 10^{6}$\\
$[C^{S,RL}_{ed}]_{1122}$&$(1.1\pm0.1) \cdot 10^{7}$&$[C^{S,RR}_{ee}]_{1331}$&$(3.1\pm0.2) \cdot 10^{7}$\\
$[C^{T,RR}_{eu}]_{1111}$&$(-1.3\pm0.1) \cdot 10^{8}$&$[C^{V,LR}_{ee}]_{1221}$&$(-5.1\pm0.4) \cdot 10^{5}$\\
$[C^{T,RR}_{ed}]_{1111}$&$(6.5\pm0.5) \cdot 10^{7}$&$[C^{V,LR}_{ee}]_{1331}$&$(-8.3\pm0.7) \cdot 10^{6}$\\
$[C^{T,RR}_{ed}]_{1122}$&$(5.1\pm0.4) \cdot 10^{7}$&~&-\\

%%%%%%%%%%%%%%%%%%%%
\hline

\end{tabular}}
}
\caption{Master Formulae for $\omega_{\rm ThO}$ and  $\omega_{\rm YbF}$  in WET in the JMS basis at the EW scale.
The $\alpha_I^{\rm ThO}$ and $\alpha_I^{\rm YbF}$ have units of (mrad/s)$\rm TeV^{2}$ and
(mrad/s)$\rm TeV$ for the 4f and dipole operators, respectively.
The $90\%$ CL upper bound on $\omega_{\rm ThO}$ and $\omega_{\rm YbF}$ are $1.3$mrad/s and 23.5mrad/s, respectively.}
\label{tab:masterThOYbF}
\end{center}
\end{table}
%%%%%%%%%

\begin{table}[H]
\begin{center}
 \renewcommand*{\arraystretch}{1.0}
 \resizebox{\textwidth}{!}{
\begin{tabular}{ |cc|cc|cc|}
\hline
\hline
\multicolumn{6}{|c|}{{\bf $\alpha_I^{p} (\mu_{ew})$ for proton EDM}}  \\
\hline 
\hline
\multicolumn{2}{|c|}{$n_f=3+1$} & \multicolumn{2}{|c|}{$n_f=4+1$} & \multicolumn{2}{|c|}{$n_f=5+3$}  \\ \hline
%%%%%%%%%%%%%

$C_{ \widetilde G}$&$(-7.6\pm3.8) \cdot 10^{-22}$&$[C_{uG}]_{22}$&$(-1.1\pm0.5) \cdot 10^{-20}$&$[C_{dG}]_{33}$&$(-1.5\pm0.8) \cdot 10^{-21}$\\
$[C_{dG}]_{11}$&$(2.1\pm1.0) \cdot 10^{-17}$&$[C_{u\gamma}]_{22}$&$(1.0\pm0.5) \cdot 10^{-21}$&$[C_{d\gamma}]_{33}$&$(-2.9\pm1.4) \cdot 10^{-23}$\\
$[C_{dG}]_{22}$&$(1.5\pm0.9) \cdot 10^{-20}$&$[C^{S1,RR}_{uu}]_{2222}$&$(7.8\pm3.9) \cdot 10^{-25}$&$[C^{V1,LR}_{uddu}]_{1331}$&$(-1.5\pm0.7) \cdot 10^{-21}$\\
$[C_{uG}]_{11}$&$(4.8\pm1.8) \cdot 10^{-17}$&$[C^{S8,RR}_{uu}]_{2222}$&$(-1.8\pm0.9) \cdot 10^{-25}$&$[C^{V8,LR}_{uddu}]_{1331}$&$(-8.1\pm2.0) \cdot 10^{-22}$\\
$[C_{d\gamma}]_{11}$&$(2.3\pm0.2) \cdot 10^{-17}$&$[C^{S1,RR}_{uu}]_{1122}$&$(5.7\pm2.9) \cdot 10^{-22}$&$[C^{V1,LR}_{dd}]_{1331}$&$(-6.1\pm3.7) \cdot 10^{-22}$\\
$[C_{d\gamma}]_{22}$&$(3.0\pm1.8) \cdot 10^{-19}$&$[C^{S8,RR}_{uu}]_{1122}$&$(3.8\pm8.2) \cdot 10^{-23}$&$[C^{V8,LR}_{dd}]_{1331}$&$(5.1\pm1.0) \cdot 10^{-22}$\\
$[C_{u\gamma}]_{11}$&$(-9.3\pm0.4) \cdot 10^{-17}$&$[C^{S1,RR}_{ud}]_{2211}$&$(2.9\pm1.3) \cdot 10^{-22}$&$[C^{V1,LR}_{dd}]_{2332}$&$(1.1\pm0.6) \cdot 10^{-24}$\\
$[C^{V1,LR}_{uddu}]_{1111}$&$(5.9\pm2.9) \cdot 10^{-22}$&$[C^{S8,RR}_{ud}]_{2211}$&$(1.2\pm0.4) \cdot 10^{-22}$&$[C^{V8,LR}_{dd}]_{2332}$&$(4.1\pm2.2) \cdot 10^{-24}$\\
$[C^{V8,LR}_{uddu}]_{1111}$&$(8.9\pm4.7) \cdot 10^{-22}$&$[C^{S1,RR}_{ud}]_{2222}$&$(1.5\pm1.4) \cdot 10^{-26}$&$[C^{S1,RR}_{dd}]_{3333}$&$(3.0\pm1.5) \cdot 10^{-25}$\\
$[C^{V1,LR}_{uddu}]_{1221}$&$(1.1\pm0.5) \cdot 10^{-21}$&$[C^{S8,RR}_{ud}]_{2222}$&$(5.1\pm2.8) \cdot 10^{-25}$&$[C^{S8,RR}_{dd}]_{3333}$&$(-6.4\pm3.1) \cdot 10^{-26}$\\
$[C^{V8,LR}_{uddu}]_{1221}$&$(1.7\pm0.8) \cdot 10^{-21}$&$[C^{S1,RR}_{uddu}]_{2112}$&$(-1.3\pm0.5) \cdot 10^{-21}$&$[C^{S1,RR}_{dd}]_{1133}$&$(3.8\pm1.9) \cdot 10^{-22}$\\
$[C^{V1,LR}_{dd}]_{1221}$&$(5.1\pm2.7) \cdot 10^{-22}$&$[C^{S8,RR}_{uddu}]_{2112}$&$(-3.6\pm5.8) \cdot 10^{-23}$&$[C^{S8,RR}_{dd}]_{1133}$&$(9.6\pm5.3) \cdot 10^{-23}$\\
$[C^{V8,LR}_{dd}]_{1221}$&$(7.8\pm4.0) \cdot 10^{-22}$&$[C^{S1,RR}_{uddu}]_{2222}$&$(-2.4\pm1.5) \cdot 10^{-24}$&$[C^{S1,RR}_{dd}]_{2233}$&$(1.3\pm0.8) \cdot 10^{-25}$\\
$[C^{S1,RR}_{uu}]_{1111}$&$(-1.3\pm0.7) \cdot 10^{-21}$&$[C^{S8,RR}_{uddu}]_{2222}$&$(-1.9\pm1.2) \cdot 10^{-24}$&$[C^{S8,RR}_{dd}]_{2233}$&$(-3.7\pm2.4) \cdot 10^{-25}$\\
$[C^{S8,RR}_{uu}]_{1111}$&$(4.1\pm2.6) \cdot 10^{-22}$&$[C^{S1,RR}_{uu}]_{1221}$&$(-1.8\pm1.2) \cdot 10^{-21}$&$[C^{S1,RR}_{dd}]_{1331}$&$(-2.3\pm1.1) \cdot 10^{-21}$\\
$[C^{S1,RR}_{dd}]_{1111}$&$(-6.2\pm3.6) \cdot 10^{-22}$&$[C^{S8,RR}_{uu}]_{1221}$&$(8.3\pm1.3) \cdot 10^{-22}$&$[C^{S8,RR}_{dd}]_{1331}$&$(5.2\pm1.6) \cdot 10^{-22}$\\
$[C^{S8,RR}_{dd}]_{1111}$&$(1.9\pm0.9) \cdot 10^{-22}$&$[C^{V1,LR}_{uddu}]_{2112}$&$(-2.5\pm0.9) \cdot 10^{-22}$&$[C^{S1,RR}_{dd}]_{2332}$&$(1.5\pm0.9) \cdot 10^{-24}$\\
$[C^{S1,RR}_{dd}]_{2222}$&$(1.0\pm0.6) \cdot 10^{-25}$&$[C^{V8,LR}_{uddu}]_{2112}$&$(-1.5\pm0.3) \cdot 10^{-22}$&$[C^{S8,RR}_{dd}]_{2332}$&$(2.9\pm1.7) \cdot 10^{-24}$\\
$[C^{S8,RR}_{dd}]_{2222}$&$(1.2\pm0.6) \cdot 10^{-25}$&$[C^{V1,LR}_{uddu}]_{2222}$&$(-8.1\pm4.9) \cdot 10^{-25}$&$[C^{S1,RR}_{ud}]_{1133}$&$(8.2\pm3.7) \cdot 10^{-22}$\\
$[C^{S1,RR}_{dd}]_{1122}$&$(-1.4\pm0.9) \cdot 10^{-22}$&$[C^{V8,LR}_{uddu}]_{2222}$&$(-2.8\pm1.6) \cdot 10^{-24}$&$[C^{S8,RR}_{ud}]_{1133}$&$(3.9\pm1.1) \cdot 10^{-22}$\\
$[C^{S8,RR}_{dd}]_{1122}$&$(5.1\pm2.3) \cdot 10^{-23}$&$[C^{V1,LR}_{uu}]_{1221}$&$(-2.9\pm18.4) \cdot 10^{-23}$&$[C^{S1,RR}_{uddu}]_{1331}$&$(-6.2\pm2.5) \cdot 10^{-21}$\\
$[C^{S1,RR}_{dd}]_{1221}$&$(-2.2\pm0.9) \cdot 10^{-22}$&$[C^{V8,LR}_{uu}]_{1221}$&$(9.4\pm0.6) \cdot 10^{-22}$&$[C^{S8,RR}_{uddu}]_{1331}$&$(-1.1\pm2.8) \cdot 10^{-22}$\\
$[C^{S8,RR}_{dd}]_{1221}$&$(6.0\pm2.6) \cdot 10^{-23}$&~&-&$[C^{S1,RR}_{ud}]_{2233}$&$(-2.1\pm1.2) \cdot 10^{-25}$\\
$[C^{S1,RR}_{ud}]_{1111}$&$(-4.8\pm2.5) \cdot 10^{-22}$&~&-&$[C^{S8,RR}_{ud}]_{2233}$&$(-6.8\pm3.3) \cdot 10^{-26}$\\
$[C^{S8,RR}_{ud}]_{1111}$&$(1.5\pm0.8) \cdot 10^{-22}$&~&-&$[C^{S1,RR}_{uddu}]_{2332}$&$(1.4\pm0.6) \cdot 10^{-24}$\\
$[C^{S1,RR}_{ud}]_{1122}$&$(-3.0\pm1.7) \cdot 10^{-22}$&~&-&$[C^{S8,RR}_{uddu}]_{2332}$&$(-1.7\pm0.9) \cdot 10^{-25}$\\
$[C^{S8,RR}_{ud}]_{1122}$&$(1.1\pm0.5) \cdot 10^{-22}$&~&-&$[C^{V1,LR}_{uddu}]_{2332}$&$(3.8\pm1.8) \cdot 10^{-25}$\\
$[C^{S1,RR}_{uddu}]_{1111}$&$(-4.9\pm2.5) \cdot 10^{-22}$&~&-&$[C^{V8,LR}_{uddu}]_{2332}$&$(-1.0\pm0.5) \cdot 10^{-25}$\\
$[C^{S8,RR}_{uddu}]_{1111}$&$(1.5\pm0.8) \cdot 10^{-22}$&~&-&~&-\\
$[C^{S1,RR}_{uddu}]_{1221}$&$(-5.1\pm1.8) \cdot 10^{-22}$&~&-&~&-\\
$[C^{S8,RR}_{uddu}]_{1221}$&$(9.7\pm5.3) \cdot 10^{-23}$&~&-&~&-\\

%%%%%%%%%%%%%
\hline
\end{tabular}
}
\caption{Master Formula for $d_p$ in WET in the JMS basis at the EW scale. The $\alpha_I^p$ have units of e-cm.$\rm TeV^{2}$ and 
e-cm.$\rm TeV$ for the four-fermion and dipole operators, respectively.}
\label{tab:masterdp}
\end{center}
\end{table}

\newpage
\addcontentsline{toc}{section}{References}

\small

\bibliographystyle{JHEP}
\bibliography{Bookallrefs}

\end{document}